\documentclass[onecolumn,journal,12pt]{IEEEtran}

\newcommand{\subparagraph}{}
\usepackage{graphicx}
\usepackage{amsfonts}
\usepackage{amsmath}
\usepackage{comment}
\usepackage{cite}
\usepackage{setspace}
\ifCLASSOPTIONonecolumn \doublespace \fi

\graphicspath{{Figures/}}

\def\urltilde{\kern -.15em\lower .7ex\hbox{\~{}}\kern .04em}

\begin{document}
\title{Quantifying Potential Energy Efficiency Gain\\ in Green Cellular Wireless Networks}
\author{\IEEEauthorblockN{Kemal Davaslioglu~\IEEEmembership{Student Member,~IEEE}} and \IEEEauthorblockN{Ender Ayanoglu~\IEEEmembership{Fellow,~IEEE}}\\
\ifCLASSOPTIONonecolumn
\IEEEauthorblockA{Center for Pervasive Communications and Computing\\
Department of Electrical Engineering and Computer Science\\
University of California, Irvine
}
\else
\thanks{The authors are with the Center for Pervasive Communications and Computing, Department of Electrical Engineering and Computer Science, University of California, Irvine, CA 92697-2625, USA.}
\thanks{This work was partially supported by the National Science Foundation under Grant No. 1307551. Any options, findings, and conclusions or recommendations expressed in this material are those of the authors and do not necessarily reflect the view of the National Science Foundation.}
\fi
}
\maketitle
\begin{abstract}
\boldmath
Conventional cellular wireless networks were designed with the purpose of providing high throughput for the user and high capacity for the service provider, without any provisions of energy efficiency. As a result, these networks have an enormous Carbon footprint. In this paper, we describe the sources of the inefficiencies in such networks. First we present results of the studies on how much Carbon footprint such networks generate. We also discuss how much more mobile traffic is expected to increase so that this Carbon footprint will even increase tremendously more. We then discuss specific sources of inefficiency and potential sources of improvement at the physical layer as well as at higher layers of the communication protocol hierarchy. In particular, considering that most of the energy inefficiency in cellular wireless networks is at the base stations, we discuss multi-tier networks and point to the potential of exploiting mobility patterns in order to use base station energy judiciously. We then investigate potential methods to reduce this inefficiency and quantify their individual contributions. By a consideration of the combination of all potential gains, we conclude that an improvement in energy consumption in cellular wireless networks by two orders of magnitude, or even more, is possible.
\end{abstract} 
\section{Introduction}\label{sec:intro}
The development of Information and Communication Technologies (ICT) within the last few decades has improved our lives tremendously, made information highly accessible, and increased productivity to unprecedented levels. It is expected that this trend will continue. However, this extraordinary improvement in our lives has a hidden cost. ICT employs computers, their peripherals, and communications equipment, all of which use energy, in many cases even when they are idle. As a result, energy consumption, and therefore the generation of greenhouse gases by this technology are already at very high levels. It is currently estimated that the ICT industry is responsible from about 2-4\% of all of the Carbon footprint generated by human activity \cite{ITU07,Prasad10}. This corresponds to about 25\% of all car emissions and is approximately equal to all airplane emissions in the world \cite[p. 80]{Prasad10}. This trend will only increase. Internet equipment manufacturer Cisco Systems Inc. estimated an annual growth rate of 70\% in global mobile data traffic in 2012 \cite{Cisco13}. Andrew Odlyzko of the University of Minnesota, who has been closely observing the growth of the Internet since the 1990s, estimated an annual Internet traffic growth rate of 40-50\% \cite{OdlyzkoJC}. With the proliferation of smart phones, video, and social networking, this rate of increase can be expected to be at least sustained for many years to come. As a result, serious concerns about the Carbon footprint impact of this development have been raised, and the topic of ``Green Communications" has been attracting attention in ICT circles (see, e.g., %\cite{RHanGreen,HBB11,BCRR12,Feng13,HBF12,WRZ12,MHLB08,MHL08,NESTFGWAN11,GreenLastMile11,MHLT12,EED12,Serrano12,ZZZ13,EEHetNet13,YIAA13,PredictiveGreen13,EEFemto13,TAEOGLTCS14}). 
\cite{RHanGreen,HBB11,LXXYZCX11,BCRR12,Feng13,HBF12,WRZ12,VC13,KKHM13,MHLB08,MHL08,NESTFGWAN11,GreenLastMile11,MHLT12,EED12,Serrano12,ZZZ13,EEHetNet13,YIAA13,PredictiveGreen13,EEFemto13,TAEOGLTCS14}).
Increasingly, more workshops, conferences, special issues of magazines and journals, and industry initiatives focus on the need to develop new approaches to communications and networking that result in drastically lower energy consumption. As the current communications and networking systems and protocols were not designed with this consideration, and since in many cases a greenfield approach needs to be taken, this effort will likely take a long time.

\begin{comment}
\begin{figure}[!t]
\begin{minipage}[b]{0.45\linewidth}
\centering
%\includegraphics[height=45mm]{wirelessdata1.eps}
\includegraphics[height=45mm]{mobile_data.eps}
\caption{Expected growth of mobile data traffic \cite{Cisco13}.}
\label{fig:wirelessdata1}
\end{minipage}
\hspace{5mm}
\begin{minipage}[b]{0.45\linewidth}
\centering
\includegraphics[height=45mm]{mobile_video.eps}
\caption{Expected growth of mobile video \cite{Cisco13}. An Exabyte is $10^{18}$ bytes.}
\label{fig:wirelessdata2}
\end{minipage}
\end{figure}
\end{comment}

In the case of wireless communications and networking, the increase in traffic is even more than the Internet as a whole. It is estimated that wireless data volume in the cellular segment alone is almost doubling annually. The study by Cisco Systems, Inc. quoted above states that the expected Compound Annual Growth Rate (CAGR) in the mobile data during 2012-2017 is 66\% \cite{Cisco13}. %This is depicted in Fig.~\ref{fig:wirelessdata1}.
As a result, global mobile data traffic in 2017 will be 13 times that of 2012. Although this growth is dramatic, even bigger growth rates have been observed in the past. For example, the United Kingdom-based service provider Telef\'{o}nica O$2$ reported that its mobile data traffic in Europe doubled every three months in 2009; Telecom Italia announced that its mobile traffic grew 216\% from mid-2008 to mid-2009; and AT\&T has reported that its mobile traffic increased 5000\% in 2008-2010 (see, e.g., \cite{Cisco10-2}).
%In this figure, a Terabyte is $10^{12}$ bytes and an Exabyte (EB) is $10^{18}$ bytes.

%As detailed in Fig.~\ref{fig:wirelessdata2}, about
About 2/3 of the traffic in 2017 is expected to be due to mobile video \cite{Cisco13}. This is somewhat surprising since technologists questioned the value of mobile video in the past. However, the development of smart phones and the unexpected evolution of social networking have changed this picture in a major way. The average smart phone user generates 50 times the amount of traffic generated by the basic-feature phone user, with tablets and laptops these are 120 and 368 times, respectively \cite{Cisco13}. Handset traffic is highest in regions with the highest smart phone penetration. In addition to smart phones and tablets, e-readers, wireless video cameras, laptop computers, and mobile phone projectors are expected to be responsible from this increase in wireless traffic. This enormous increase in traffic will place a similarly enormous burden on the wireless service infrastructure. New approaches for coming up with new technologies will certainly be necessary to address these expected needs.

Conventional designs of mobile wireless networks mainly focused on ubiquitous access and large capacity, or high throughput to the user, without any considerations of power or energy efficiency. But, with the inevitable constant use of ICT in our lives, there needs to be a serious thought given to increasing the energy efficiency in these technologies. All major ICT manufacturers, such as Cisco, %Systems, Inc.,
Ericsson, and Huawei, realize the importance of this problem and are planning to take steps in this direction.
%Examples are Cisco Systems, Inc. \cite{CiscoGreen}, Ericsson \cite{EricssonGreen}, Huawei \cite{HuaweiGreen}.
Alcatel-Lucent is leading a consortium with partners from industry and academia to generate a plan to reduce power consumption in ICT drastically, by three orders of magnitude, by the year 2015 \cite{GreenTouch}. Alcatel-Lucent states this is the result of an internal, not publicly available, study, which is based on Shannon theory and which shows four orders of magnitude reduction can be realized.

It may be possible to come up with a number of simple protocol changes to achieve energy savings such as introducing sleep cycles into existing protocols. An important example along those lines is the IEEE 802.3 Ethernet protocol. This protocol is commonly used to network personal computers, servers, and various peripherals at speeds of 10 or 100 Mb/s, or 1 or 10 Gb/s. In its original version, it was designed to transmit physical layer signals even when there is no information to be transferred. With the highly bursty nature of data communications, this is unnecessary and wasteful. The IEEE 802.3az version of the standard changed that by sending a Low-Power-Idle (LPI) indication signal for a specified time and consequently allowing the transmit chips in the system to be turned off. LPI is sent periodically to refresh the sleep mode. When there is data to transmit, a normal idle signal is sent to wake the transmit system up before data is due to be sent. The data link is considered to be always operational, as the receive signal circuit remains active even when the transmit path is in sleep mode. The standard was approved as IEEE Standard 802.3az-2010 at the September 2010 IEEE Standards Board meeting \cite{802.3az} and products are available from various manufacturers. The savings will be small at the onset due to the presence of legacy equipment but they are expected to rise to \$410M/year for the U.S. and over \$1B/year worldwide, exceeding the cost of current energy used by Ethernet physical layer components \cite{Christensen10}. In an attempt to introduce power savings for 802.11 Wi-Fi Wireless Local Area Networks (LANs), reference \cite{Haratcherev09} notices that most Wi-Fi Access Points (AP) are always on. It also makes the estimation that an active daily session for a home AP is about 91 minutes. It introduces a low-power sleep mode concept and shows through a prototype that average power could be reduced from 3.36 Watts to 2.48 Watts between the interface being on and off. A lot more power reductions are possible when there are extended periods of no use, e.g., when the residents are not home or when they are sleeping. Reference \cite{Haratcherev09} recognizes this and targets a 90\% power reduction as future work.

In this paper, we address the need for significant energy reduction in cellular wireless networks. After a literature survey, we first identify and discuss sources of potential and significant energy reduction in such networks. We discuss the application of a new access technique and a different network topology from the literature to eliminate existing major power losses. We also discuss a dynamic network architecture that will adapt to use patterns in order to reduce energy consumption. The dynamic nature of this architecture allows it to sense the environment and adapt. We present a physical layer technique to operate power amplifiers efficiently. We also discuss multiple antenna techniques for significant energy reduction in power amplifiers of Radio Base Stations (RBSs) and Remote Terminals (RT).

The rest of the paper is organized as follows. In Section~\ref{sec:survey}, we provide a brief review of the existing literature on this subject. In particular, we summarize a number of prior surveys and review books. In addition, in Section~\ref{sec:survey}, we briefly specify how our paper differs from the existing surveys. In Section~\ref{sec:losssources}, we discuss the sources of energy inefficiency in typical cellular wireless networks of today. Then, in Section~\ref{sec:methods}, a number of examples for hardware, software, and algorithmic changes and improvements to remedy the energy losses discussed in Section~\ref{sec:losssources} are given. In Section~\ref{sec:future} we provide a discussion of the potential challenges, lessons learned in the process of this research, and future work for the engineering community regarding energy-efficient next generation cellular wireless networks. Finally, in Section~\ref{sec:conclusion}, we provide a set of concluding arguments. We would like to note that, in this paper, we will use the terms cellular wireless and cellular interchangeably.  
\section{Surveys on Green Cellular Networks}\label{sec:survey}
The subject of energy efficiency in ICT industry, and in particular, in cellular communications, has been an active area of study and research since about 2007 or so, after the international community realized the large Carbon footprint of this industry \cite{ITU07}. Since then, a number of surveys and tutorials have been published. In this section, we will provide a summary of these, attempting to briefly point out their salient points. Although the period we will cover is quite short, we will attempt to keep this discussion historical by going from earlier works to later ones in sequence.

One of the first papers that presented a review of energy-efficient cellular networks is \cite{RHanGreen}. The authors of the paper are affiliated with a number of the universities in the United Kingdom, participating in the Mobile Virtual Centre of Excellence (VCE) Green Radio project, established in 2009 \cite{VCE}. The program has the ultimate goal of achieving two orders of magnitude reduction in power consumption for cellular networks, although it is not clear why two orders of magnitude was chosen. Reference \cite{RHanGreen} is a more modest attempt in this direction, targeting only a 50\% reduction in the power required to operate an RBS. The paper defines two measures, or metrics, of energy efficiency. One is in Joules per bit and equals the peak power divided by the maximum data throughput for an RBS. This measures energy efficiency of an RBS in absolute terms, standing on its own. A relative metric compares the energy consumption by a particular implementation against a standard one while the pertinent parameters such as throughput are kept equal. The paper introduces three case studies with potential to provide gains in energy efficiency. The first discusses a tradeoff in terms of bandwidth: In order to achieve a given throughput, one can choose a lower size constellation but employ a larger bandwidth, rather than keeping the bandwidth the same and increasing the size of the constellation thereby increasing the power consumption. The second case study has to do with interference cancellation. The authors show that different interference cancellation techniques have different values of energy consumption gain, or the second metric introduced above. Finally, the third case study shows the use of relays can aid in the energy efficiency of an RBS.

Reference \cite{HBB11} provides a survey of the field of energy-efficient cellular networks, or green cellular networks, with an emphasis on solutions employing cognitive radio and cooperative relaying. This paper provides a discussion of how to measure energy efficiency. It provides three main categories of energy efficiency metrics. These are categorized as facility-level, equipment-level, and network-level metrics.  The paper specifies two potential metrics in the first category, six in the second category, and two in the third category, stating that the list is non-exhaustive. Among the alternatives in a category, the paper does not provide any conclusive statements of which are preferred. The paper discusses RBS architectures for energy efficiency. First, it briefly discusses power amplifier alternatives. Then it specifies the sleeping modes for the 802.16e Mobile Worldwide Interoperability for Microwave Access (Mobile WiMAX) standard, and the Discontinuous Transmission (DTX) and Discontinuous Reception (DRX) alternatives for Long Term Evolution (LTE) systems. Besides these protocol changes, it also provides a discussion of cooperative RBS power management schemes. One technique is based on offloading the user traffic of heavily loaded RBSs to the neighboring ones and, conversely, turning off RBSs without much traffic; while another is based on changing the RBS cell size according to the traffic conditions\footnote{These techniques are sometimes termed cell breathing or cell zooming, but the terminology as to which is which is not consistent throughout the literature.}. The paper provides a discussion of a hierarchical network structure, however, unlike what we will present in the sequel, it does not discuss this concept on the basis of user mobility. A number of other topics are also studied, such as using renewable energy sources, especially in off-grid locations, low-energy spectrum sensing, energy-aware medium access control and routing, cross-layer design, user cooperation, cooperative relays, etc.

Reference \cite{LXXYZCX11} is unique in that it concentrates on relays for energy efficiency. It begins with a description of energy-efficient design of MIMO systems and relay systems. It concentrates on a discussion of the future directions for relays and cooperative relays for energy efficiency. It points out to the importance of additional overhead for relays considering both the additional time and the energy used. It brings up the point that most existing work focuses on point-to-point transmission. It states that point-to-multipoint and multipoint-to-point are important and makes the point that these topics are not as widely investigated as point-to-point. Another point this reference makes is the importance of bidirectional relays. It poses the research question of how to design an energy-efficient bidirectional relaying system.

Reference \cite{BCRR12} is a general classification of green networking, or specifically, green networking research. It is more general than green cellular or green wireless networking, but its general approach remains applicable to green wireless communications. The paper specifies four general areas various green networking efforts may fall under, a classification which it calls a {\em taxonomy.\/} It articulates each of the four parts of this taxonomy. It then identifies, for 22 different publications in networking, which of these areas they belong to. A particular publication may have aspects of more than one such class. The first of the four classes is on adaptive link rate. The authors of \cite{BCRR12} make a distinction between reaching a zero rate value, or a sleeping mode, versus a changing set of rates, which they call as a rate switch. They call the second class interface proxying. This can be thought of as subdividing a given task such that, instead of being performed at a central processor, it can be carried out at a number of peripheral units so that the power-hungry central processor is avoided. The remaining two classes cover potential implementations that do not fall under the first or the second classes discussed above. The third class involves hardware implementations or infrastructure while the fourth involves software implementations or applications. If adopted by the research community, this taxonomy will provide a standard way of classifying different attempts towards green networking. It applies to wireline or wireless networks.

Another survey on energy-efficient wireless communications appeared in \cite{Feng13}. This survey is different than \cite{RHanGreen,HBB11,LXXYZCX11,BCRR12} in a number of aspects. First, this survey discusses a number of international research projects for energy-efficient wireless communications, namely Green Radio, EARTH, OPERA-Net, and eWIN. These European research projects are analyzed in a table in \cite{Feng13} in terms of their energy metrics and models, their hardware, architecture, and resource management. In a similar fashion to \cite{RHanGreen} and \cite{HBB11}, this paper provides a discussion of metrics for energy efficiency. In this regard, the paper reaches at two conclusions. In the first, the paper states that the conventional approach to energy efficiency considers only transmit power consumption and criticizes this approach on the basis of other sources of power consumption such as circuit power consumption. In fact, as we will show in the sequel, even if one considers transmit power only, minimizing power and minimizing energy can lead to different conclusions. We will show that what matters is energy, and that minimization of power in a transmission system can result in a suboptimal solution for energy. The paper also states, based on its earlier observation that transmit power is not the only source of power use in a communication system, that a metric that covers the overall system is needed. The paper then discusses energy-efficient radio resource management and energy-efficient network deployment. In the former, the paper emphasizes, in addition to traffic load variations, the importance of Quality-of-Service (QoS) considerations. We agree on the importance of QoS, however, due to space limitations, in the development we will not discuss QoS in detail. In a similar fashion to \cite{HBB11}, this reference discusses hierarchical cell structures but does not elaborate on mobility considerations. Another strength of \cite{Feng13} is its discussion of future work in all of its sections.

The subject of energy-efficient cellular wireless communications is currently very active. To that end, in addition to the survey papers discussed above, a number of books on the subject have appeared, e.g., \cite{HBF12,WRZ12}. These two references are compilations of a number of research papers. This field is currently at its infancy. It can be expected that as the subject matures, there will be more surveys and books on this subject.

On the technical topics it covers, our paper provides more technical detail than \cite{RHanGreen,HBB11,LXXYZCX11,BCRR12,Feng13} which tend to provide higher level discussions. We first discuss sources of energy inefficiency and then their potential solutions. We summarize a number of algorithms and their quantitative results from the literature. We make a number of specific suggestions and recommendations, and point out potential challenges. On the other hand, we believe the technical discussion in our paper is more condensed than the edited contributions in \cite{HBF12,WRZ12}. Furthermore, our paper has the advantage of being more recent and we believe it provides a more up-to-date discussion of the subject compared to \cite{RHanGreen,HBB11,LXXYZCX11,BCRR12,Feng13,HBF12,WRZ12}. In passing, we would also like to point out to two surveys that concentrate on the energy efficiency of the RT. These are \cite{VC13} and \cite{KKHM13}. Since most of the energy efficiency of a cellular wireless network is in the RBSs, we concentrate our attention mostly on them. However, we will discuss an important issue regarding the energy efficiency of RTs, called the energy trap, in Section~\ref{sec:future}. 
\section{Sources of Energy Inefficiency in Cellular Networks}\label{sec:losssources}
\begin{comment}
\begin{figure}[!t]
\begin{minipage}[b]{0.3\linewidth}
\centering
%\includegraphics[height=38mm]{MtCO2e02.eps}
\includegraphics[height=33mm]{MtCO2e02.eps}
\vspace{1mm}
\caption{Percentage of MtCO$_2$e contributions in 2002.}
\label{fig:MtCO2e02}
\end{minipage}
\hspace{1.6mm}
\begin{minipage}[b]{0.3\linewidth}
\centering
%\includegraphics[height=38mm]{MtCO2e20.eps}
\includegraphics[height=33mm]{MtCO2e20.eps}
\vspace{0.2mm}
\caption{Percentage of MtCO$_2$e contributions in 2020.}
\label{fig:MtCO2e20}
\end{minipage}
\hspace{5mm}
\begin{minipage}[b]{0.3\linewidth}
\centering
%\mbox{\hspace{-5mm}}\includegraphics[height=46mm]{PAPower.eps}
\mbox{\hspace{-3mm}}\includegraphics[height=40mm]{PAPower.eps}\\
\vspace{-3mm}
\caption{Distribution of power consumption in a cellular RBS.}
\label{fig:PAPower}
\end{minipage}
\end{figure}
\end{comment}
In 2002, telecommunications infrastructure and devices (not including data servers, PCs, and peripherals) were responsible from 151 Metric tonne (ton) CO$_2$ equivalent (MtCO$_2$e) emissions, with cellular networks contributing 64 Mt (42\%) and mobile phones contributing 16 Mt (11\%). %, as depicted in Fig.~\ref{fig:MtCO2e02}.
In 2020, the telecommunications infrastructure and services contribution will more than double and reach 349 MtCO$_2$e with mobile cellular networks contributing 179 Mt (51\%) and mobile phones 22 Mt (6\%)
%as depicted in Fig.~\ref{fig:MtCO2e20}
\cite{Barth09}. This shows three things. First, the contribution due to cellular networks and mobile phones will more than double by 2020 in absolute terms. Second, this contribution is actually increasing in relative terms. And third, the contribution of cellular networks is far greater than mobile phones. Reference \cite{Barth09} states 80\% of the total power in cellular networks is consumed at RBSs. Furthermore, \cite{Barth09} states that, in a cellular RBS, the Power Amplifier (PA), together with its feeder, consumes 50-80\% of the total power, the remainder going to cooling (10-25\%), signal processing (5-15\%), and power supply (5-10\%).
%, as shown in Fig.~\ref{fig:PAPower}.
The distributions quoted by a number of other sources are similar (see, e.g., \cite{Grant09}). As a result of the discussion %and the quoted figures
above, it makes sense to target energy efficiency in cellular RBSs and particularly for issues related to the PA, as the highest priority in research in this area. Conventionally, in cellular networks, RBSs are designed with the goal of maximizing capacity or user throughput, and reducing capital expenses and operating expenses (commonly abbreviated as CapEx and OpEx, respectively). However, because of the increasing amount of traffic, OpEx are constantly increasing. The critical point is that increasing the user throughput makes conventional RBSs operate in an inefficient mode. This is a problem that needs to be solved. To make matters a lot worse, newer generations of cellular wireless technologies are promising increasingly higher transmission speeds at higher distances from the RBS. In order to satisfy these needs, RBS PAs need to be operated in a linear mode where they are highly inefficient and consume a lot more power. As a result, it is quite clear that there is a need for a major change in the design of base RBSs of the future.

There have been a number of generations of cellular wireless communications technologies. The 1st Generation (1G) was for analog voice. For the 2nd Generation (2G), two major standards were developed, one based on Time Division Multiple Access, known as Global System for Mobile Communications (GSM), and the other based on Code Division Multiple Access, known as CDMA. For both standards, the 3rd Generation (3G) has evolved through several intermediate steps with a number of different names. In this process, while CDMA-based technologies were still known in general as CDMA, GSM evolved into a technology known as Wideband CDMA (WCDMA). Originally, 2G was for digital voice only and 3G technologies offered relatively low data transmission rates of about 1-2 Mb/s. In time, both have evolved to incorporate data transmission with increasingly higher rates, with most sophisticated versions of 3G offering tens of Mb/s, although being restricted in terms of the number of simultaneous users and further evolution.
Known as 4th Generation (4G), two new standards have emerged to replace 2G and 3G. Both of these standards have very aggressive goals for transmission rates to be offered to users. The first of these is developed around the IEEE 802.16 standard and is known as WiMAX. The first version of WiMAX can provide up to 40 Mb/s to the user. The next version
is poised to offer up to 1 Gb/s fixed (not mobile) access. The second standard, developed through an industry consortium, is known as LTE. The term it stands for, Long Term Evolution, is intended to mean the evolution of the GSM/WCDMA technology, which has the largest user base in the world.
LTE is designed to provide up to 50 Mb/s on the uplink and 100 Mb/s on the downlink. Its advanced version LTE-Advanced is planned to provide up to 1 Gb/s fixed access. In order to provide these kinds of transmission rates, both WiMAX and LTE employ sophisticated techniques with modulation schemes employing large constellation sizes and
%, like WiMAX,
a broadband wireless technology known as Orthogonal Frequency Division Multiplexing (OFDM). OFDM removes the need for long training and equalization periods to overcome the limitations of a frequency selective channel and is therefore without competition when broadband wireless services are to be offered. However, the current direction of offering OFDM via legacy RBSs covering a range of a few kms radius and with aggressive transmission rates of 10s or even 100s of Mb/s to the users is questionable, especially from an energy consumption viewpoint. There is no doubt that this push will enormously increase energy consumption in cellular wireless systems, which is already at alarmingly high levels.

The bulk of the energy inefficiency in cellular wireless networks is in RBSs. Some of these problems are due to inefficient cooling or the use of energy sources outside the energy grid, as discussed in \cite{HBB11}. This paper concentrates on the sources of inefficiency in communication protocol layers and therefore such problems and their solutions are outside the scope of this paper. In the rest of this section, we first investigate the problems due to the physical layer. We study the impact of modulation techniques on PAs in Section~\ref{sec:papr} and identify the critical problems. We then investigate a broad range of problems on the mismatch of the cellular wireless system to the instantaneous traffic. The traffic changes with time and with mobility, while currently the cellular wireless system is designed to operate at peak traffic rates at every part of the system. Inevitably, this is a major cause of inefficiency. We discuss this topic in Section~\ref{sec:deploymenttraffic}. Finally, these inefficiencies cause problems not only for the environment and therefore the public at large, but also, for the service provider. We elaborate on this topic in Section~\ref{sec:rbsconsumption}.

\subsection{The Impact of Modulation on Power Amplifiers}\label{sec:papr}
OFDM transmits signals over multiple subcarriers simultaneously. The advantage of this can be thought of as dividing a broadband spectrum into smaller bands, each with a lot less frequency response variation. As a result, the requirement of equalizing the broadband frequency spectrum is substantially simplified. However, the multi-carrier nature of OFDM also has drawbacks. One of these drawbacks is the increased Peak-to-Average Power Ratio (PAPR). A peak in the signal occurs when all, or most, of the subcarriers are aligned in phase. The worst-case value of the PAPR is directly proportional to the number of carriers and is given by $\textrm{PAPR}=10 \log N$ where $N$ is the number of subcarriers \cite{Wight01,Tellambura}. For example, in the commonly used IEEE 802.11g Wi-Fi wireless LAN standard, there are $N=64$ subcarriers. In reality only 52 are used for data transmission, but for this argument and for simplicity we can assume all $N=64$ are. This means a worst-case PAPR of about 18 dB results. The PA must be able to accommodate this peak power as well as lower power levels, at least by an amount equal to PAPR. Typically, the operating point of the PA is set at the average level and the peak point is set at a level higher by an amount corresponding to PAPR in dB. This is the power level the PA loses its linear behavior. This is called power backoff. Backing off the power in this manner causes PAs to be more expensive since they should be able to accommodate a wider power range in their linear operating regions. On the other hand, linear power amplification is highly power-inefficient. As a result, power consumption of the PA increases substantially, resulting in poor power efficiency of the overall system.

\begin{figure}[!t]
\vspace{1mm}
\begin{center}
\ifCLASSOPTIONonecolumn
\begin{minipage}[b]{0.6\linewidth}
\else
\begin{minipage}[b]{1.0\linewidth}
\fi
\centering
\includegraphics[width=60mm]{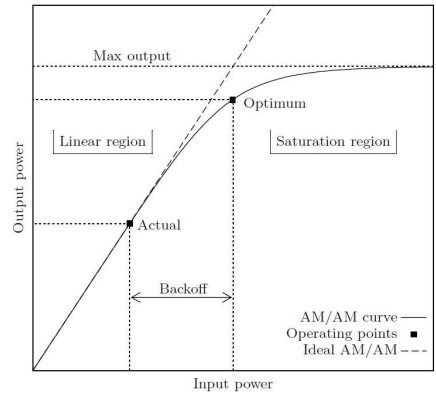}
\caption{Power amplifier transfer function and power backoff \cite{Thompson05}.}
\label{fig:Backoff}
\end{minipage}
\end{center}
\vspace{-7mm}
\end{figure}

A representation of backoff is depicted in Fig.~\ref{fig:Backoff} \cite{Thompson05}. Ideally, the output of the PA is equal to the input times a gain factor. In reality, the PA has a limited linear region, beyond which it saturates to a maximum output power level. Fig.~\ref{fig:Backoff} shows a representative input/output curve, known as the AM/AM conversion. The most efficient operating point is actually the saturation point of the PA, but for signals with large PAPR, the operating point must shift to the lower power region on the left, keeping the amplification linear. The average output power is reduced by an amount equal the power backoff value. To keep the peak power of the input signal less than or equal to the input saturation level, the power backoff must at least be equal to the PAPR. As discussed in \cite{Thompson05}, for an OFDM system with $N=16$, the PAPR and therefore the required power backoff is about 9.5 dB, excluding the subcarriers that do not carry data (or about 12 dB including all of $N=16$ subcarriers). At this point, the efficiency of a Class A PA is less than 6\%. This is going to get much worse when the number of subcarriers increases. We would like to note that both WiMAX and LTE have incorporated larger numbers of subcarriers, e.g., $N=2048$ for a $20$ $\mathrm{MHz}$ system bandwidth. This increases PAPR to approximately 33 dB. Although the worst case PAPR is uncommon and practical PAPR values are several dBs below the worst-case value of $10\log N$, the change in practical PAPR values is still linear with the number of subcarriers $N$ \cite{Goldsmith}, and therefore the increase in the number of subcarriers from 64 to 2048 will have a substantial negative impact on power efficiency of OFDM PAs, especially at the RBSs.

In addition, PAPR also increases slightly with increasing the constellation size \cite{Tellambura}. However, increasing the constellation size increases the power consumption in a PA. For example, it is well-known that for bandlimited channels, for modulation techniques such as Quadrature Amplitude Modulation (QAM), and for the same Bit Error Rate (BER), as the constellation size increases, a larger value of bit energy is needed \cite{Proakis}. Therefore, increasing the constellation size requires an increased power level at the receiver. The received power in a wireless system is proportional to the transmitted power, based on the distance between the transmitter and the receiver as well as a descriptor of the channel known as path loss. As a result, increasing the constellation size increases the power consumption, and therefore the power inefficiency, at the transmitter. In most cases, the transmission rate from the RBS to the RT will be higher. Furthermore, as discussed earlier, the power inefficiency at the RBS is the more significant problem. QAM is employed in each subcarrier of an OFDM system and therefore the conclusion that increasing the constellation size increases the power inefficiency holds for OFDM. The OFDM techniques used in WiMAX and LTE employ channel coding. With channel coding, the conclusion that an increase in constellation size increases the power inefficiency still holds. In order to see this, we refer the reader to simulation results of a system similar to that used in WiMAX as well as LTE, including channel coding, in \cite{Ayanoglu01}, or in \cite{Ayanoglu00} with clearer plots.

We will discuss in the sequel that one of the potential remedies of traffic increase is to make the cells smaller. This makes cells more power-limited than bandwidth-limited \cite{Proakis}.
Communication theory indicates that in the case of power-limited channels, modulation techniques optimum in the case of bandwidth-limited channels, such as QAM or Phase Shift Keying (PSK) are actually not optimal. In this case, the optimum modulation technique is known as orthogonal signaling where each message is assigned a distinct signal, all of these signals are orthogonal to each other, and only one signal is active at a given time \cite{Proakis}. Although the word orthogonal appears in the expansion of the acronym OFDM, OFDM is not orthogonal in this particular sense because all of its orthogonal signals are active simultaneously. An example of achieving this orthogonality is Frequency Shift Keying (FSK) where the signals are sinusoids with carrier frequencies spaced apart so that they are orthogonal over a symbol period. One can employ $M$ orthogonal signals to generate a modulation technique known as MFSK. The resulting signal will have continuous amplitude (or envelope) and it is desirable to make such signals have continuous phase as it leads to better performance when nonlinear, efficient PAs are used to transmit them. The tradeoff for this power advantage is the increased bandwidth due to orthogonality.  In wireless sensor networks where transmission distances are very short, of the order of meters, and transmission rates are very low, of the order of bits per second, interesting tradeoffs exist that can favor MFSK together with transmission time tradeoffs and with or without channel coding (see, e.g., \cite{CGB05}). These techniques can indeed be applicable to wireless sensor networks due to short distance transmission and therefore the possibility of reusing the frequency spectrum over space as well as low bit rates. However, for cellular transmission, the use of MFSK would be inefficient, even prohibitive, because of the bandwidth tradeoff. However, a constant envelope version of OFDM could be an interesting choice. There are recent developments in a constant amplitude version of OFDM as we will discuss in Section~\ref{sec:CEOFDM}.

\subsection{Deployment, Frequency Plan, Mobility, and Traffic Patterns}\label{sec:deploymenttraffic}
Reference \cite{CZXL11} states a fundamental observation in realizing energy-efficient networks is the tradeoff between energy and deployment. %We agree that this
This is a very important issue. One of the most significant inefficiencies in cellular networks today has to do with RBS deployment. Therefore, we will next summarize the way cellular networks are deployed (for more information, see, e.g., \cite{Rappaport}).
\begin{comment}
\begin{figure}[!t]
\begin{center}
\begin{minipage}[b]{0.35\linewidth}
\centering
\mbox{\hspace{1mm}}\includegraphics[width=50mm]{reuse.eps}
\caption{Frequency reuse.}
\label{fig:reuse}
\end{minipage}
\hspace{15mm}
\begin{minipage}[b]{0.35\linewidth}
\centering
\mbox{\hspace{-5mm}}\includegraphics[width=50mm]{sector.eps}
\caption{Sectoring.}
\label{fig:sector}
\end{minipage}
\end{center}
\end{figure}
\end{comment}
%Early mobile telephony systems consisted of a single, high powered transmitter with an antenna mounted on a tall tower. This provided wide coverage but could only serve a small number of channels because it did not allow for the reuse of the spectrum at different physical locations in a wide area. The concept of a cellular network was introduced to solve this problem. This concept replaces a single, high power transmitter (which can be interpreted as a large cell or a mega cell) with many low power transmitters (or small cells), each providing coverage to a small part of the service area. Every cell has an RBS at its center.
Today's mobile telephony systems are deployed in a cellular structure. In this structure, every cell has an RBS at its center. Each RBS is allocated a portion of the total number of channels so that all the available channels are assigned to a small number of neighboring RBSs. Neighbor RBSs are assigned different groups of channels so that the interference between RBSs (and RTs in the area they control) is minimized. As a result, all of the available channels are reused throughout a coverage region.
%Each RBS is allocated a group of channels to be used within a small geographic area, technically called a {\em cell.\/} Fig.~\ref{fig:reuse} illustrates the concept of frequency reuse. Each of the hexagonal cells labeled from A through E is allocated a set of different frequency channels.
The distances are chosen such that the interference among cells with the same frequency designation can be kept to a minimum.
%The hexagonal cell shapes are conceptual and in reality frequency planning is carried out, taking the effects of the terrain and shadowing due to buildings.
In order to reduce interference among cells further, a technique known as
{\em sectoring\/} is employed. In this technique, each cell is further subdivided into sectors. %, as shown in Fig.~\ref{fig:sector} with 120 degrees.
Each RBS employs a directional antenna transmitting and receiving only within the sector angle. As a result, interference among RBSs is reduced.
%For example, in the 120$^\circ$ configuration shown in Fig.~\ref{fig:sector}, sectoring reduces the number of interfering RBSs from 6 to 2.

At this point, we would like to emphasize that the ideas of cellular networks and frequency reuse are solid as far as efficient energy consumption is concerned. The problem, especially looking into the future, lies with the cell radii, available PA technology, intended transmission rates, and user mobility. Although transmission rates demanded by users in a fixed location or moving at pedestrian speeds may be large, transmission rates demanded by users moving at vehicular speeds will always be limited. Users who are in public transport vehicles such as trains, buses, or planes can be served by in-vehicle APs with high-rate connections to the backbone via wireline or, for example, satellite connections.
%Users driving cars cannot demand transmission rates as high as multiple Mb/s or higher for safety concerns. As a result, trying to serve a limited number of users in automobiles demanding very high rates using the cellular network can be avoided. Such users can also be served via satellite, removing a big burden on the cellular network.
Users driving cars cannot demand transmission rates as high as multiple Mb/s or higher for safety concerns. There may be other users in cars who demand high transmission rates, but due to the potential cost of the infrastructure necessary to offer this service, it is likely that their numbers will be much smaller as compared to users who are stationary or are moving at pedestrian speeds demanding such transmission rates. As a result, trying to serve a large number of users in automobiles demanding very high rates using the cellular network can be avoided.  Instead, most users demanding high transmission rates can be served by an overlay network covering only areas of such demand, or hot spots. As an example, we would like to quote the two-tier network, distinct for highly mobile users (in macrocells) and less mobile users (in microcells or picocells), as in Fig.~\ref{fig:umbrella} \cite{Rappaport}. Although this idea is appreciated by the cellular research and development community, it has actually not been fully developed or implemented.
\ifCLASSOPTIONonecolumn
\begin{figure}[!t]
\begin{center}
\begin{minipage}[b]{0.45\linewidth}
\centering
\includegraphics[width=80mm]{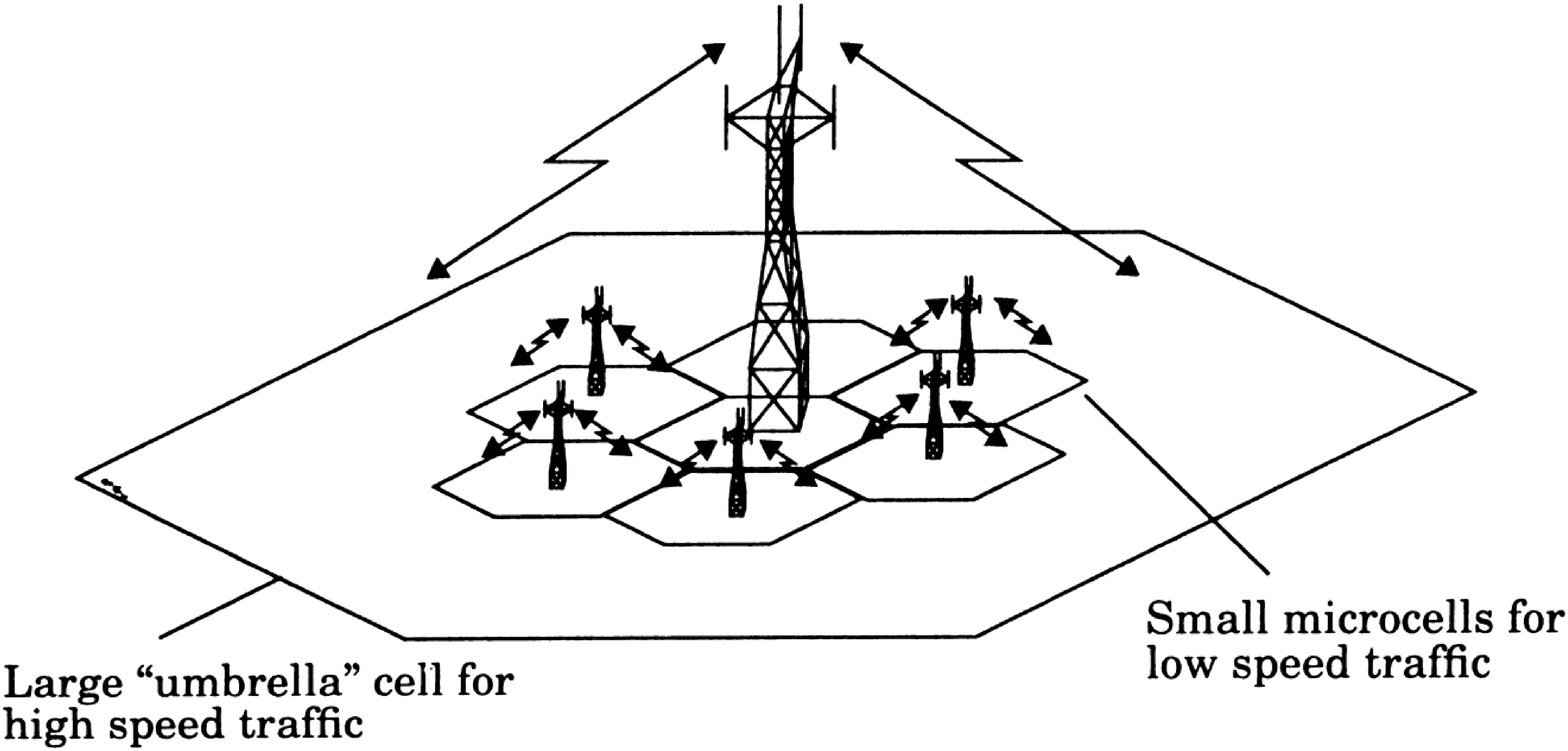}
\vspace{10mm}
\caption{Umbrella network \cite{Rappaport}.}
%\includegraphics[width=80mm]{umbrella.eps}
%\includegraphics[width=80mm]{umbrella_network.eps}
%\vspace{10mm}
%\caption{Umbrella network.}
\label{fig:umbrella}
\end{minipage}
\hspace{5mm}
\begin{minipage}[b]{0.45\linewidth}
\centering
\includegraphics[width=80mm]{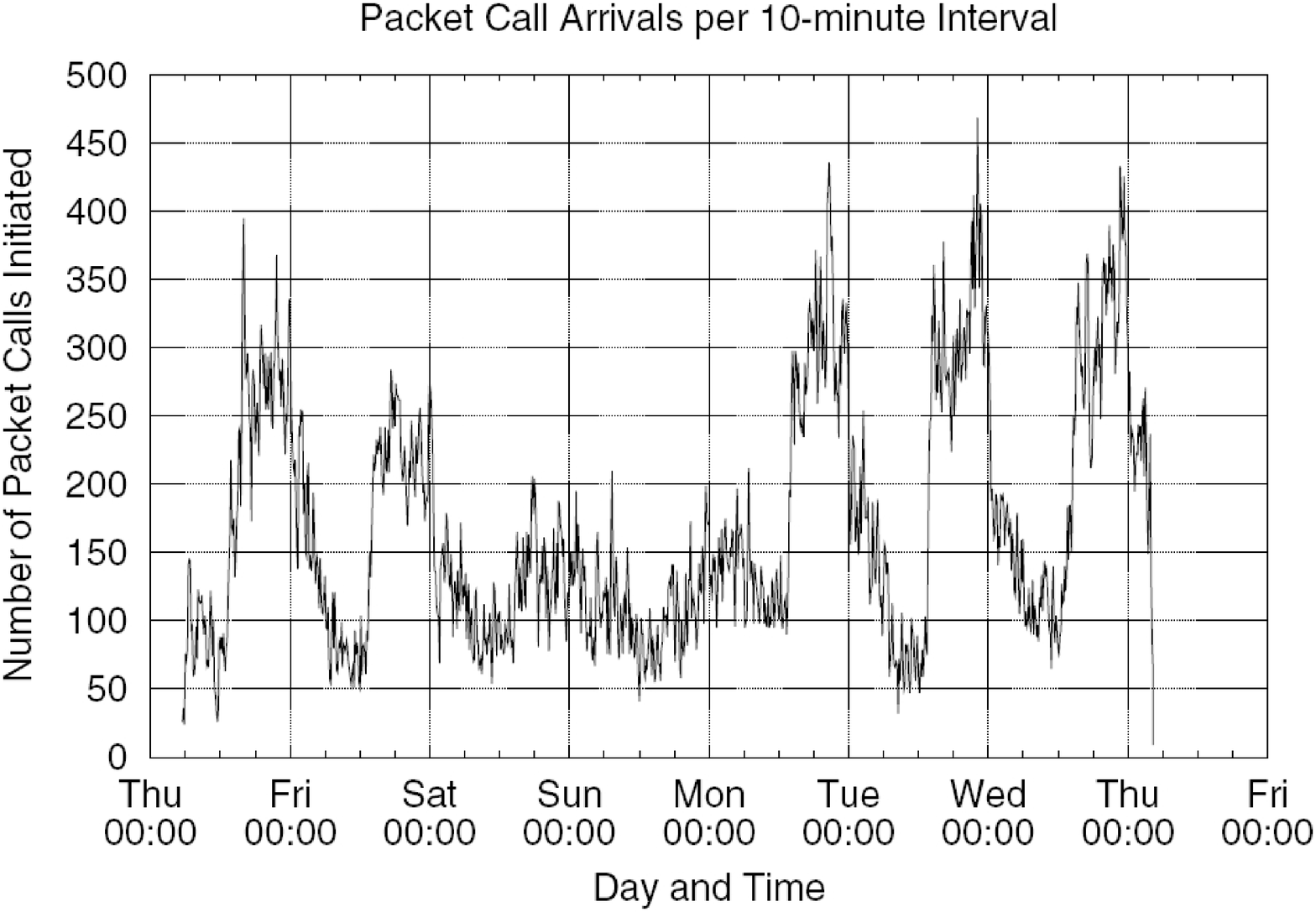}
\caption{Weekly traffic pattern at a cellular RBS \cite{WHSW05}.}%
\label{fig:traffic}
\end{minipage}
\end{center}
\end{figure}
\else
\begin{figure}
\vspace{2mm}
\begin{center}
\begin{minipage}[b]{1.0\linewidth}
\centering
\includegraphics[width=80mm]{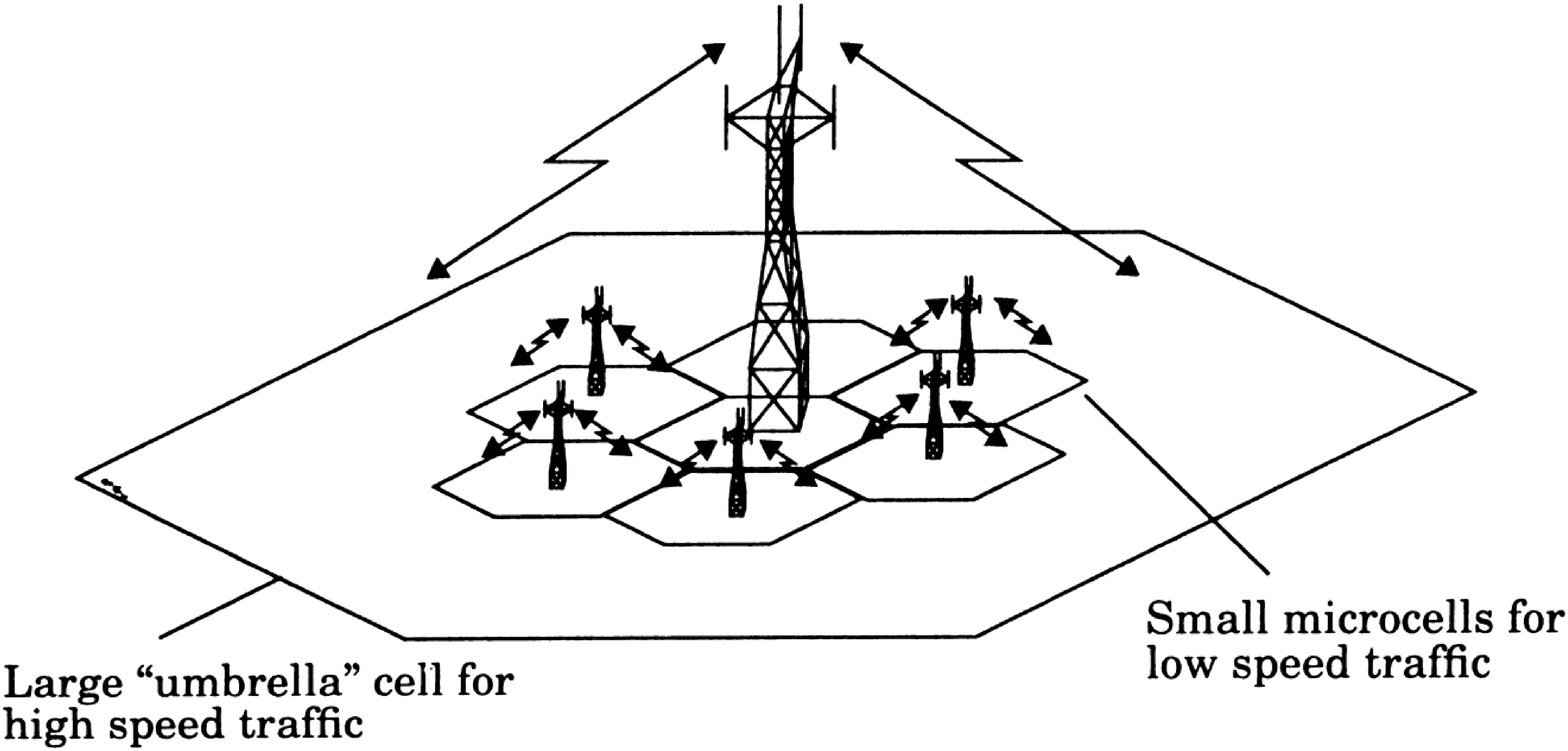}
\caption{Umbrella network \cite{Rappaport}.}
%\includegraphics[width=80mm]{umbrella.eps}
%\includegraphics[width=80mm]{umbrella_network.eps}
%\caption{Umbrella network.}
\label{fig:umbrella}
\end{minipage}
\end{center}
\vspace{-1mm}
\end{figure}
\begin{figure}
\vspace{-1mm}
\begin{center}
\begin{minipage}[b]{1.0\linewidth}
\centering
\includegraphics[width=80mm]{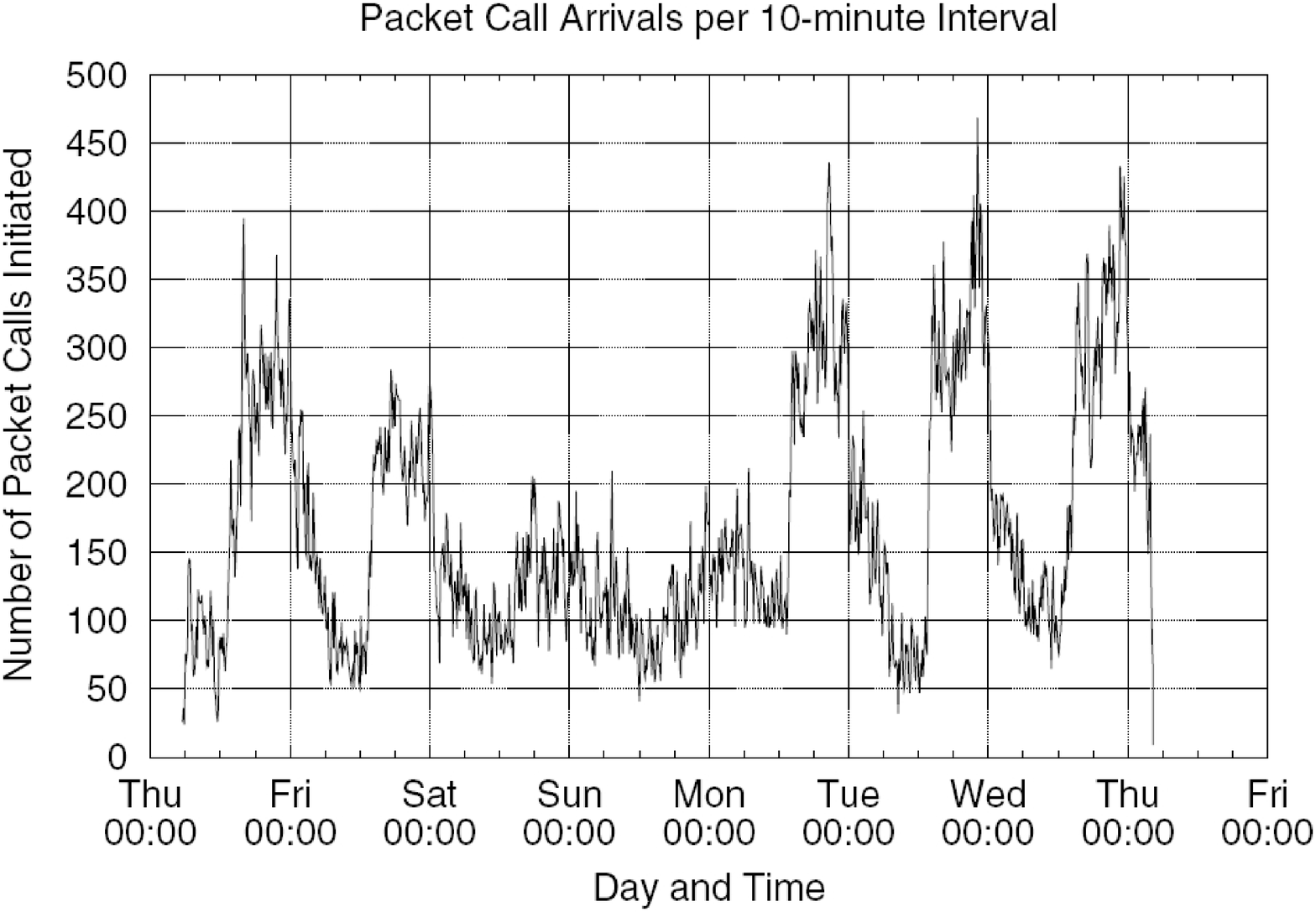}
\vspace{-2.5mm}
\caption{Weekly traffic pattern at a cellular RBS \cite{WHSW05}.}%
\label{fig:traffic}
\end{minipage}
\end{center}
\vspace{-2mm}
\end{figure}
\fi

There is some data available on cellular network traffic, e.g., \cite{WHSW05,TRKP10,Correia10,HowMuchEnergy11,OKLN11}, \cite[p. 250]{HT09}. As can be observed in Fig.~\ref{fig:traffic}, cellular network traffic shows a great deal of variation during the day and also, according to the day of the week. The traffic is time-varying, with heavier traffic occurring during weekdays as compared to weekends. Overall, there are significant periods of low network utilization. Heaviest traffic occurs during late afternoon and evening hours. This is different than the wired Internet where traffic more closely aligns with normal working hours. The ratio of the heaviest traffic volume to the lightest is of the order of about 2 in \cite{WHSW05,HT09}, and significantly higher (about 6) in \cite{TRKP10}. Currently, this is not exploited, and RBSs remain idle during these periods. In idle periods, RBSs keep consuming substantial power. %A more detailed version of Fig.~\ref{fig:PAPower} is given in Fig.~\ref{fig:grant}. This figure
Fig.~\ref{fig:grant} shows that 19\% of the power consumed is wasted in transmitter idling. This is potentially a substantial source of improvement.
\ifCLASSOPTIONonecolumn
\begin{figure}[!t]
\begin{center}
\begin{minipage}[b]{0.6\linewidth}
\centering
\hspace{6mm}\includegraphics[width=80mm]{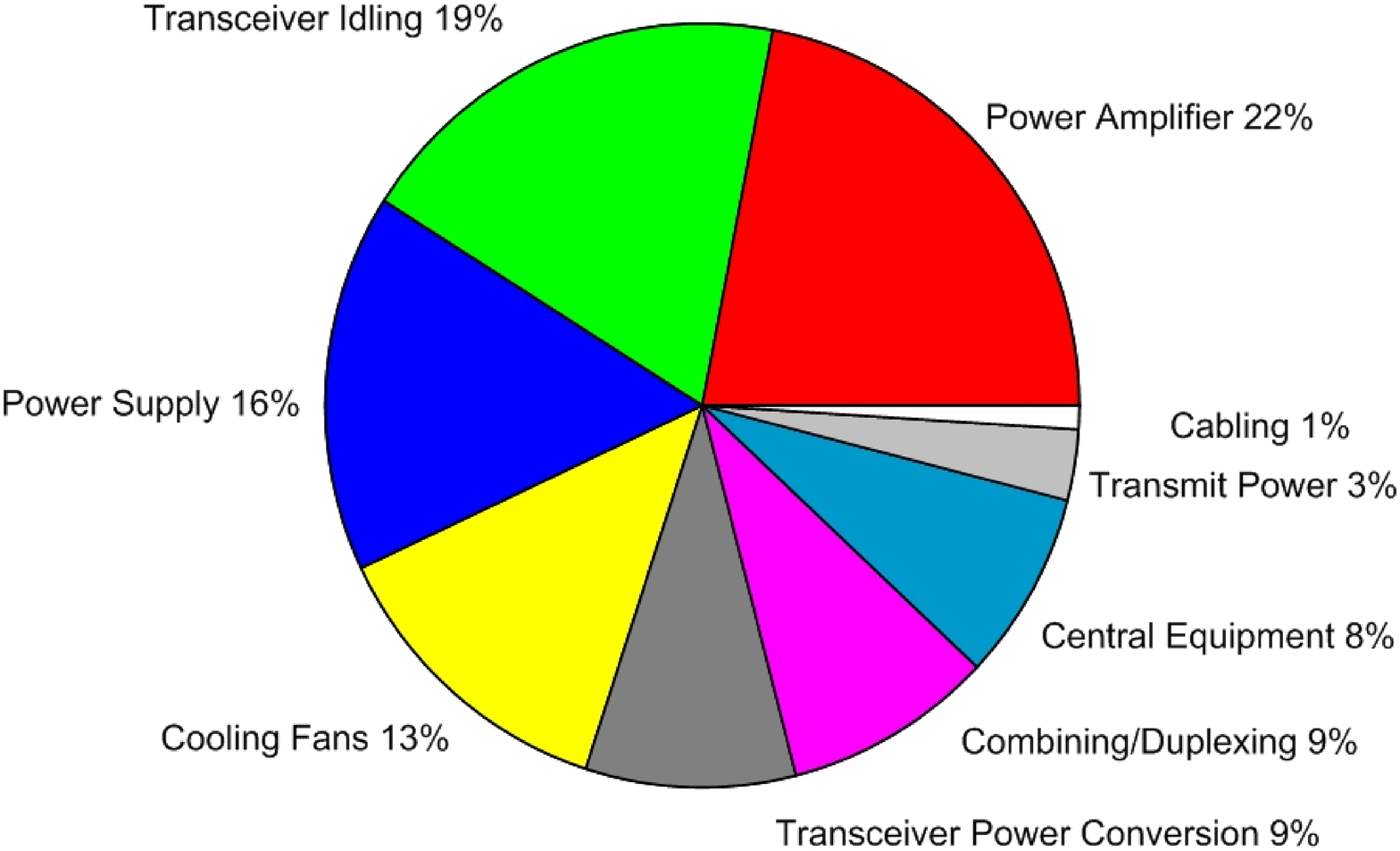}
\caption{Distribution of power consumption in a cellular RBS \cite{Grant09,Karl03}.}
\label{fig:grant}
\end{minipage}
\end{center}
\end{figure}
\else
\begin{figure}[!t]
\vspace{1mm}
\begin{center}
\begin{minipage}[b]{1.0\linewidth}
\centering
\includegraphics[width=80mm]{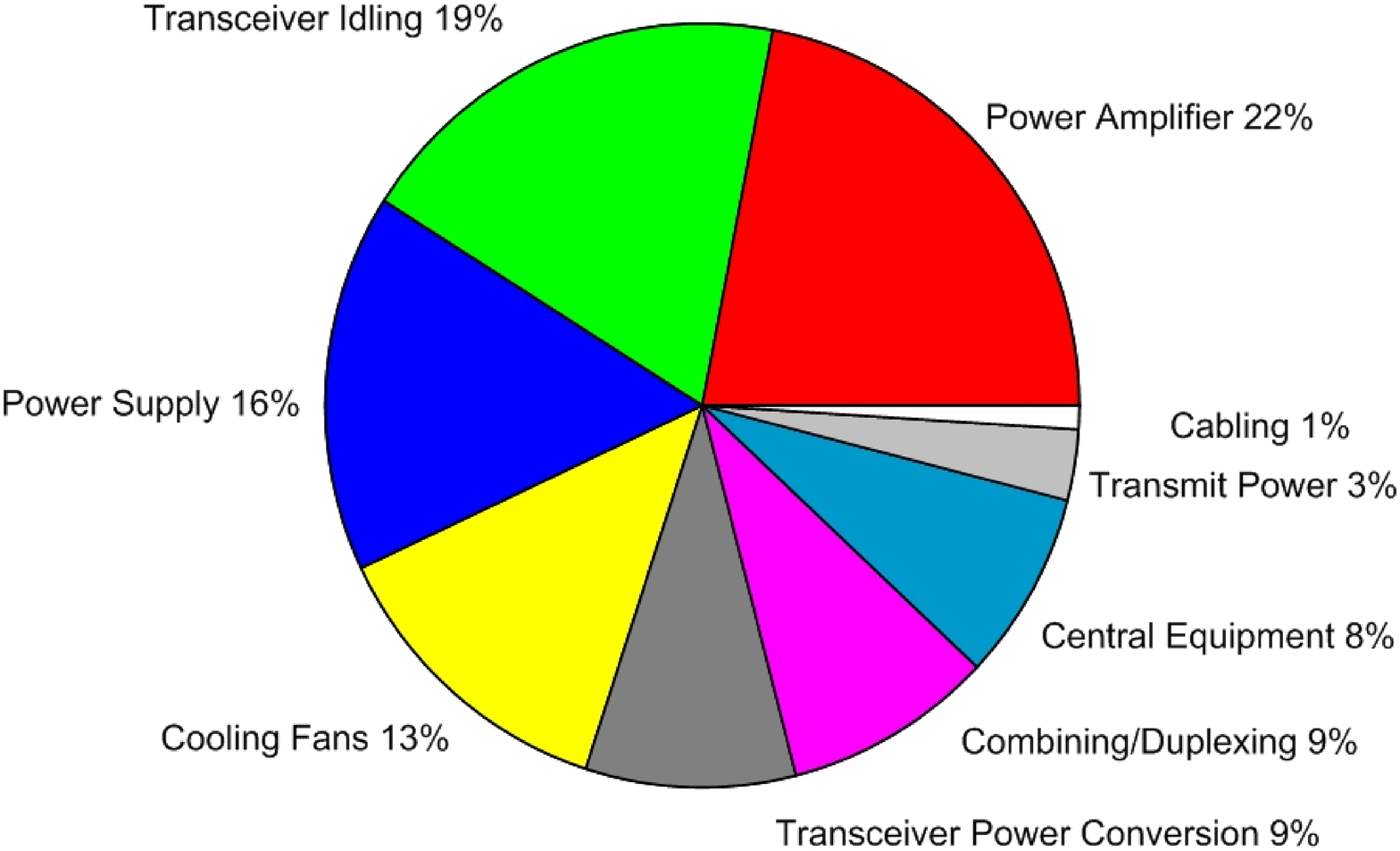}
\caption{Distribution of power consumption in a cellular RBS \cite{Grant09,Karl03}.}
\label{fig:grant}
\end{minipage}
\end{center}
\vspace{-5mm}
\end{figure}
\fi

In addition, there is traffic variation with respect to location as well. In this regard, there is a large savings potential, and not only for periods of low traffic. According to an Alcatel-Lucent study, typically 10\% of the sites carry 50\% of all traffic; in addition, 50\% of sites are lightly loaded, carrying only 5\% of the traffic \cite{Barth09}. Currently, this difference is also not exploited.

\subsection{RBS Energy Consumption and Its Effects for the Service Provider}\label{sec:rbsconsumption}
The cellular RBS is the most energy-consuming component in a cellular network. A typical 3G RBS is highly inefficient. For about 40 W output power, it consumes about 500 W of power, corresponding to only 8\% efficiency \cite{Grant09}. Consequently, the annual energy consumption of an RBS is around 4.4 MWh. It has been reported that there were approximately 52,500 (52.5K) RBS sites at the end of 2009 in the United Kingdom \cite{MOA}. This results in an energy consumption of 230 GWh per year by the United Kingdom cellular network, corresponding to 165K MtCO$_2$e, or the CO$_2$ emissions equivalent of 31.5K cars \cite{epa}.
The numbers are larger for countries with larger populations. For example, in China, there are 500K GSM/WCDMA and 200K 3G CDMA RBSs \cite{Grant09}. This results in 3.1 TWh, or 2.2M MtCO$_2$e, or the CO$_2$ emission equivalent of 421K cars. The number for the United States is similar, about 750K \cite{TRKP10}. In addition to CO$_2$ emissions, energy consumption increases the system OpEx for service providers. Service providers will clearly want to improve the efficiency figure of 8\% for 3G RBSs. This figure is actually likely to go further down with 4G RBSs due to the introduction of OFDM.

\ifCLASSOPTIONonecolumn
\begin{figure}[!t]
\begin{center}
\begin{minipage}[b]{0.6\linewidth}
\hspace{12mm}\includegraphics[width=80mm]{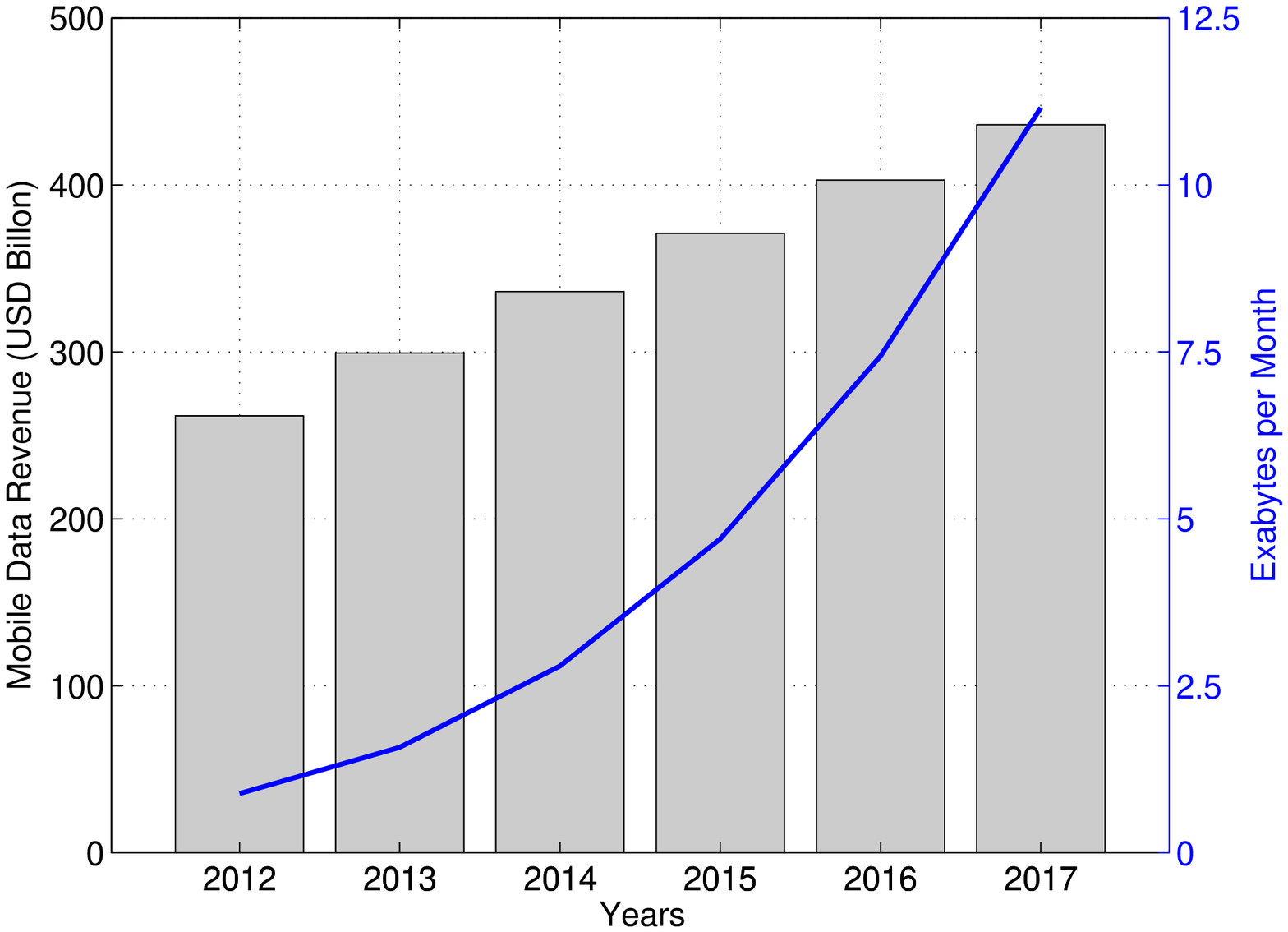}
\caption{Projected data traffic (blue line) and revenue growth (gray bars) 2012-2017.}
%\caption{Projected data traffic (red line) and revenue growth (blue bars) 2009-2014.}
\label{fig:cagr}
\end{minipage}
\end{center}
\end{figure}
\else
\begin{figure}[!t]
\begin{center}
\vspace{4mm}
\begin{minipage}[b]{1.0\linewidth}
\hspace{7mm}\includegraphics[width=65mm]{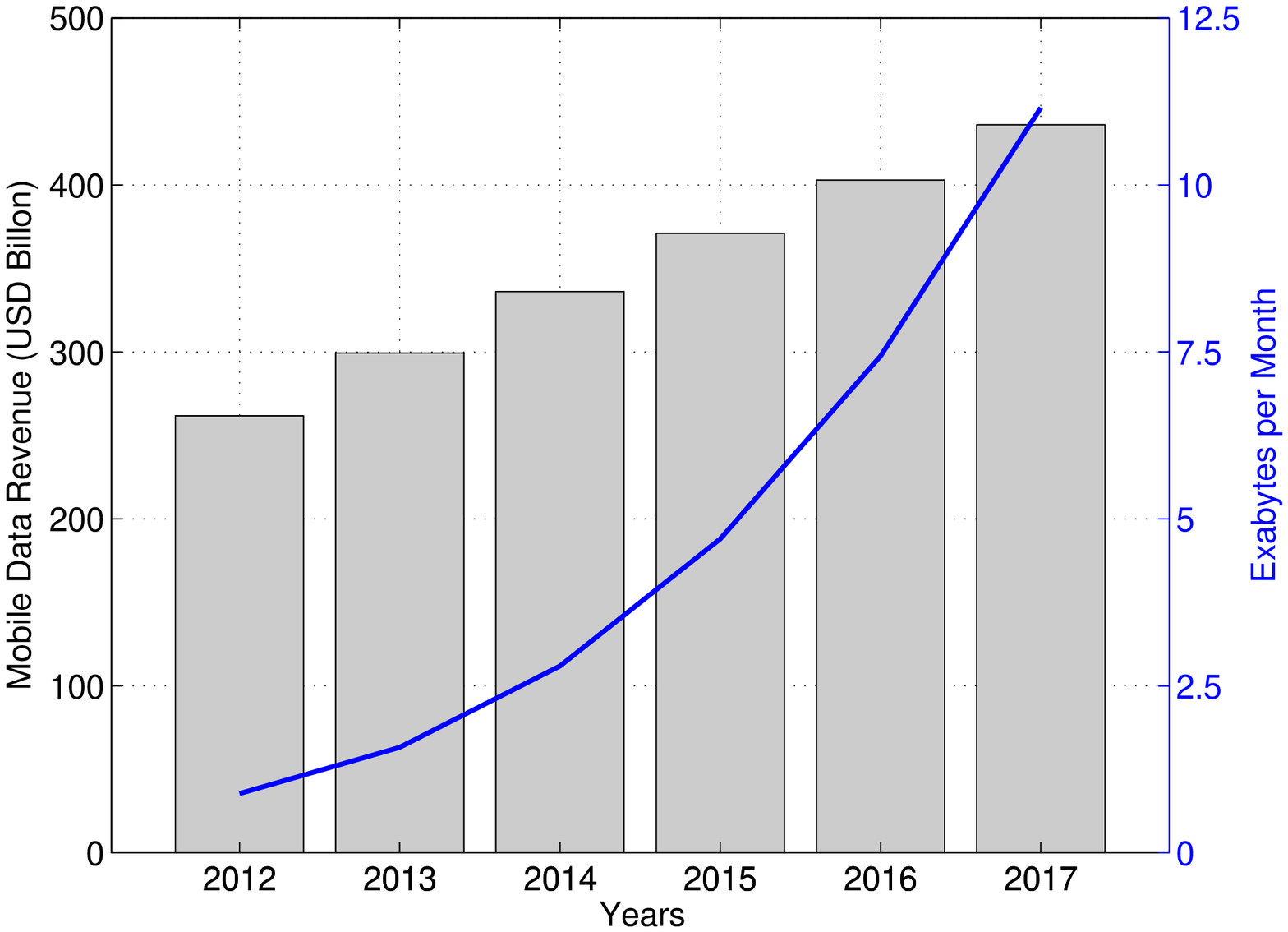}
\vspace{1mm}
\caption{Projected data traffic (blue line) and revenue growth (gray bars) 2012-2017.}
%\caption{Projected data traffic (red line) and revenue growth (blue bars) 2009-2014.}
\label{fig:cagr}
\end{minipage}
\end{center}
\vspace{-8mm}
\end{figure}
\fi

The expected growth rate for wireless data volume is 66\%. The projections for global wireless data revenues for service providers are much less, typically around 6-11\% CAGR \cite{AnalysisMason,InsightCorp,MGTHLKHW11}. These projections result in an increase of 13 times for wireless data traffic in 2017 as compared to that of 2012 \cite{Cisco13}, whereas only 1.67 times for data revenue (corresponding to a 11\% CAGR) for the same period \cite{AnalysisMason}. We depict the forecasted mobile data growth per month and mobile data revenue between the same period in Fig.~\ref{fig:cagr}. Note that the growth in data demand is exponential whereas the mobile data revenues are linearly increasing for network operators. Wireless voice CAGR values are actually being projected as becoming negative \cite{MGTHLKHW11}. As a result, reducing OpEx values is imperative for service providers. Since the same service providers will have to accommodate a tremendous increase in traffic, managing OpEx becomes very important. It can be argued that projections of data revenue CAGR were made without insights into explosion in data traffic. Even though this argument is somewhat valid, the fact remains that wireless service providers are under heavy competitive pressure to keep subscription fees low while at the same time are facing exponentially increasing traffic. So, reduction of service provider OpEx will certainly be extremely important. % \cite{Smith10,SNL,ABI}

\section{Methods for Energy Efficiency in Cellular Networks}\label{sec:methods}
In this section we will discuss methods we consider promising for improving the energy efficiency of cellular networks. We identified the main sources of energy efficiency in Sections~\ref{sec:papr} through \ref{sec:rbsconsumption}. In the following Sections~\ref{sec:CEOFDM} through \ref{sec:mobiles}, we investigate the methods to address these sources of energy inefficiency. How Sections~\ref{sec:CEOFDM} through \ref{sec:mobiles} relate to Sections~\ref{sec:papr} through \ref{sec:rbsconsumption} is provided in Table~\ref{tbl:sectionrelations}.
\begin{table*}[t!]
\begin{center}
\caption{Relations of subsections of Section~\ref{sec:losssources} to Section~\ref{sec:methods}.}
\label{tbl:sectionrelations}
\small
\begin{tabular}{|l|l|}
\hline
\multicolumn{1}{|c|}{Section~\ref{sec:losssources}}&\multicolumn{1}{|c|}{Section~\ref{sec:methods}}\\
\hline
\it A. Impact of Modulation on Power Amplifiers &\it A. Constant Envelope OFDM\\
&\it B. New Classes of Power Amplifiers\\
&\it C. Energy-Efficient Transmission Modes and Link Rate Adaptation\\
\hline
\it B. Deployment, Frequency Plan, Mobility, &\it D. Traffic-Adaptive Cells\\
\quad\;\it and Traffic Patterns&\it E. Relays and Cooperation\\
&\it F. Multiple Antenna Techniques\\
&\it G. Sleeping Mode for the Radio Base Station\\
&\it H. Reduced Cell Size and Hierarchical Cells\\
&\it I. Energy Efficiency of Mobile Units\\
\hline
\it C. RBS Energy Consumption and Its Effects&\it A--H.\\
\quad\;\it for the Service Provider& \\
\hline
\end{tabular}
\end{center}
\end{table*}
\begin{figure*}[t!]
%\begin{wrapfigure}{r}{120mm}
%\vspace{-8mm}
\begin{center}
\includegraphics[width=170mm]{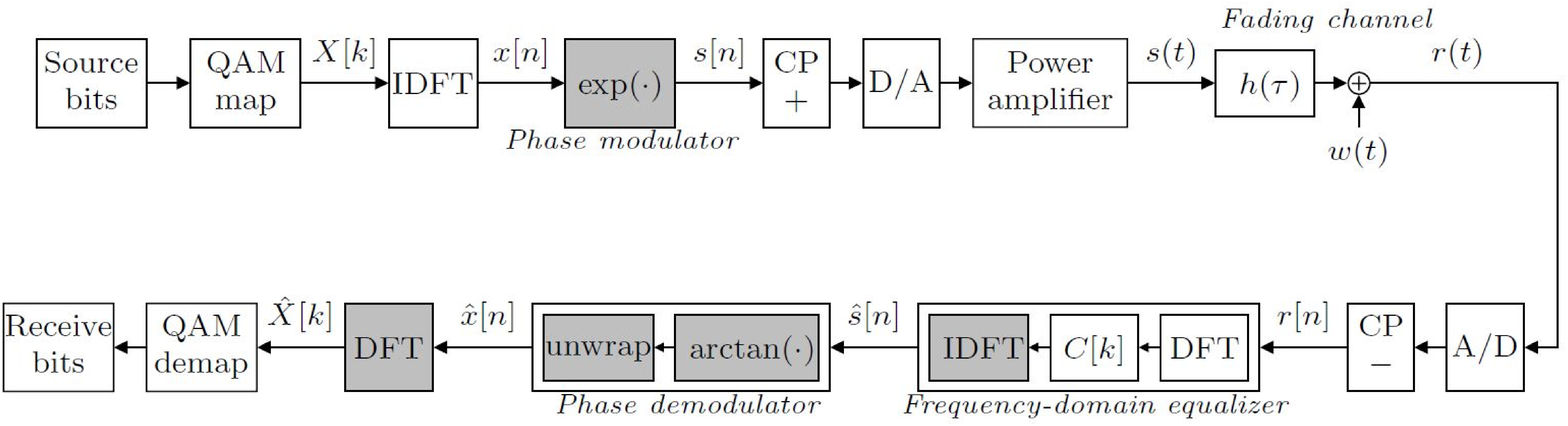}
%\vspace{-3mm}
\caption{Baseband block diagram of constant envelope OFDM \cite{TAPZG08}.}
\label{fig:ce-ofdm}
\end{center}
\vspace{-5mm}
%\end{wrapfigure}
\end{figure*}

\subsection{Constant Envelope OFDM}
\label{sec:CEOFDM}
While due to its large PAPR values OFDM is highly wasteful of power, it is possible to come up with a version of OFDM that has constant amplitude (or constant envelope). Fig.~\ref{fig:ce-ofdm} shows the block diagram of an OFDM system that achieves this. In this figure, the unshaded blocks represent a conventional OFDM system. The shaded blocks can be added to an existing OFDM implementation. In this uncoded version, the source bits are mapped through QAM modulation to generate a complex-valued sequence $X[k]$ of QAM constellation points. A conjugate symmetric version of this sequence is generated,  which, after going through an Inverse Discrete Fourier Transform (IDFT) block, generates a real-valued sequence $s[n]=\exp({jCx[n]})$, where $C$ is a scaling constant and $j=\sqrt{-1}$. Then, a short Cyclic Prefix (CP) sequence is added and the signal is transmitted. At the receiver, CP is first removed (this operation ensures frequency domain equalization can be achieved). The DFT of the signal is generated, and equalization in the frequency domain is carried out. At this point, if the block marked as the phase modulator  were not present at the transmitter, a sequence close to $X[k]$ would be obtained. In order to undo the  effect of the phase modulator, the sequence of blocks consisting of another IDFT (oversampled) an ${\rm arctan(}\!\cdot\!{\rm )}$ operation, phase unwrapping, and a DFT (with subsequent subsampling) are performed. The resulting signal has constant amplitude and we will abbreviate it as CE-OFDM.

What is happening can be interpreted as in Fig.~\ref{fig:constenv}. The members of the sequence $X[k]$ are from a finite alphabet of QAM symbols. After the IDFT, the OFDM sequence $x[n]$ has varying magnitudes. However, after the transformation $s[n]=\exp({jCx[n]})$, the complex-valued sequence $s[n]$ has unit magnitude. The information content is transformed into the phase of $s[n]$. The sequence $x[n]$ has large PAPR, whereas the sequence $s[n]$ has 0 dB, or no PAPR.

\ifCLASSOPTIONonecolumn
\begin{figure}[t!]
%\begin{wrapfigure}{l}{55mm}
%\vspace{-8mm}
\begin{center}
\begin{minipage}[b]{0.6\linewidth}
\centering
\includegraphics[width=80mm]{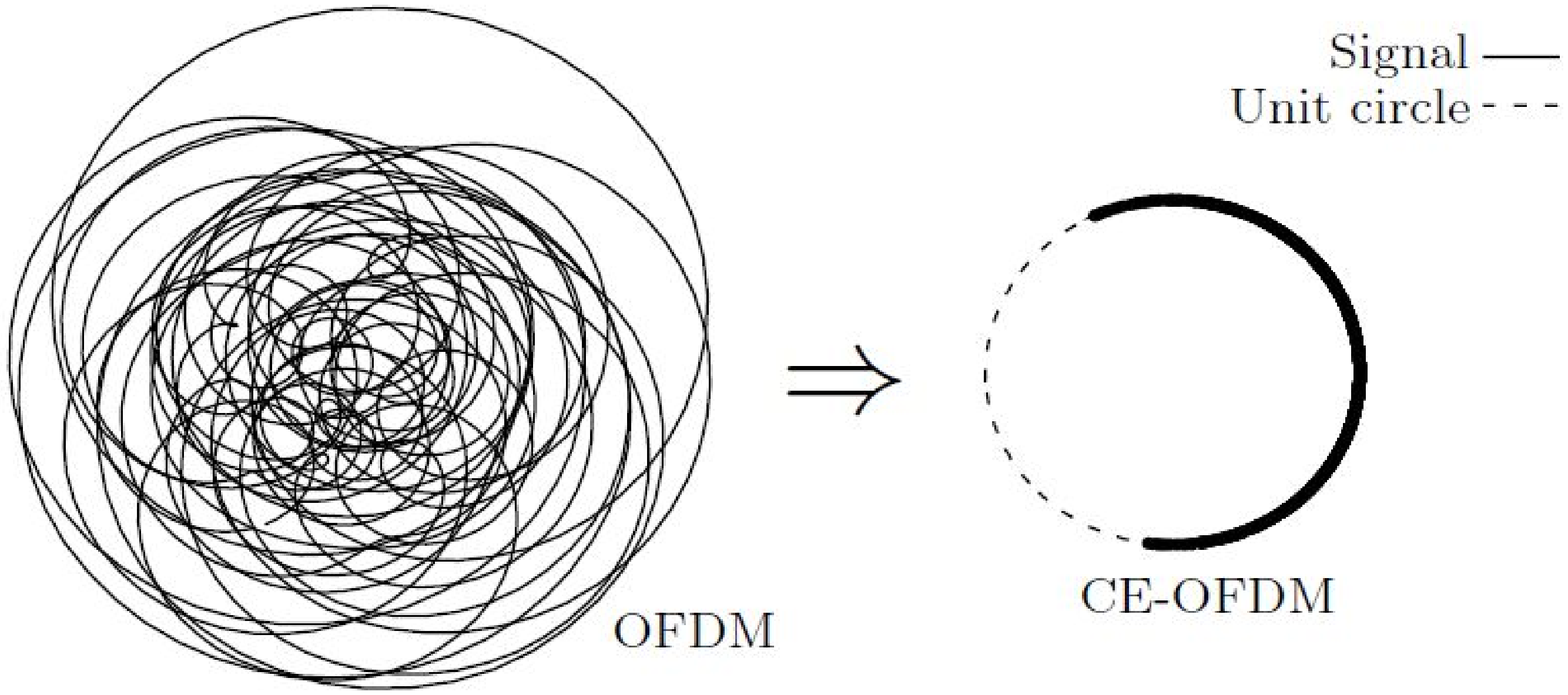}
%\vspace{-3mm}
\caption{OFDM to CE-OFDM mapping \cite{Thompson05}.}
\label{fig:constenv}
\end{minipage}
\end{center}
\end{figure}
\begin{figure}[t!]
\begin{minipage}[b]{0.5\linewidth}
\centering
\includegraphics[width=65mm]{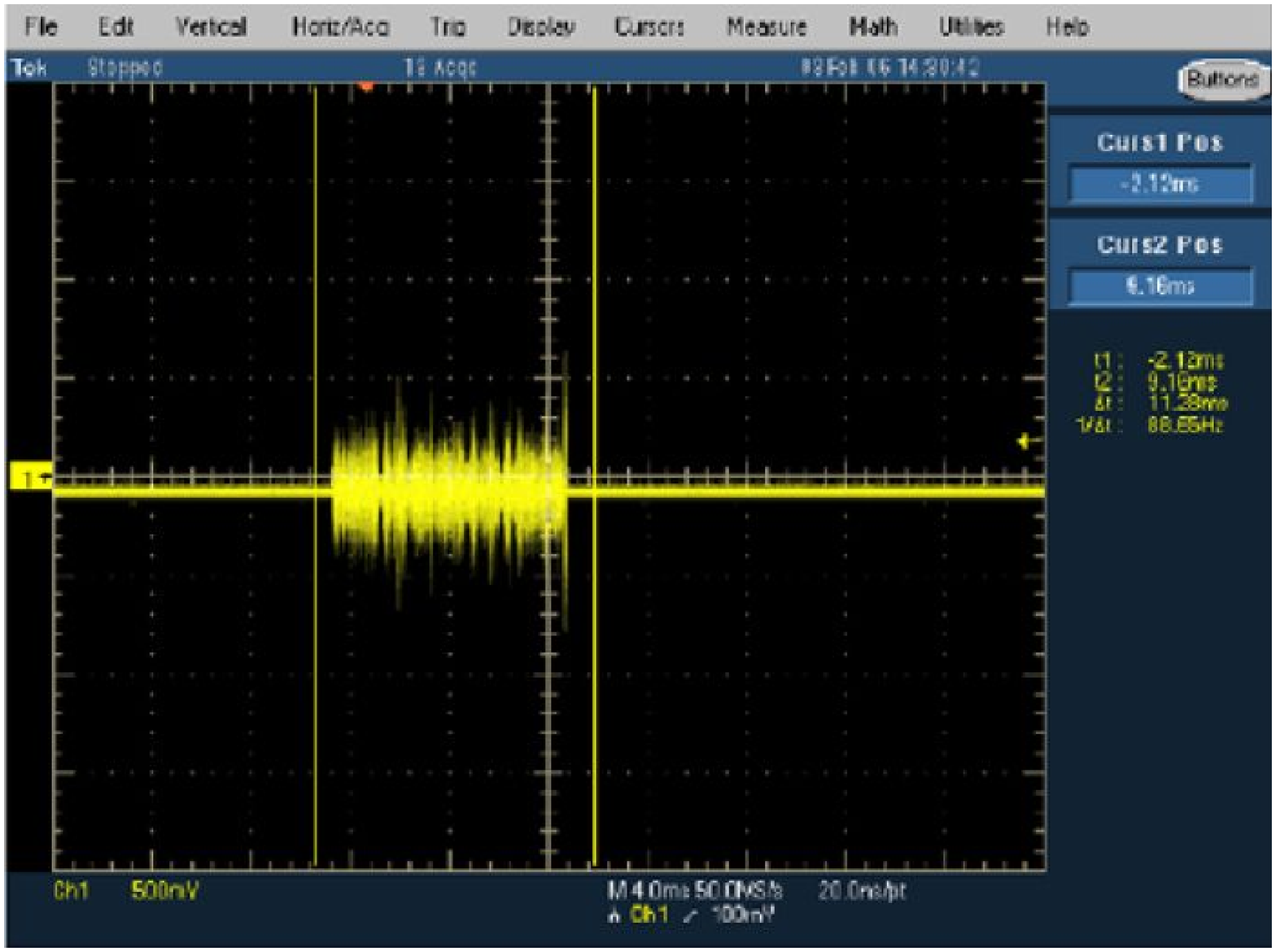}
%\vspace{-1mm}
\caption{OFDM signal output with high PAPR \cite{TAPZG08}.}
\label{fig:ofdmsdr}
\end{minipage}
\hspace{1.6mm}
\begin{minipage}[b]{0.5\linewidth}
\centering
%\vspace{3mm}
\includegraphics[width=65mm]{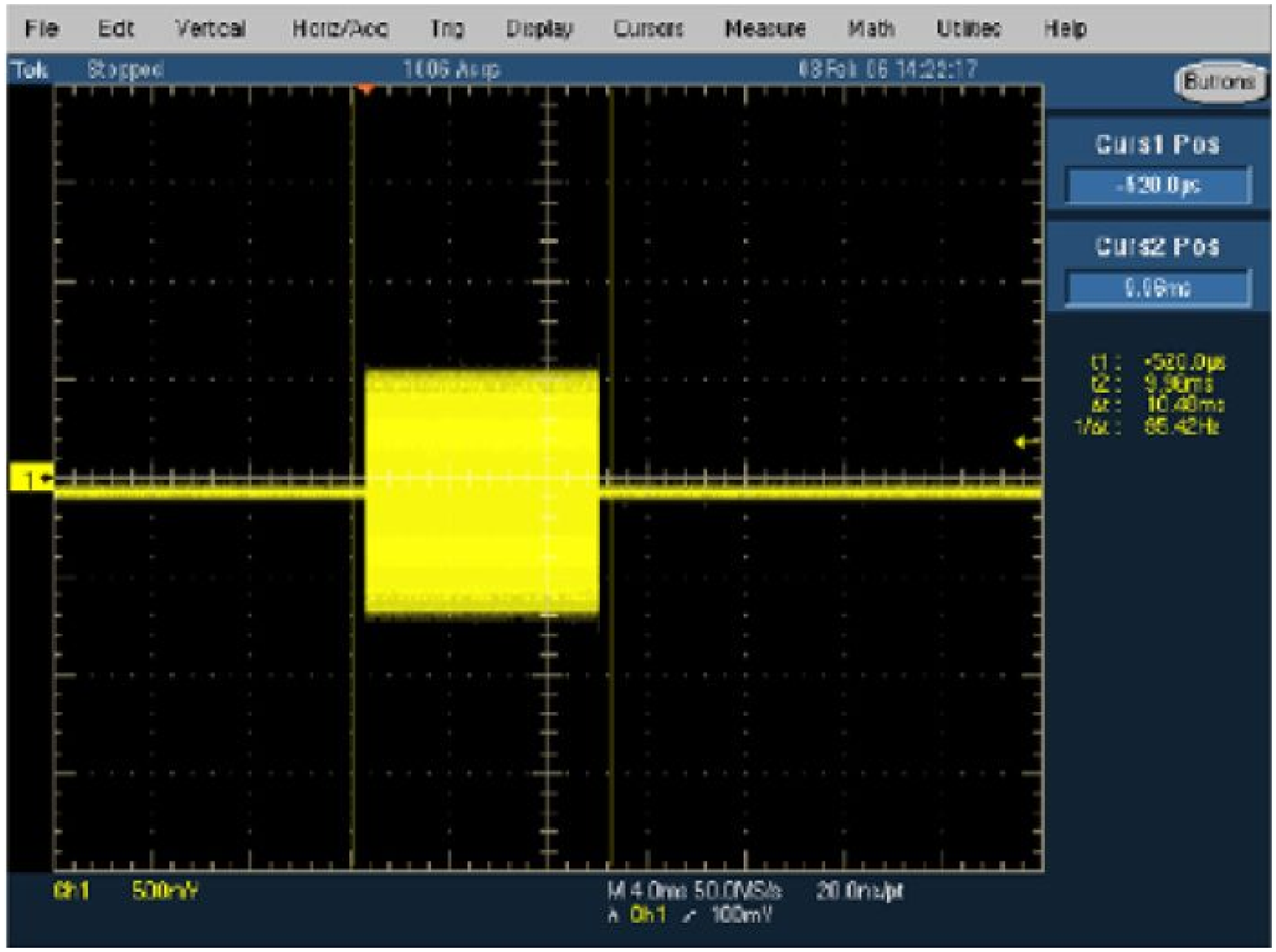}
%\vspace{-1mm}
\caption{CE-OFDM signal output with no PAPR \cite{TAPZG08}.}
\label{fig:ceofdmsdr}
\end{minipage}
%\vspace{-3mm}
%\end{wrapfigure}
\end{figure}
\else
\begin{figure}[t!]
\begin{center}
\begin{minipage}[b]{1.0\linewidth}
\centering
\includegraphics[width=80mm]{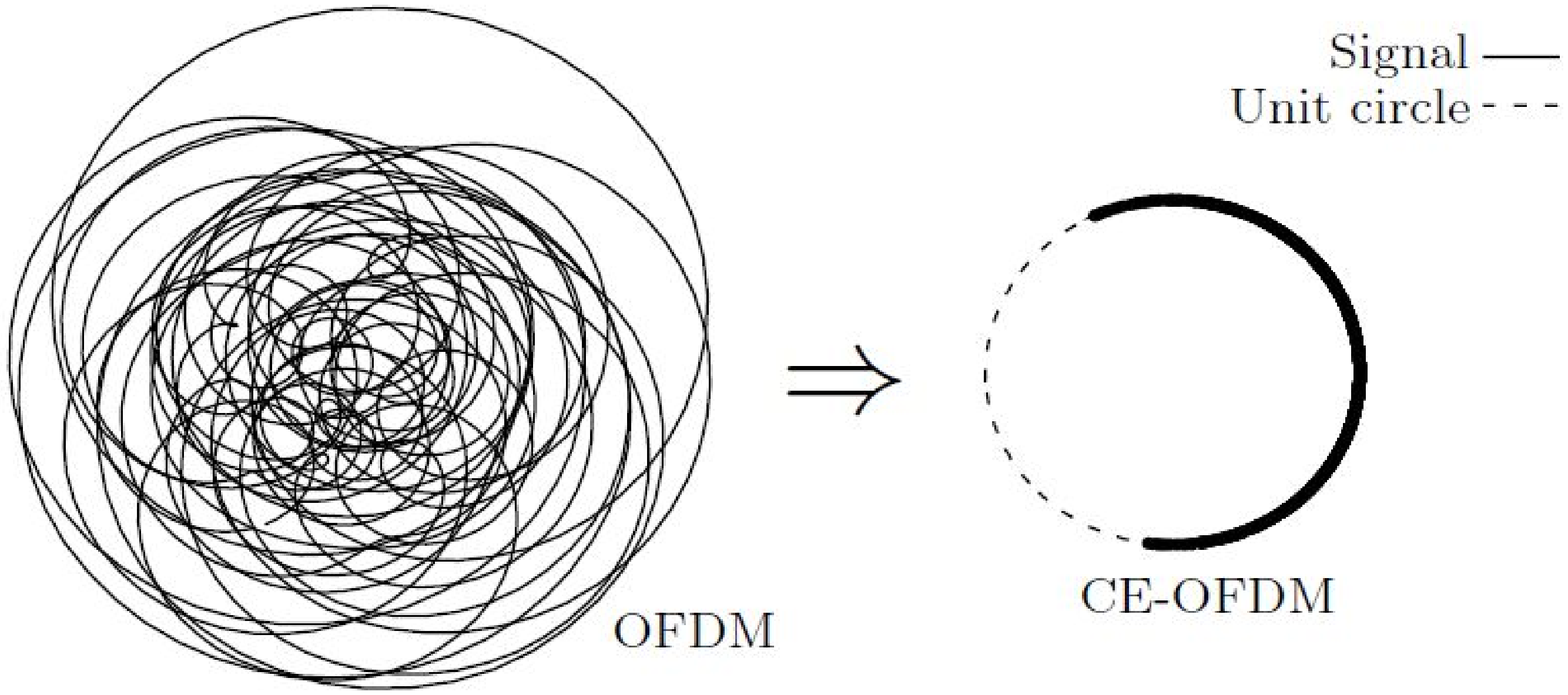}
\caption{OFDM to CE-OFDM mapping \cite{Thompson05}.}
\label{fig:constenv}
\end{minipage}
\end{center}
\vspace{-7mm}
\end{figure}
\fi

The idea of transmission of OFDM signals through FM radios was first explored in \cite{CL91} without consideration for PAPR. The PAPR advantage was recognized and the basic system was first described in \cite{CC99}, followed by \cite{TS02,KMCSB05}, and by a group at the University of California San Diego in a string of papers that were finalized in \cite{Thompson05,TAPZG08}.

An implementation of CE-OFDM was carried out in \cite{TAPZG08} using Field Programmable Gate Arrays (FPAs), a microprocessor, and a Digital Signal Processing (DSP) chip. CE-OFDM was implemented by adding the required operations to an existing OFDM implementation.
Fig.~\ref{fig:ofdmsdr} shows the OFDM signal with high PAPR whereas Fig.~\ref{fig:ceofdmsdr} shows the CE-OFDM signal. % with 0 dB or no PAPR and with the desirable constant envelope property.
Clearly, the signal in Fig.~\ref{fig:ofdmsdr} has large variations in its envelope while that in Fig.~\ref{fig:ceofdmsdr} has constant envelope. This translates into a very high value of PAPR for the signal in Fig.~\ref{fig:ofdmsdr}, while the signal in Fig.~\ref{fig:ceofdmsdr} has no PAPR (0 dB PAPR). As discussed earlier in Section~\ref{sec:papr}, this fact will result in a forced operation of the power amplifier with a large value of power backoff for the signal in Fig.~\ref{fig:ofdmsdr}, while for the signal in Fig.~\ref{fig:ceofdmsdr}, no power backoff is needed. Since power backoff translates into substantial energy inefficiency, the advantage of CE-OFDM should be clear.
\ifCLASSOPTIONtwocolumn
\begin{figure}[t!]
\vspace{3mm}
\begin{center}
\begin{minipage}[b]{1.0\linewidth}
\centering
\includegraphics[width=65mm]{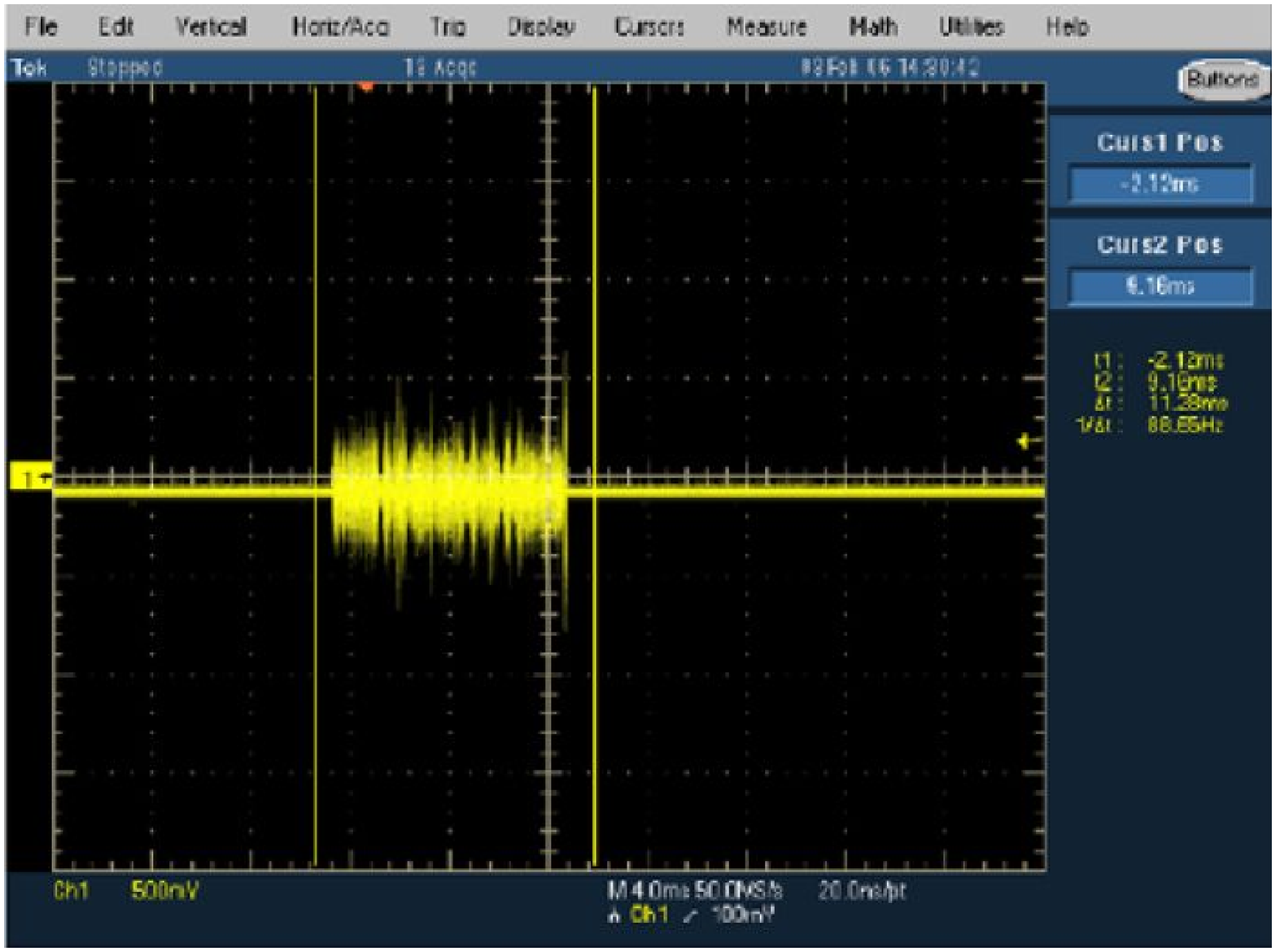}
\caption{OFDM signal output with high PAPR \cite{TAPZG08}.}
\label{fig:ofdmsdr}
\end{minipage}
\end{center}
\begin{center}
\begin{minipage}[b]{1.0\linewidth}
\centering
\includegraphics[width=65mm]{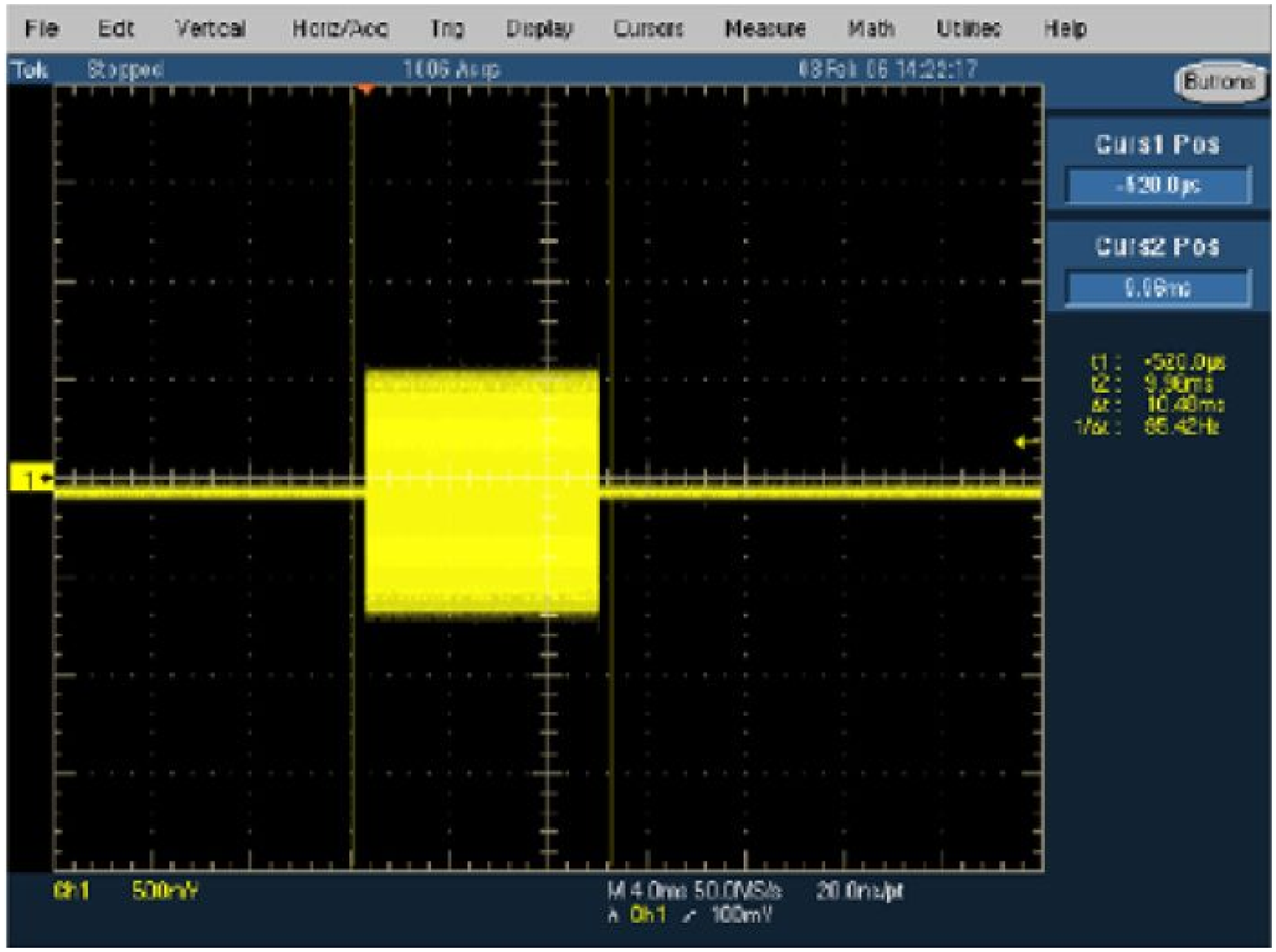}
\caption{CE-OFDM signal output with no PAPR \cite{TAPZG08}.}
\label{fig:ceofdmsdr}
\end{minipage}
\end{center}
\vspace{-7mm}
\end{figure}
\fi

It has been shown via simulations that CE-OFDM has better Bit Error Rate (BER) performance than OFDM when used with realistic power amplifier models and with realistic backoff values for OFDM in additive white Gaussian noise (AWGN) and fading channels. Also, it is shown that CE-OFDM has better fractional out-of-band power as compared to OFDM. CE-OFDM is certainly promising. However, there is more work that needs to be carried out. First, CE-OFDM has a bandwidth tradeoff which needs to be clearly quantified. Second, phase unwrapping in the presence of noise worsens its BER performance. This could be improved with the oversampling factor but this has complexity implications which need to be understood. Third, it may be possible to improve the decoding performance at low SNR. Fourth, continuous-phase modulation techniques known in general as CPM can have good spectral efficiency and BER performance at the same time. The use of CPM for this purpose has not been investigated at all. Fifth, all of the research on CE-OFDM remains for uncoded systems whereas practical systems would require channel coding. This is yet another open research direction. Finally, there are a number of conventional techniques for combating PAPR (e.g., see \cite{Litsyn} and its references). Each of these techniques has its disadvantages but a thorough study of the advantages and disadvantages of these techniques against the CE-OFDM approach is needed.
\subsection{New Classes of Power Amplifiers}\label{sec:classj}
PAs are categorized with respect to the way they are biased as well as the way their power loading is set up. Class A PA has its bias point in the middle point of the linear operating region of the power transistor. Class B has two transistors, %one has
with one having its bias point placed such that it amplifies the positive swing of the input signal while the other transistor amplifies the negative swing. Class AB is in between. Classes A, B, and AB are considered linear or close to linear. Class C has a tuned circuit as its power load and is nonlinear. These four classes are conventional power amplification techniques and have been around since the days of vacuum tubes. More recently, newer types have been invented. Most recently, a new class, Class J, has been introduced \cite{Cripps06,CTCLB09}. This new class eliminates a number of harmonics at its loading stage and as a result, can achieve good amplification and efficiency over a very wide range of frequencies. It is sometimes quoted as having very high efficiency such as larger than 60\% efficiency over a 140 MHz range and therefore is recommended for energy reduction in cellular networks (e.g., \cite{Grant10}). However, in closer examination, one can see that this PA needs to be backed off by the PAPR value as well. For a signal with the corresponding PAPR (of 8.51 dB), the resulting power efficiency figure is closer to what a Class AB PA would provide \cite{Litsyn,WLBTC09}, namely about 30\%, while the PA gain is over 10 dB and it can be sustained over a frequency range of 1.7-2.7 GHz \cite{WLBTC09}.

Highly efficient PAs are an essential part of a wireless system to achieve, small, reliable, low-cost, and power-efficient transmitters. An excellent review of this subject can be found in \cite{Cripps06}. This reference actually introduced Class J amplifiers, as well as discussing a number its alternatives such as the Doherty amplifier. In terms of efficiency, bandwidth, and linearity, Class J amplifiers are very attractive. On the other hand, their implementation so far requires the use of GaN semiconductor technology. This technology is currently becoming available for RF amplifiers in high-end commercial applications. Although currently they cannot compete with existing CMOS approaches in terms of cost, area, or yield; due to their many advantages, Class J amplifiers can be a candidate to become the RF amplifier of choice in the RBSs of the future, especially if energy efficiency considerations dominate the choice.
\subsection{Energy-Efficient Transmission Modes and Link Rate Adaptation}
\label{sec:EETM}
\begin{figure}[!t]
\begin{center}
\ifCLASSOPTIONonecolumn
\includegraphics[width=100mm]{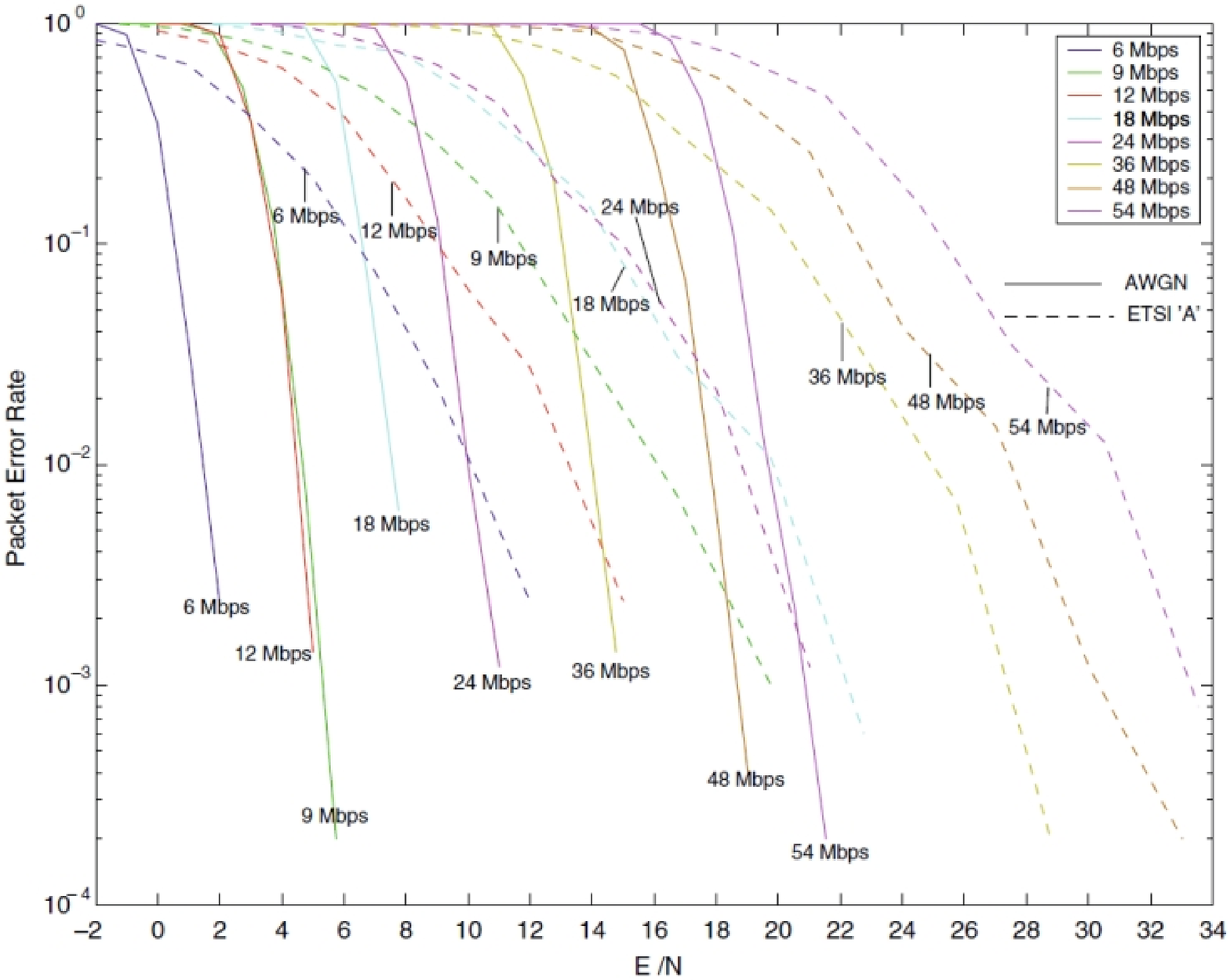}
\else
\includegraphics[width=80mm]{per.eps}
\fi
\caption{Packet error rates vs $E_s/N_0$ in 802.11a \cite{CJT05}.}
% This curve is from Chevillat, Jelitto, and Truong \cite{CJT05}
\label{fig:per}
\end{center}
\vspace{-7mm}
\end{figure}

A wireless network offers a number of different transmission rates. For example, IEEE 802.11g or 802.11a can operate at 6, 12, 18, 24, 48, and 54 Mb/s. For a given transmission rate and a given Signal-to-Noise Ratio (SNR) in the channel, there is an average error rate incurred by the channel, where the error can be measured in terms of BER or Packet Error Rate (PER) and where the packet is of a standard length, e.g., 1000 bytes. For example, Fig.~\ref{fig:per} shows the PER performance against $E_s/N_0$, a measure of the SNR in the channel \cite{CJT05}. The solid lines are for an additive white Gaussian noise (AWGN) channel, whereas the dashed lines correspond to a fading channel with multipath, known as the ETSI-A channel. To operate at higher rates with an acceptable loss rate, the SNR should be larger than a threshold. As a result, higher rates are only possible when the transmitter and the receiver are in a given proximity. In conventional wireless networks, an algorithm determines the user transmission rate. This is achieved by an operation known as {\em link rate adaptation.\/} The goal of most common link rate adaptation algorithms is to maximize the network throughput. As a result, most of these algorithms are designed to give the user the highest rate they can communicate with at their SNR level. However, as can be deduced from Fig.~\ref{fig:per}, higher rates consume more energy. Therefore, for the link adaptation algorithm in an energy-efficient cellular network, energy consumption of the candidate rates must be taken into consideration.

A number of link rate adaptation algorithms have been proposed for wireless networks, e.g., \cite{CJT05,QCS02,WYLB06}. Some of these algorithms take power consumption into account, and some also have additional power control, such as \cite{CJT05}. For example, the goodput values for various modes of 802.11a are shown in Fig.~\ref{fig:QCS02a} \cite{QCS02}. In this figure, obtained by the numerical evaluation of an analysis for a link model, the channel is AWGN and the packet size is 2000 bytes. Based on the values in Fig.~\ref{fig:QCS02a}, the best overall effective goodput and the outcome of a link adaptation algorithm are given in Fig.~\ref{fig:QCS02b} \cite{QCS02}. Note that mode 2 is not part of the adaptation algorithm, since mode 2 results in a smaller effective goodput than mode 3 under all SNR conditions. Also note that, the outcome may be different for a wireless channel other than AWGN, such as shown in Fig.~\ref{fig:per} with dashed lines against solid lines.
\ifCLASSOPTIONonecolumn
\begin{figure}[!t]
\begin{minipage}[b]{0.42\linewidth}
\centering
\includegraphics[height=65mm]{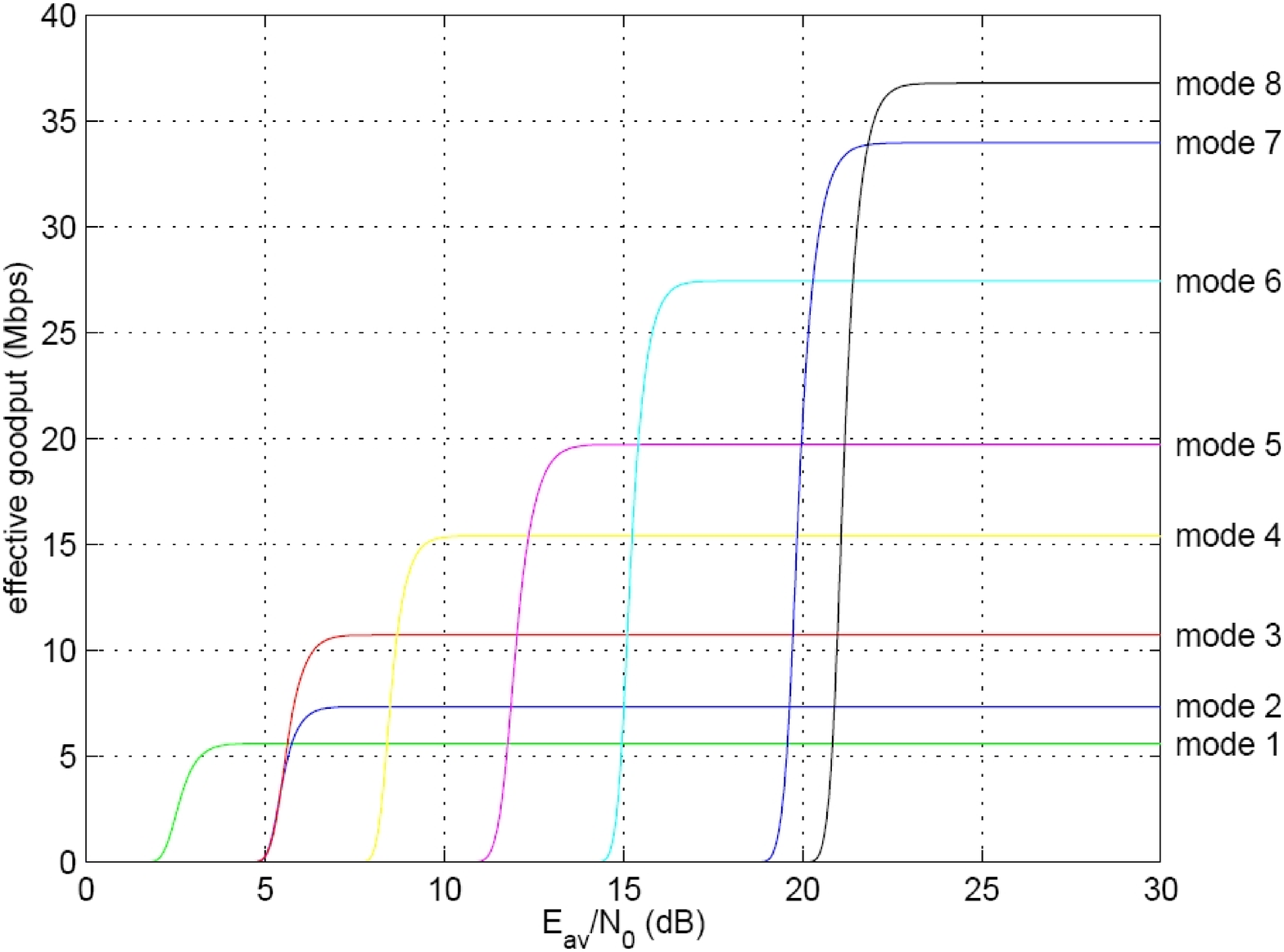}
\caption{Effective goodput vs. SNR \cite{QCS02}.}
\label{fig:QCS02a}
\end{minipage}
\hspace{20mm}
\begin{minipage}[b]{0.42\linewidth}
\centering
\includegraphics[height=63mm]{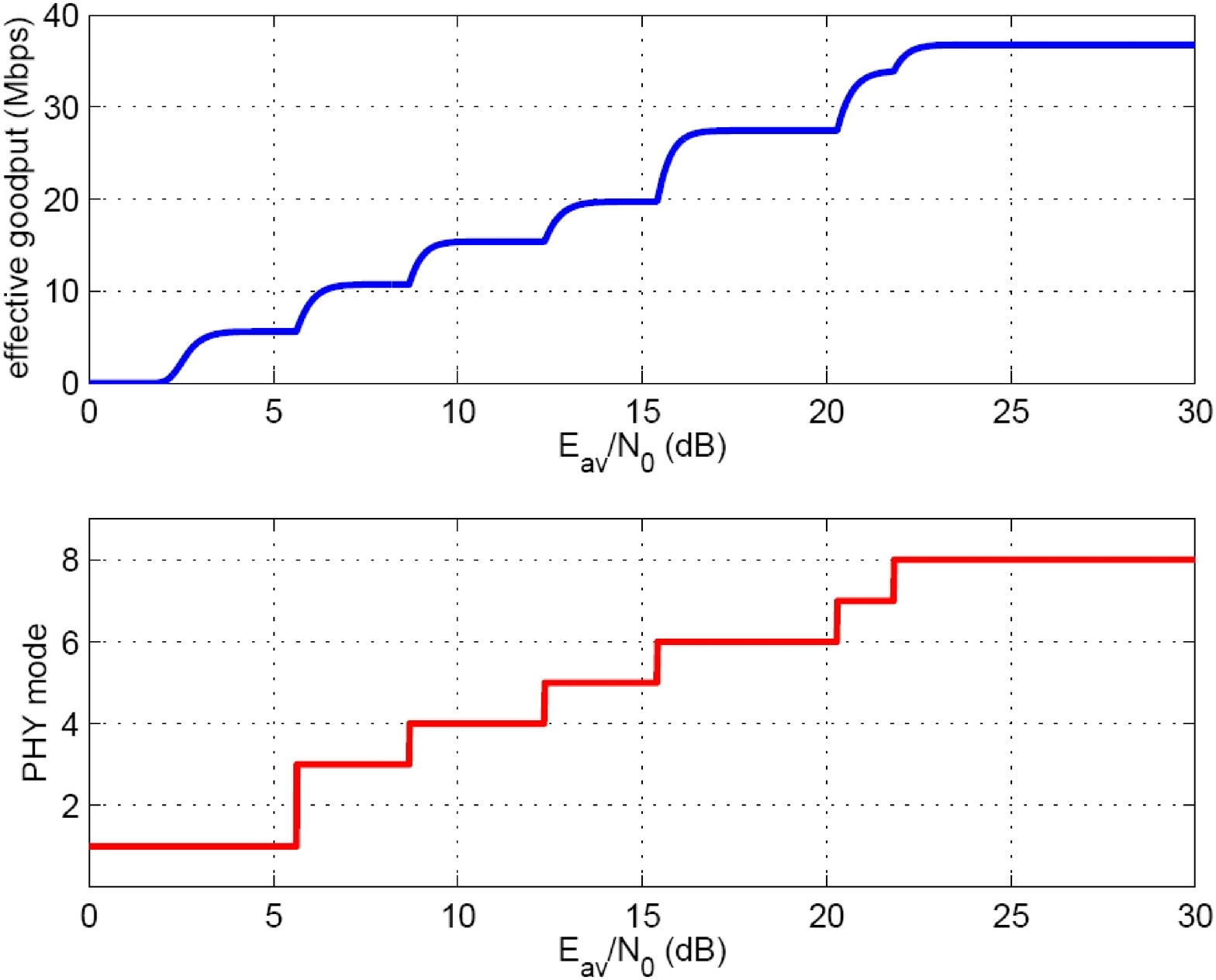}
\caption{Best effective goodput and the link adaptation algorithm \cite{QCS02}.}
\label{fig:QCS02b}
\end{minipage}
% These curves are from \cite{QCS02}
\end{figure}
\else
\begin{figure}[!t]
\begin{center}
\begin{minipage}[b]{1.0\linewidth}
\centering
\includegraphics[height=63mm]{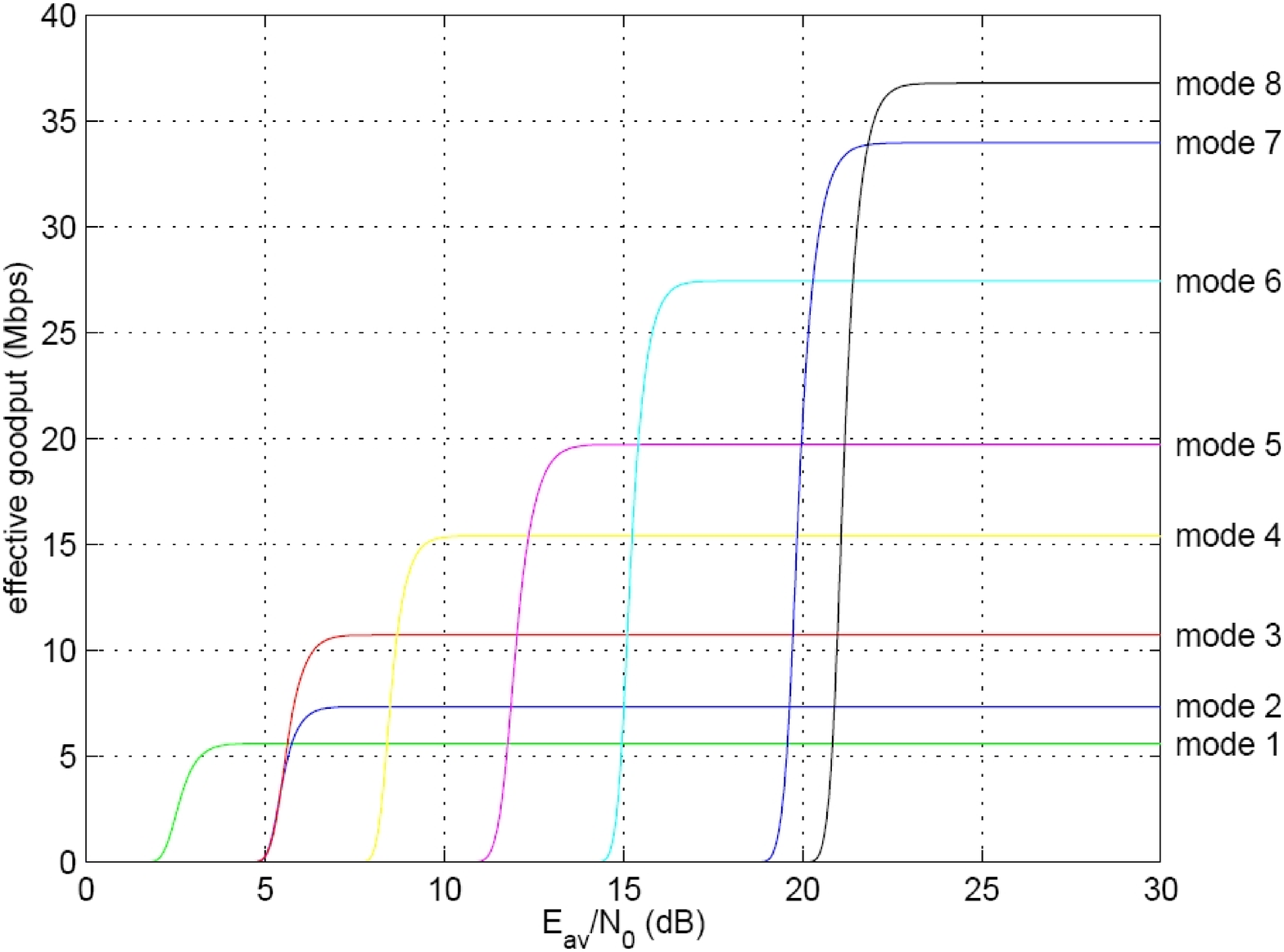}
\caption{Effective goodput vs. SNR \cite{QCS02}.}
\label{fig:QCS02a}
\end{minipage}
\end{center}
\vspace{-1mm}
%\end{figure}
%\begin{figure}[t!]
\begin{center}
\begin{minipage}[b]{1.0\linewidth}
\centering
\includegraphics[height=65mm]{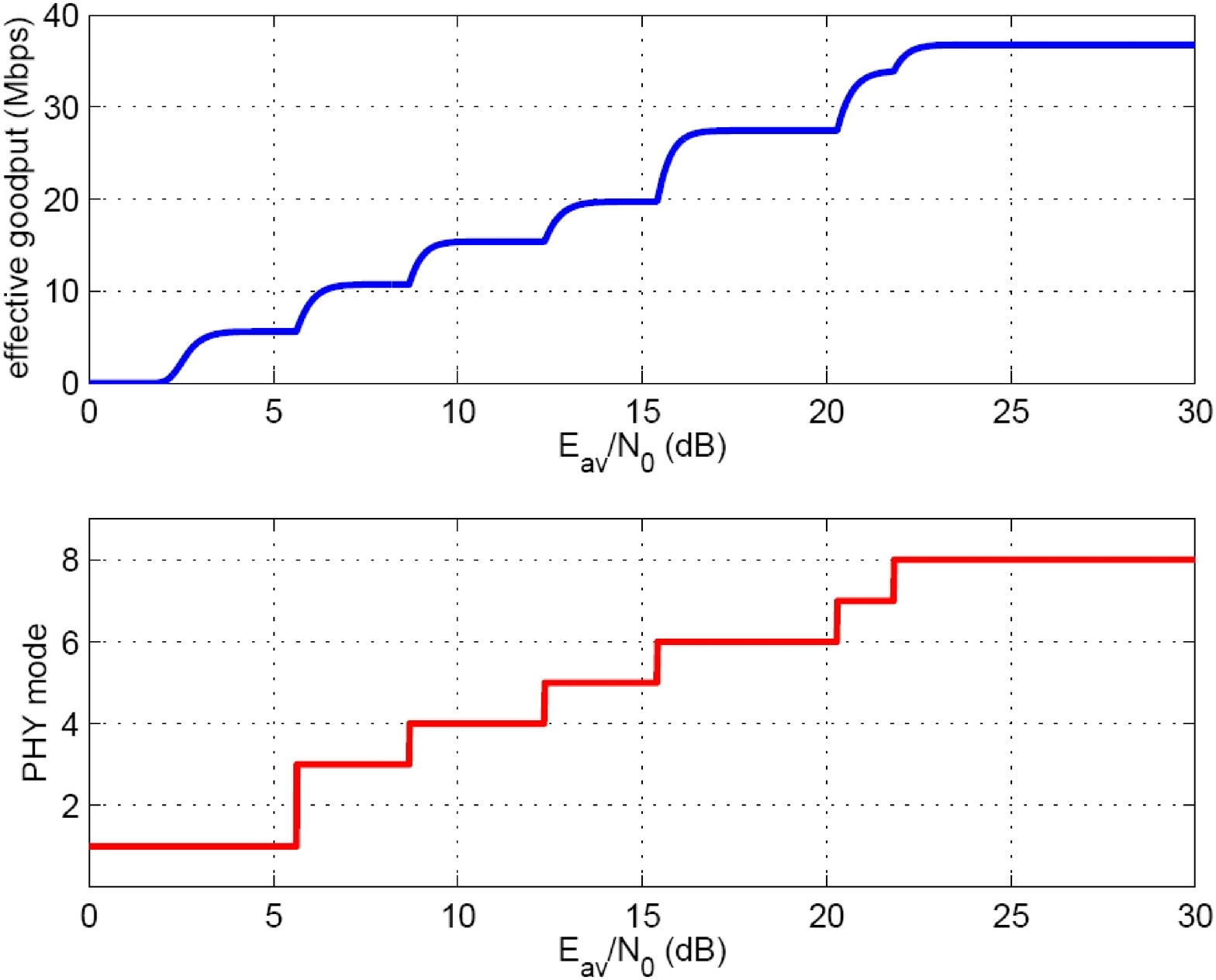}
\caption{Best effective goodput and the link adaptation algorithm \cite{QCS02}.}
\label{fig:QCS02b}
\end{minipage}
\end{center}
\vspace{-7mm}
% These curves are from \cite{QCS02}
\end{figure}
\fi

Although energy consumption has been taken into consideration in the literature on sensor networks, in cellular networks, sometimes more emphasis is placed on power. However, what matters is the energy. The two are not proportional when there are different transmission rates involved and care should be exercised in link rate adaptation. To see this point, consider the $E_s/N_0$ values corresponding to PER $=10^{-2}$ in Fig.~\ref{fig:per}. This PER value is considered to be the maximum acceptable in 802.11 networks. Reference \cite{CJT05} states the relative linear values of $E_s/N_0$ corresponding to bit rates $\{6,9,12,18,24,36,48,54\}$ Mb/s are approximately $\{1,2,2,4,8,16,32,64\}$. Reference \cite[Table 3]{Heegard} lists the $E_s/N_0$ values corresponding to a PER value of $10^{-2}$, and it can be verified that these two sets of values indeed match. The number of {\em channel\/} bits per {\em channel symbol\/}, in the same format, is $\{1,1,2,2,4,4,6,6\}$. The channel code rates associated with the transmission rates specified above, in the same format, are $\{1/2,3/4,1/2,3/4,1/2,3/4,2/3,3/4\}$. Let us now compare transmission at rate 6 Mb/s versus 36 Mb/s. The first impression may be that, due to Fig.~\ref{fig:per}, 36 Mb/s has larger power and perhaps also more energy consumption. Surprisingly, this is not so. First, note that the relative time it takes to transmit a piece of data at 36 Mb/s vs 6 Mb/s is $1/6$. In terms of $E_s/N_0$, we have a ratio of 16/1. Taking the number of channel bits per channel symbol as well as the channel code rates for the two transmission, for 6 Mb/s against 36 Mb/s, the SNR per bit $E_b/N_0$ becomes $8/3$. Therefore, the ratio of the energy use in the 36 Mb/s channel vs 6 Mb/s is $4/9=44.4\%$. This analysis shows that the 6 Mb/s channel rate is actually energy-inefficient, all other rates are more efficient than this rate. The 9 Mb/s and 12 Mb/s rates are more energy-efficient than only the 54 Mb/s rate, and the 12 Mb/s rate has the same energy efficiency as the 24 and 48 Mb/s rates. In a one-by-one comparison of the remaining transmission rates of 24, 36, 48 Mb/s, and 54 Mb/s, the lower rate channel is always more energy-efficient than the higher rate channel. Although the analysis above ignores packet sizes and the header and Medium Access Control (MAC) overhead, these can be easily incorporated as in \cite{CJT05,QCS02}. However, most of the results of this analysis are still going to hold even with the more detailed calculations.

There are additional considerations in designing an energy-efficient link rate adaptation algorithm. First, most user applications require a minimum transmission rate, although some, such as e-mail, messaging, or file transfers can accommodate  best effort rates. Therefore, the user required transmission rate, if any, will be one of the parameters. Second, as discussed in \cite{GCJM09}, existing user rates, or the occupancy in the channel, or utilization, is another parameter. In general lower rates are preferable for energy efficiency, (with considerations as in the paragraph above), however, with lower rates, the user data will take longer to transmit. This is acceptable at low utilization values, but at higher utilizations, it will be important to serve users as quickly as possible. Therefore, the algorithm should choose lower rates at low utilizations but should gradually allow higher rates as the utilization increases. Third, the algorithm will improve if the user SNR and the channel characteristics are available. This data can be extracted directly from signal and noise strength measurements or indirectly from packet error rates or acknowledgement (ACK) messages employed in the protocol, e.g., as in \cite{CJT05}. Finally, the particular MAC used is an important factor. For example, in the case of 802.11, considerations for the hidden terminal problem, backoff times, retransmissions, etc. should be incorporated into the design process.

An energy-efficient link rate adaptation algorithm should be designed using the considerations we have discussed above. As it requires the use of data belonging to individual users as well as all of the users utilizing the channel, and adapts its parameters according to the changes in these, it can be considered operating on the principle of a cognitive radio.

\begin{comment}
An example of how the optimization problem can be cast is as maximizing the energy efficiency while keeping the transmission rate above a constraint
\begin{equation}
\max_{E_T,\textrm{BW}} \frac{(1-\textrm{PER})\cdot L}{E_T\cdot \textrm{BW}}\ \textrm{(bits/J/Hz)}\hspace{5mm}
\textrm{subject to}\hspace{3mm}(1-\textrm{PER})=(1-\textrm{BER})^L,\ \frac{(1-\textrm{PER})\cdot L}{T}\ge R_0,\label{eq:seven}
\else
\max_{E_T,\textrm{BW}} \frac{(1-\textrm{PER})\cdot L}{E_T\cdot \textrm{BW}}\ \textrm{(bits/J/Hz)}\\
\textrm{subject to}\hspace{3mm}(1-\textrm{PER})=(1-\textrm{BER})^L,\ \frac{(1-\textrm{PER})\cdot L}{T}\ge R_0,\label{eq:seven}
\fi
\end{equation}
where $E_T$ is the total energy consumed to transmit $L$ bits during $T$ seconds using a bandwidth of BW, at a packet error rate PER and bit error rate BER, while $R_0$ is the minimum acceptable rate. The objective function and the constraint equation are such that the problem of solving (\ref{eq:seven}) can be cast as a Geometric Programming (GP) problem. This approach, together with the definition of a GP problem is available in \cite{KD10}. Furthermore, \cite{KD10} states that this GP problem can be efficiently solved using convex optimization theory. In this process, the problem formulation results in the minimization of two objectives. The first is the minimization of the total energy used per successfully received bit, and the second is the maximization of throughput \cite{KD10}.
\end{comment}
%
The following example demonstrates how a general link adaptation for MIMO systems can be cast as an optimization problem. It maximizes the energy efficiency, while keeping the transmission rate above a constraint. It is formulated as \cite{KD10}
\ifCLASSOPTIONonecolumn
\begin{equation}
\max_{E_T,\textrm{BW}} \frac{(1-\textrm{PER})\cdot L}{E_T\cdot \textrm{BW}}\ \textrm{(bits/J/Hz)}\hspace{5mm}
\textrm{subject to}\hspace{3mm}(1-\textrm{PER})=(1-\textrm{BER})^L,\ \frac{(1-\textrm{PER})\cdot L}{T}\ge R_0,\label{eq:seven}
\end{equation}
\else
\begin{align}
\max_{E_T,\textrm{BW}} & \frac{(1-\textrm{PER})\cdot L} {E_T\cdot \textrm{BW}}  \ \textrm{(bits/J/Hz)}\label{eq:seven}\\
\textrm{subject to}& \hspace{3mm}(1-\textrm{PER}) =(1-\textrm{BER})^L,\nonumber\\
& \frac{(1-\textrm{PER})\cdot L}{T} \ge R_0,\nonumber
\end{align}
\fi
where $E_T$ denotes the total energy consumed to transmit $L$ bits during $T$ seconds using a bandwidth of BW, at a packet error rate PER and bit error rate BER, while $R_0$ is the minimum acceptable rate. The system dependent parameters $E_T$, $L$, $T$, BER, and PER are functions of the following variables: the transmit power $P_T$, number of transmit antennas $N_T$, number of receive antennas $N_R$, number of spatial streams $N_{SS}$, and the bandwidth BW. The study in \cite{KD10} presents energy consumption, throughput, and BER models relating these variables to the system dependent parameters. These relations are omitted in this paper for brevity and to present a general framework. Note that the optimization problem in (\ref{eq:seven}) can be translated into a Geometric Programming (GP) problem, as shown in \cite{KD10}, which can be solved efficiently using convex optimization theory. In this process, the problem formulation results can be cast as the optimization of one of two objectives. The first is the minimization of the total energy used per successfully received bit, and the second is the maximization of throughput \cite{KD10}.
Simulations of the resulting system provide very large gains, close to an order of magnitude. However, it should be recognized that this gain is in terms of the transmitted and received energy; it excludes the bulk of energy consumed by the system in an inefficient manner. Nevertheless, it shows the importance of judiciously choosing transmit parameters for energy, and not power, efficiency.
\subsection{Traffic-Adaptive Cells}
%\subsection{Automated Frequency Planning}
\label{sec:AFP}
Ideal locations for RBSs and the allocation of different frequency bands involve understanding the electromagnetic wave propagation conditions in a given environment, together with user traffic patterns. For cellular networks, this task requires a detailed engineering study. It is usually carried out by professionals with specialized training, using tools that provide a description of the terrain, electromagnetic wave propagation within that terrain (including parameters such as {\em path loss\/} which describes  how the electromagnetic waves are attenuated within a particular location), user population, and measured or estimated traffic patterns. The goal of this effort is to increase the capacity of a given network, or the total user throughput. In addition, the problem may be formulated as increasing revenue for the service provider. This part of the design of a cellular network is usually considered to be the most challenging and time-consuming activity of the overall effort.

There have been attempts to automate this process. For example, considering that in an enterprise, the deployment of a wireless LAN is the responsibility of information technology professionals, who often lack the knowledge and skills needed to complete a frequency plan, Cisco Systems, Inc. produced a tool to aid in the deployment of a wireless LAN \cite{CiscoSwan}. With this tool, system administrators can conduct assisted site surveys (a walk through in order to determine propagation and interference conditions within the enterprise, can be restricted to specified areas only). As a result, wireless LAN settings including optimal AP locations, frequency selection, transmit power, and cell coverage are achieved. The system can provide self-healing reconfiguration of the network in the case of AP failures \cite{CiscoSwan}. An approach for wireless LANs based on sophisticated propagation models such as ray tracing has been described in \cite{WSLWH04}. Since \cite{CiscoSwan}, a number of similar systems have been developed. A desired capability in such systems is that, in addition to the original network design and recovery in the case of failures, they adapt the frequency plan according to the traffic demand, potentially turning off APs not needed during off-peak hours, or turning on inactive APs when there is increased traffic. There are many scenarios such a system can be useful, some examples being university campuses, malls, etc. Such a system adaptively responds to the changes in traffic and can be called a traffic-adaptive system, and its cells traffic-adaptive cells.

For cellular networks, automatic frequency planning has been investigated in, e.g., \cite{HBW00,AHTW01,PNN01,LANMP08}. However, none of these approaches target energy efficiency as its main goal.
In a similar fashion to what we described for wireless LANs above,
the use of automatic frequency planning in cellular networks can be envisioned in two stages. In both cases, the interest is on cells with small size as described above, as well as a two-tier structure for low-mobility and high-mobility users. During the first stage, the goal is to determine locations of RBSs, operating frequencies, etc. Both propagation models and field measurements can be employed to complete this task. In the second stage, automatic frequency planning can be employed continuously, and the frequency plan can be evolved dynamically, according to user traffic and mobility patterns. As we specified in the previous paragraph, we call this mode of operation traffic-adaptive cells.

\begin{figure}[!t]
\begin{center}
\ifCLASSOPTIONonecolumn
\includegraphics[width=110mm]{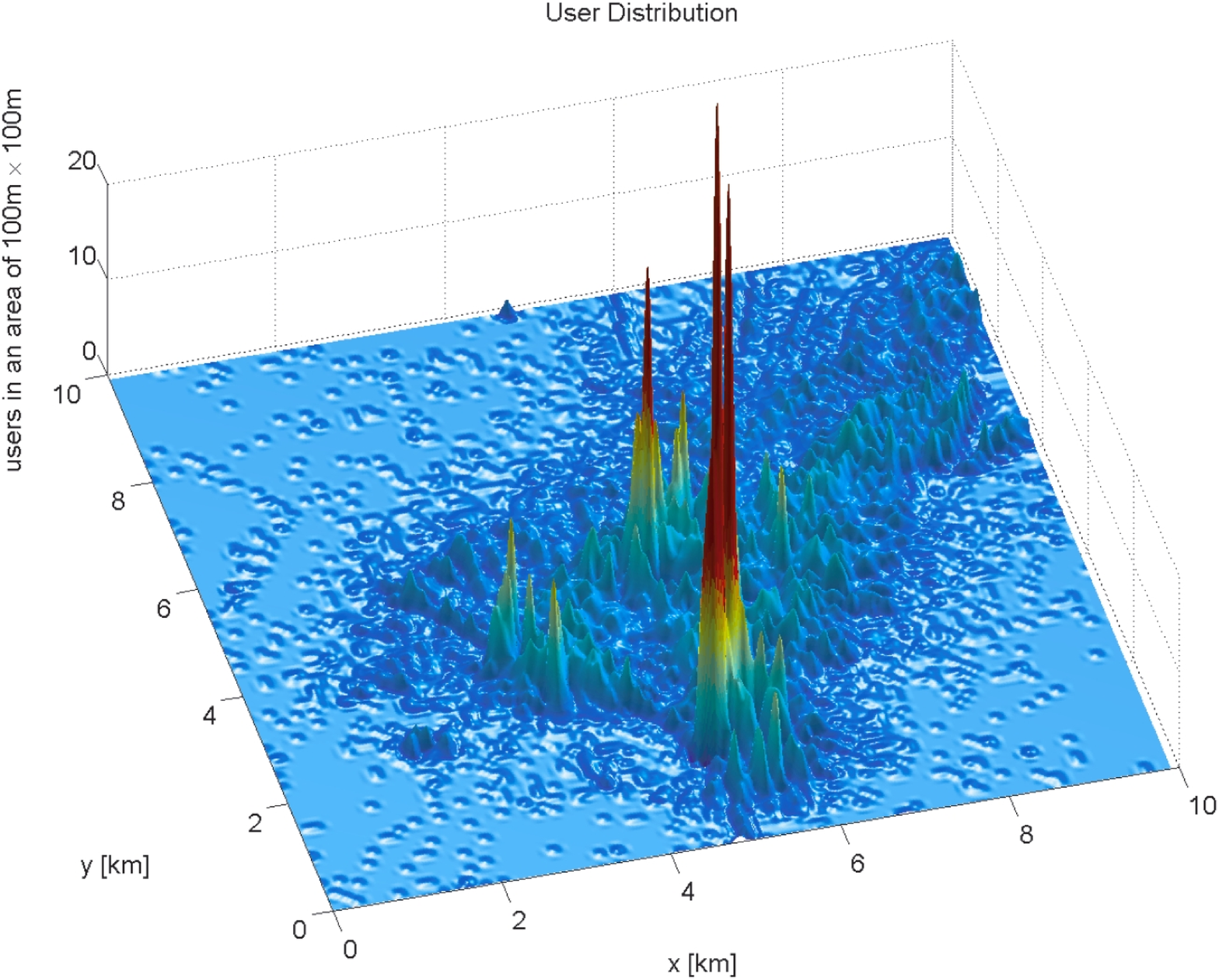}
\else
\includegraphics[width=90mm]{Wellington2.eps}
\fi
\caption{Distribution of cellular traffic in Wellington, New Zealand \cite{CWP08,Claussen09}.}
\label{fig:wellington}
\end{center}
\vspace{-6mm}
\end{figure}

Fig.~\ref{fig:wellington} shows traffic distribution in a $10\ {\rm km}\times 10\ {\rm km}$ area in the city of Wellington, New Zealand \cite{CWP08,Claussen09}. According to this figure, there are parts of the city, corresponding to the $(x, y)$ coordinates in km of approximately (4, 2) and (4, 4), for example, where the user traffic shows a ratio of about 20 to 1 or more as compared to most of the rest of the city. We note that there are other peaks with smaller ratios of user traffic. We note that this distribution also varies with time. We would like to point out that this figure illustrates the enormous potential that can be exploited in terms of the traffic variation in space. There is a similar potential in terms of the traffic variation in time. Combined, the potential for improving the energy consumption is tremendous.

\begin{comment}
We will provide an example of the kind of mathematical optimization needed, below. Reference \cite{BBDFGHHMPRSWW05} supplies a formulation for dynamic optimization in networks. In \cite{BBDFGHHMPRSWW05}, there are multiple RBSs that can be transmitting data to a particular RT. The optimization problem is based on minimizing power, particularly, the power load of the most heavily loaded cell. %The index $m$ represents the mobile (RT) and the index $c$ represents the cell (RBS).
The quantity $p_{mc}\ge 0$ is the power cell $c$ provides to mobile $m$, $\alpha_{mc}$ is the propagation loss, $\gamma$ is the target Signal-to-Interference-plus-Noise Ratio (SINR), and $\eta$ is the noise floor. Then, the optimization can be formulated as
\begin{equation}
\min_{p_{mc}}\max_{c}L_c\hspace{5mm} \textrm{subject to}\ \sum_c \alpha_{mc}p_{mc}\ge \gamma (\eta+\sum_c\alpha_{mc}L_{mc})\ \textrm{for all }m\label{eq:3}
\end{equation}
where $L_{mc}=\sum_{n\neq m}p_{nc}$, $L_{c}=\sum_{n}p_{nc}$. This is a linear program (LP). It can be solved by a standard solver or by a distributed approximation as in \cite{BBDFGHHMPRSWW05}.
\end{comment}

The following example illustrates a mathematical framework to incorporate load balancing in multiuser multicell CDMA systems. Reference~\cite{BBDFGHHMPRSWW05} formulates a joint cell selection and power allocation problem. The objective of the optimization problem is to minimize the load of the most heavily loaded cell. The quantity $p_{mc} \geq 0$ denotes the transmit power of RBS $c$ to mobile $m$, $\alpha_{mc}$ is the propagation loss, $\gamma$ is the target Signal-to-Interference-plus-Noise Ratio (SINR), and $\eta$ is the noise floor. The authors incorporate the soft-handover scenarios in CDMA systems such that multiple RBSs can transmit to the same user in the overlapping coverage areas. Then, the problem can be cast as a linear program (LP) optimization as
\ifCLASSOPTIONonecolumn
\begin{align}
\min_{p_{mc}}\max_{c}\hspace{.05in} L_c\hspace{5mm} \textrm{subject to}\ \sum_c \alpha_{mc} \hspace{.05in}p_{mc}\ge \gamma (\eta+\sum_c\alpha_{mc} \hspace{.05in}L_{mc})\ \textrm{for all }m\label{eq:3}
\end{align}
\else
\begin{align}
\min_{p_{mc}}\max_{c} & \hspace{.05in} L_c\hspace{5mm}\label{eq:3}\\
 \textrm{subject to}\ & \sum_c \alpha_{mc} \hspace{.05in}p_{mc}\ge \gamma (\eta+\sum_c\alpha_{mc} \hspace{.05in}L_{mc})\ \textrm{for all }m\nonumber
 \end{align}
\fi
where $L_{mc}=\sum_{n\neq m}p_{nc}$, $L_{c}=\sum_{n}p_{nc}$. Notice that the constraint enforces the power control, while the objective balances the load such that the varying user distribution and traffic conditions can be handled. This problem can be solved by a standard solver or by a distributed approximation as presented in \cite{BBDFGHHMPRSWW05}.

Reference \cite{GGFPTKC11} provides an example of exploiting the nonuniform distribution of traffic as in Fig.~\ref{fig:wellington}. This reference considers a conventional urban or rural cellular network, in which the problem is to place several small base stations to reduce the overall use of energy. It is formulated as finding the optimal number and position of base stations so that the long term total energy consumption is minimized. This problem is cast as a {\em facility location problem\/} \cite{AMO93} from the field of operations research. In this problem, the goal is to find the facility locations that minimize the cost of installation (fixed cost) and the unitary cost of transporting products from these facilities to each customer (variable costs). Reference \cite{GGFPTKC11} proposes a Mixed Integer Programming (MIP) model for the solution of the problem. The proposed scheme can provide optimal allocations considering different user patterns. Different traffic distributions in the coverage area at different times of the day can be taken into account. We note that simulation results in \cite{GGFPTKC11} show that the power reductions of 96\% or more can be achieved for the example scenario that was considered in the paper.
\subsection{Relays and Cooperation}\label{sec:RC}
\begin{figure}[!t]
\ifCLASSOPTIONtwocolumn
\vspace{3mm}
\fi
\begin{center}
\includegraphics[width=70mm]{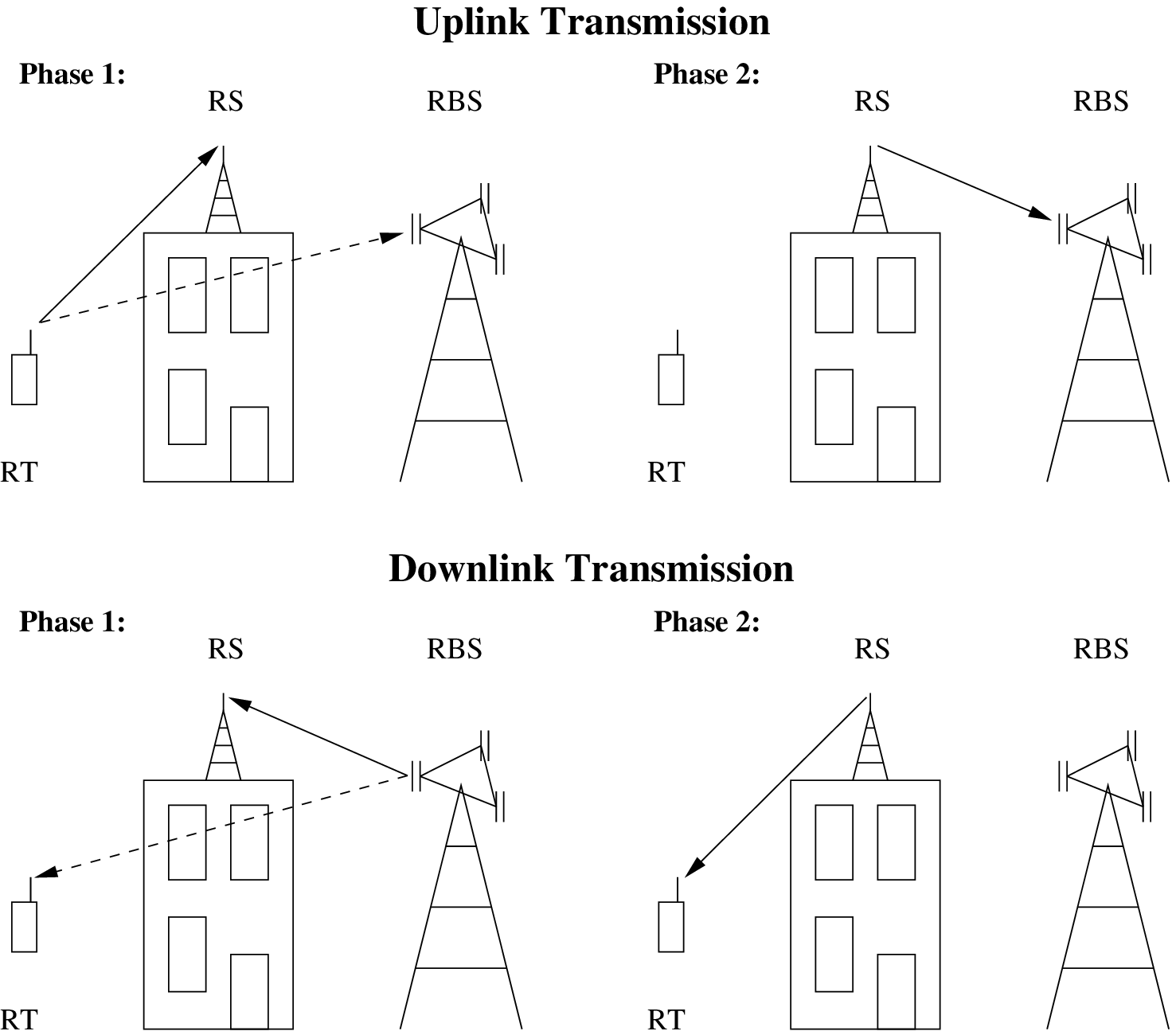}
\caption{Relay (RS) cooperating with RT and RBS for improved communication.}
\label{fig:relay}
\end{center}
\vspace{-5mm}
\end{figure}

It has been shown that cooperation in communication networks leads to improvement in performance, e.g., \cite{SEA03,LTW04}. Cooperation via relays has been standardized in both WiMAX \cite{PH09} and LTE systems \cite{36814}. Relay deployments in cellular networks are not common at present, but it is possible that they would be deployed in the future. The relays can be dedicated network elements, placed at certain planned or unplanned locations in a cell to help forward the message in the uplink or the downlink as shown in Fig.~\ref{fig:relay}. It can be assumed that they are half-duplex, or they cannot simultaneously receive and transmit, due to the limitations of the wireless medium. As illustrated in Fig.~\ref{fig:relay}, relays can be placed on rooftops, lamp posts, or building walls. There are a number of studies in the literature that investigate the employment of user terminals for relaying traffic that belongs to others. There has been a significant amount of research on cooperation in wireless networks via relays since the publication of the first papers on the subject \cite{SEA03,LTW04}. Although relays are typically used by network operators for cost-effective ways of providing cell coverage and capacity enhancement, recent studies in \cite{RDynamicResourceAssignment,RMadanEnergyEfficient,RHanzo,RHanGreen} show that they can also be used to increase the energy efficiency and achieve significant energy savings.

There are several factors that determine the performance of cooperative relay systems. The number and location of relay nodes, the channel conditions of source to relay and relay to user links, relay selection methodologies, relay capabilities, resource assignment and sharing protocols between source and relays nodes are just to name a few. In \cite{RDynamicResourceAssignment}, the downlink performance of decode-and-forward cooperative relays with a dynamic resource assignment method is studied along with comparisons to the single relay selection method. The key point is that when relays cooperate, they can provide significant gains over selecting only a single relay. For various simulation scenarios of urban macrocell and urban microcell models, \cite{RDynamicResourceAssignment} shows that relay cooperation can provide four times more throughput for the cell edge users and $11\%$ improvement on average cell throughput compared to the case without relays. Therefore, \cite{RDynamicResourceAssignment} shows that cooperative relay is an effective technique to further improve cell coverage and cell edge user experience.

As we mentioned above, relay selection and the number of cooperative relays closely determine the energy savings that can be achieved. In \cite{RMadanEnergyEfficient}, the energy efficiency of different relay selection methods are compared with practical considerations including the signaling energy costs for obtaining channel state information (CSI). In terms of implementation, there is a tradeoff between decreasing
data transmission energy to use more relays and decreasing the overhead energy for CSI acquisition to use fewer relays. Therefore, the question arises as to which and how many relays need to cooperate. Three different relay selection methods are investigated in \cite{RMadanEnergyEfficient}. The first is called single relay selection where the relay with the highest relay to user channel gain is selected. Second relay selection method is called $(M-1)$ relay selection where the best $(M-1)$ relays, where $M$ denotes a constant number, are selected. The third is called the optimal selection where the number and the subset of cooperative relays vary from time to time. The results in \cite{RMadanEnergyEfficient} depict that the energy savings up to $16\%$ are achievable with optimal relay cooperation compared to single relay selection which accounts for the effects of CSI energy overhead, but unfortunately, do not provide the gains compared to the case without relays. The findings in \cite{RMadanEnergyEfficient} show that the optimal number of relays for cooperation varies in time and is a function of system parameters. Furthermore, the relay selection strategy and the cost of training and feedback of CSI are significant in determining the achievable energy savings.

Although the focus of our interest is in the downlink, cooperative relays provide significant gains in the uplink as well. The uplink energy efficiency of amplify-and-forward type cooperative relays is investigated in an LTE deployment in \cite{RHanzo}. The benefits of relay deployments provide additional frequency and spatial diversity. The results in \cite{RHanzo} indicate that $9.7$ $\mathrm{dB}$ power reduction can be achieved compared to the non-cooperation case with multi-user relay selection. The study in \cite{RHanzo} is extended in \cite{RHanGreen}, which studies multi-user relay selection scheme under different shadowing parameters varying between $0$ and $8$ $\mathrm{dB}$. The performance of a system with multi-user relay selection is compared to a system without relays in \cite{RHanGreen}, and the results indicate that at low target rates, the cooperative relay systems can achieve $8$ times less power consumption compared to the system without relays while providing the same data rates. %Note that this is a very significant power savings in order to extend the battery life of a user terminal.
However, as the target rates of the users increase, the power savings vanish and the power consumption of the system with relays approaches to the system without relays \cite{RHanGreen}.

Use of relays for improving energy efficiency could be counterintuitive since they add energy consumption to the network. However, publications such as \cite{RDynamicResourceAssignment,RMadanEnergyEfficient,RHanzo,RHanGreen} have established that a gain can exist, while estimating the amount of the gain within limits based on a number of assumptions. When it comes to where to place the relays and how to operate them for maximum energy efficiency, less work has been performed or reported. An example of such a study is \cite{YLS10}, where optimum relay location and optimum relay selection algorithms have been developed for energy-efficient cooperative cellular networks. The work was developed for asymmetric traffic. Reference \cite{YLS10} succeeds in developing a joint uplink and downlink relay selection algorithm. We consider the work in \cite{YLS10} a successful start but believe that a much broader undertaking is needed. It is especially important to test the algorithms to be developed in simulations with realistic channel and network conditions.

Recently, a new relaying approach where the RTs function as relays, called Device-to-Device (D2D) communications, is under consideration. %The close proximity of RTs enables the D2D communications that overlay the cellular networks to achieve significantly high bit rates, low delays, low power consumption, and better resource utilization~\cite{D2DDesignAspects12}.
Studies have demonstrated that D2D communications can increase system throughput~\cite{D2DDoppler09Mag,D2DResourceSharing,D2DInterferenceAware,D2DModeSelection}, provide larger coverage~\cite{D2DDesignAspects12,D2DModeSelection}, and improve energy efficiency~\cite{D2DDesignAspects12,D2DAdvances}. The work in \cite{D2DDoppler09Mag} shows that D2D communications increases system throughput up to 65$\%$ compared to the case where only cellular network communication is used. Reference \cite{D2DInterferenceAware} identifies that an increase in the median cell capacity of 2.3 times can be achieved. Potential energy efficiency improvements of D2D communications are presented in \cite{D2DDesignAspects12} as a function of the maximum D2D pair distance. For short distances, 10 times energy efficiency is claimed to be possible \cite{D2DDesignAspects12}. Although these results are impressive, D2D brings a number of challenges. First, when to use the D2D or cellular link needs to be identified. This is known as the mode selection problem. Energy-efficient measurement and reporting mechanisms to be employed during mode selection, robust as well as scalable for dense networks, need to be developed. The time scale and speed of updating these mechanisms need to be quantified.
%References \cite{D2DModeSelection,D2DAdvances,D2DDesignAspects12} location and scheduling algorithms, power control of D2D links, and identifying the roles of the network, in particular the RBS, in D2D communications.
References \cite{D2DModeSelection,D2DAdvances,D2DDesignAspects12} demonstrate some progress, but more needs to be done. Further research directions are the role of the network or, in particular, the RBS, in D2D communications, and resource allocation, scheduling, and power control of the D2D links.
\subsection{Multiple Antenna Techniques}
\label{sec:DA}
\begin{figure}[!t]
\ifCLASSOPTIONtwocolumn
\vspace{4mm}
\fi
\begin{center}
\includegraphics[width=70mm]{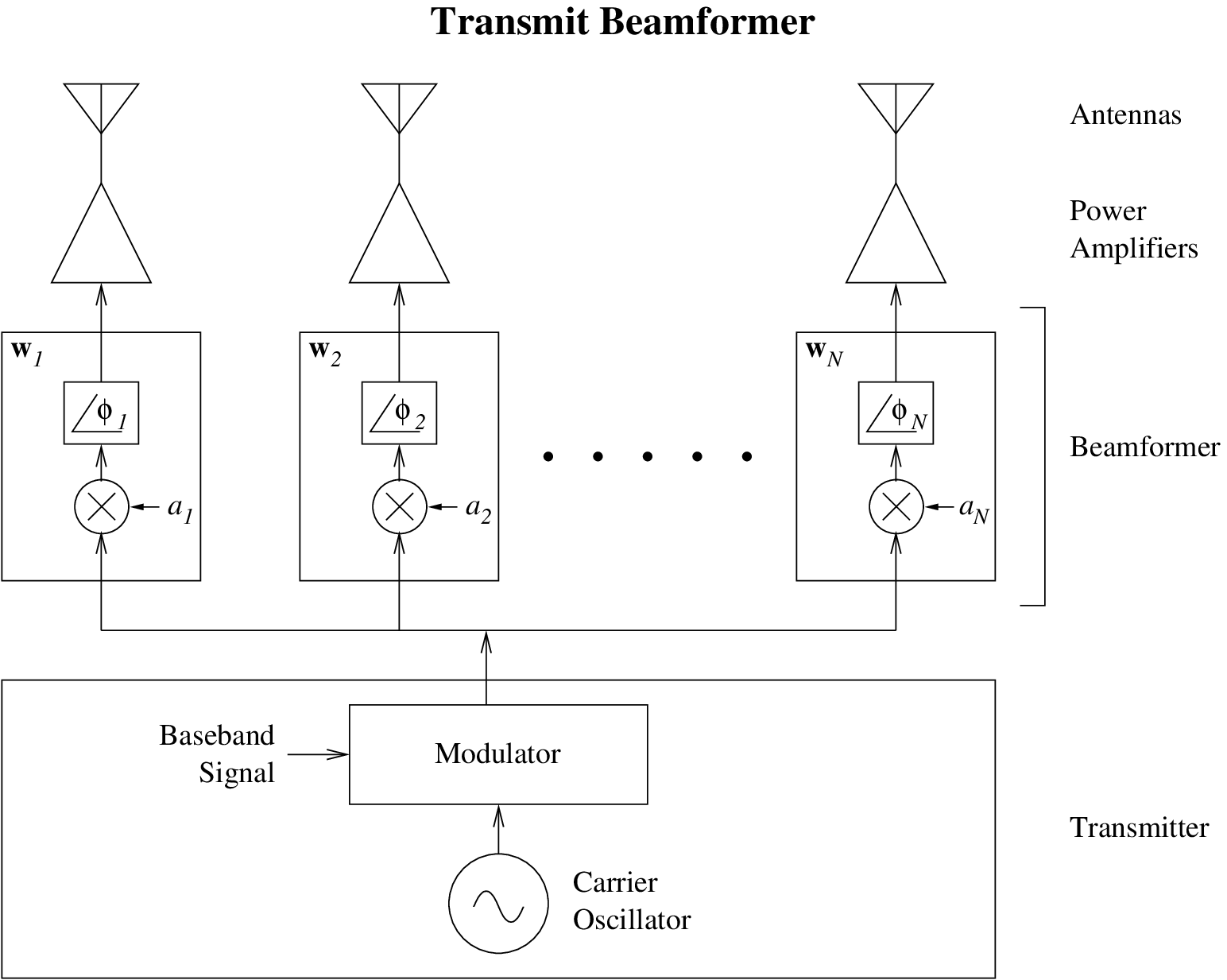}%\\[-3mm]
\ifCLASSOPTIONonecolumn
\hspace{5mm}
\else
\vspace{2mm}
\fi
\includegraphics[width=70mm]{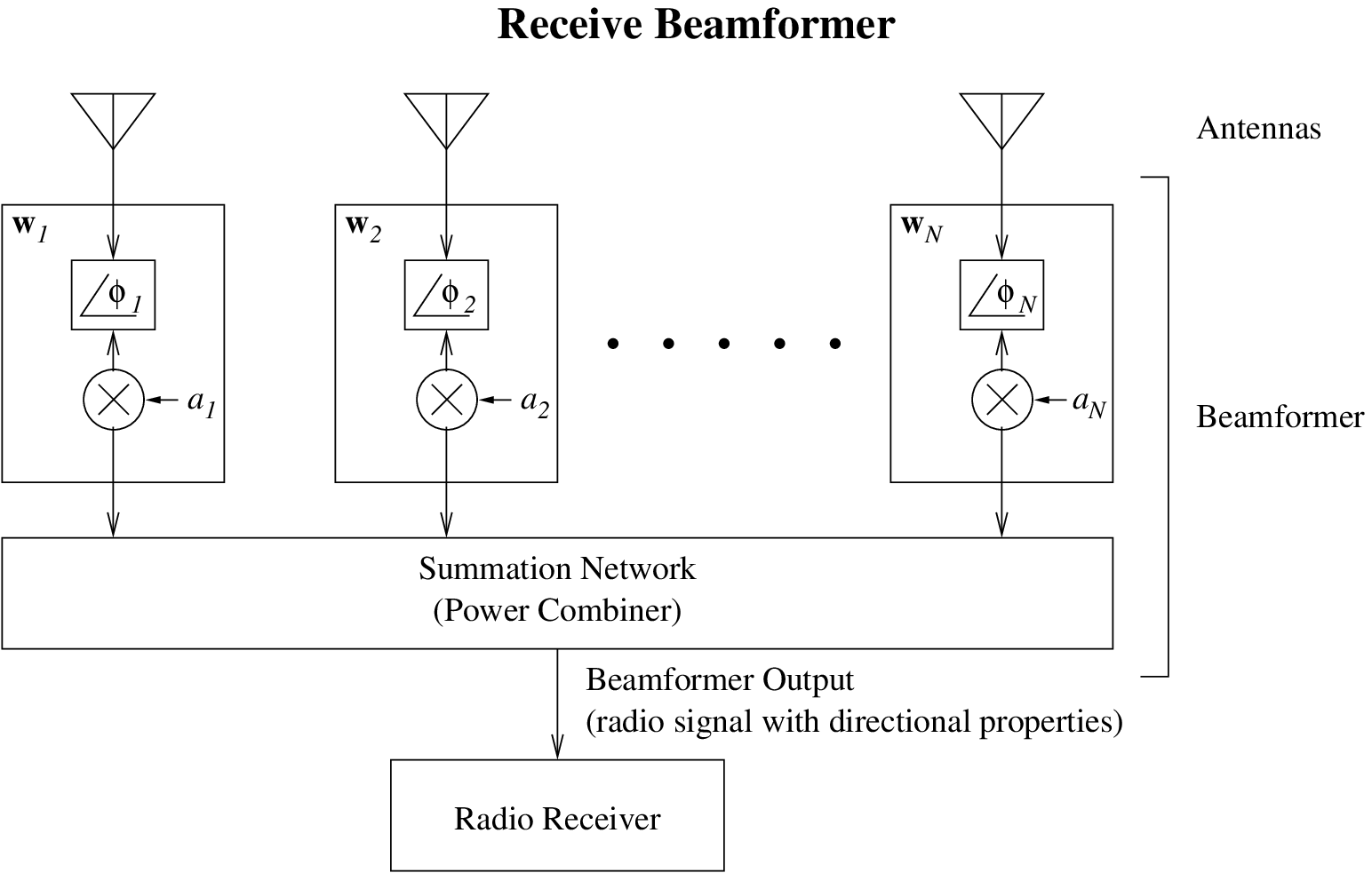}%\\[-2mm]
\caption{Sample implementations of transmit and receiver beamforming for realizing directional antennas.}
\label{fig:beamform}
\end{center}
\vspace{-5mm}
\end{figure}
There have been efforts to employ directional antennas in cellular networks since the 1990s. This can be effectively achieved by using multiple antennas. This research area was known as ``smart antennas'' \cite{PP97,Godara04}. It is related to Multi-Input Multi-Output (MIMO) systems, a technique for increasing the capacity and reliability of wireless communications systems through the use of multiple antennas
at the transmitter or receiver \cite{MM01,MM02,MM03,MM04,Paulraj04}. MIMO systems are used {\em i)\/} to increase the diversity order of the system, which means improving the BER vs SNR performance, or {\em ii)\/} establishing spatial multiplexing, which means generation of independent data streams and therefore increasing throughput. As shown in Fig.~\ref{fig:beamform}, beamforming or beamsteering can be employed for an increase in SNR, without any diversity or spatial multiplexing gain, or, in an existing MIMO system, they can be employed to provide additional gain \cite{CiscoBeamform,SP08,DPSB08,KC04,Callisch11}. This additional gain can be used for increased power or energy efficiency. Beamforming or beamsteering techniques can be employed at the RBSs \cite{CiscoBeamform,SP08,DPSB08,KC04,Callisch11,AS92,PBF06} or at the RTs \cite{YZS10,Yu10}. Due to antenna reciprocity, beamsteering techniques provide gain during transmission as well as reception. Therefore, RBS beamforming can improve power efficiency for the RBS in the downlink direction. It can improve power efficiency for the RT in the uplink direction. Similarly, RT beamforming would improve power efficiency for the RT in the uplink, and it would improve power efficiency for the RBS in the downlink. With or without MIMO, employing beamforming at the RBS has been discussed in the literature, but employing beamforming at the RT is not as widely studied \cite{YZS10}. It has been shown in \cite{YZS10} that using beamsteering in RTs with 4 antennas can improve power efficiency by 55\%. Although form factors or sizes for RTs is an important consideration, it is actually possible to place 4 antennas on RTs at the frequencies of interest. A power efficiency improvement figure of 55\% is very significant. However, as stated in \cite{YZS10}, there are a number of challenges to overcome, before energy efficiency gains can become a reality, to employ beamsteering at the RTs.

While it is known that the power efficiency increases with the number of antenna elements, a question remains as to how many antennas should be employed at RT and RBS. For RT, with today's technology and with today's operating frequencies, space on the terminal is the basic limitation against increasing the number of antennas since the minimum space is a fraction of the wavelength, usually considered to be half the wavelength for maximum advantage \cite{TV05}. Reference~\cite{Vainikainen09} states that, for future applications, multiple radios will likely be needed in RTs. It states about five antennas will be used for each radio. Reference~\cite{Quintel} discusses how many antennas are needed for the future, and although concludes that the number may increase with innovation, with today's technology one antenna can be replaced with about five antennas. On the other hand, at RBS, a large number of antennas can be employed. Recently, the possibility of using a very large number of antennas, perhaps in the hundreds, is being considered from a signal processing viewpoint, with substantial improvements in power efficiency \cite{MM05,MM06,MM07}. We will discuss this important subject in a subsection in the sequel. But, since it was proposed earlier, we will discuss another extension of basic MIMO first.
\subsubsection{Coordinated Multipoint (CoMP) Transmission and Reception}\label{sec:comp}
MIMO wireless systems are now part of current communication standards such as 802.11, LTE, and WiMAX, and are deployed throughout the world. The improvement in spectral efficiency using regular MIMO, while important, has been relatively modest. A different approach, known originally as Network MIMO, was developed to alleviate that fact.

The concept of Network MIMO is based on the following observations. As in conventional cellular wireless systems, MIMO techniques employ frequency reuse within each cell, but are still subject to high levels of interference from other cells. In order to achieve significant achievements in spectral efficiency, such intercell interference will have to be addressed. It can be observed that, for example in the uplink, intercell interference is a superposition of signals intended for other RBSs that have been collected at the wrong place. If, however, the different RBSs were cooperating, they can resolve the interfering signals as long as sufficient degrees of freedom exist. Making the RBSs coordinate can be challenging, but if they all share a common backbone network, it is possible. A similar observation can be made for the downlink. In this mode of operation, the frequency reuse approach is actually not needed, contributing to significant increases in spectral efficiency. Also, in this model, the uplink and downlink channels are not interference channels, the uplink channel becomes a multiaccess channel, while the downlink channel becomes a broadcast channel. In both cases, the channels are simultaneously used by multiple users. This concept, with the name Network MIMO, was first developed in \cite{KFVY06,KFV06,FKV06}. It has been adopted by the LTE community very quickly and is now part of the LTE-Advanced standard. In this process, its name has become Coordinated Multipoint, abbreviated as CoMP, see e.g., \cite{MF11}.

CoMP is a range of different techniques that enable the dynamic coordination of transmission and reception over a variety of different RBSs. The goal is to improve the overall quality for the user, and at the same time, improving the utilization of the network. CoMP achieves this goal by turning the Inter-Cell Interference (ICI) into a useful signal. This is especially critical for users at the cell boundaries where performance by conventional approaches may be degraded.

In its standardized form, CoMP has two versions of implementation. The first of these is called {\em Joint Processing,\/} and the second is called {\em Coordinated Scheduling or Beamforming.\/} These techniques are applicable to both the downlink and the uplink, however, their implementations differ between the downlink and the uplink. Joint processing employs multiple RBSs for simultaneous transmission or reception of data, as in Network MIMO. This form of coordinated multipoint transmission or reception places a high demand on the backhaul network because the data traverses the network for each of the RBSs. The second implementation of CoMP employs one RBS for transmission or reception, but for simultaneous transmission to multiple RTs, those RBSs employed for transmission or coordinated via scheduling or beamforming so that the effect of their interference is minimized. Unlike joint processing, this form of CoMP does not require substantial data transfer in the backhaul network. However, there is still coordination or beamforming data that needs to be transferred over the backhaul network.

\ifCLASSOPTIONtwocolumn
\begin{figure}[!t]
\vspace{1mm}
\begin{center}
\begin{minipage}[b]{1.0\linewidth}
\centering
\includegraphics[height=60mm]{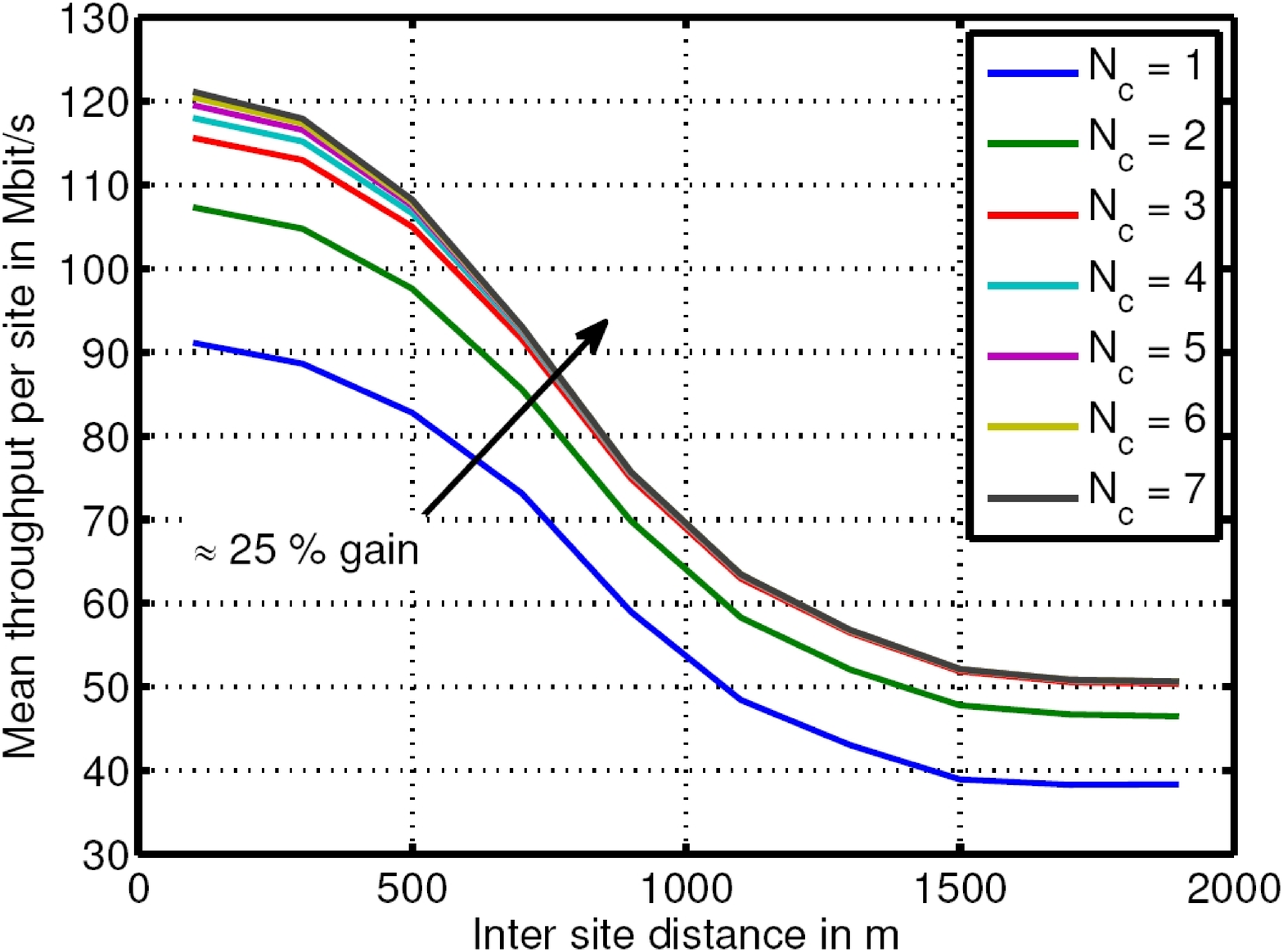}
\caption{Throughput per site for different inter site distances \cite{FMF10}.}
\label{fig:FMF10a}
\end{minipage}
\end{center}
\begin{center}
\begin{minipage}[b]{1.0\linewidth}
\centering
\includegraphics[height=60mm]{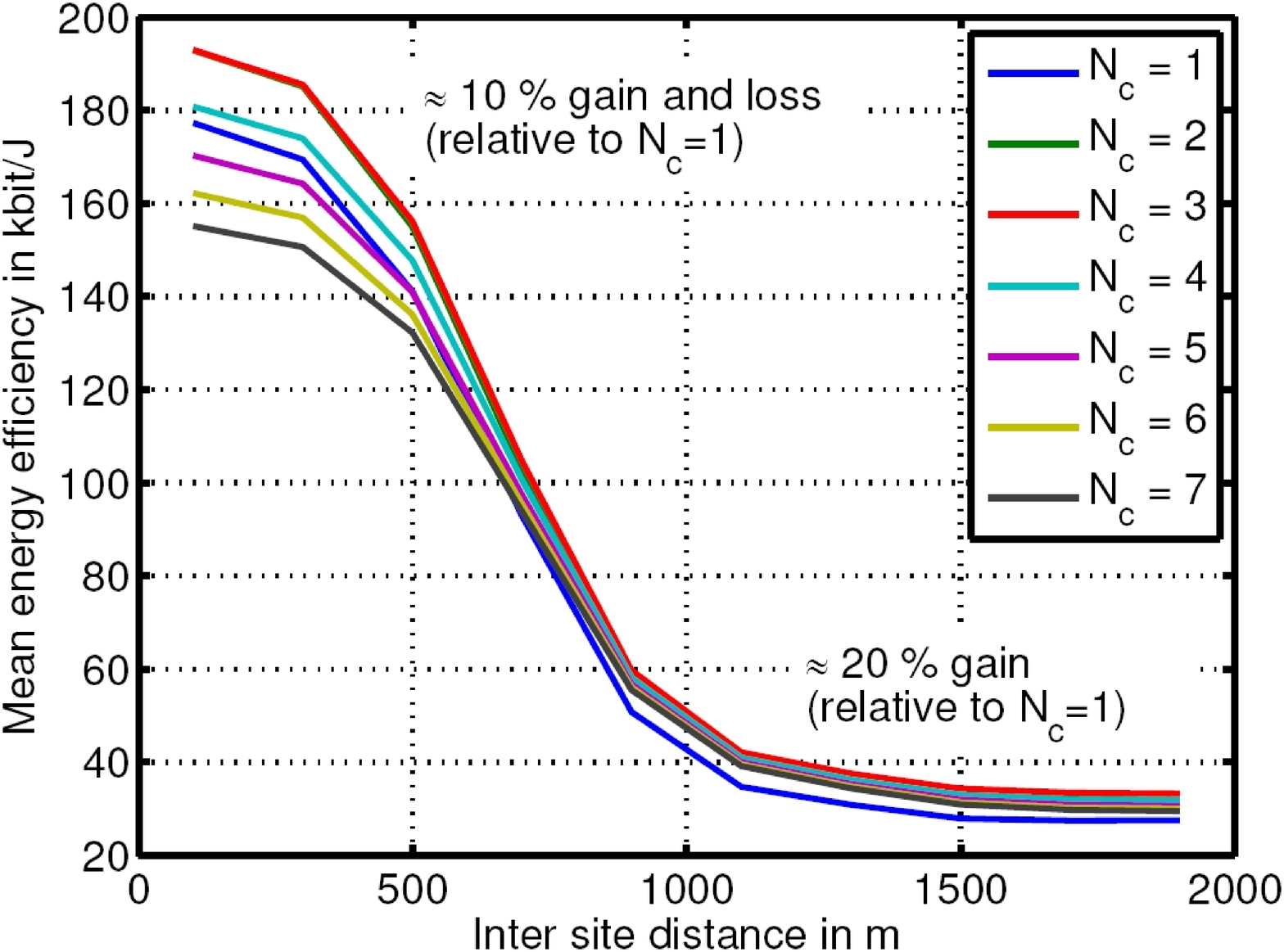}
\caption{Bit per Joule efficiency of CoMP schemes for different inter site distances \cite{FMF10}.}
\label{fig:FMF10b}
\end{minipage}
\end{center}
\vspace{-5mm}
% These curves are from \cite{FMF10}
\end{figure}
\fi
With simultaneous transmissions or backhaul data transfer, a natural question arises as to the energy efficiency of CoMP. This subject is discussed in \cite[Section 15.4]{MF11}, based on \cite{FMF10}. The work analyzes CoMP with joint processing as discussed earlier. An idealized hexagonal cellular structure and a number of models for the channel, propagation, backhauling, and energy consumption are assumed. The results are presented in terms of the inter distances of the RBSs and parametrized in terms of the number of cells participating in coordination. Two sets of results from \cite{FMF10} are replicated in Fig.~\ref{fig:FMF10a} and Fig.~\ref{fig:FMF10b}. Not surprisingly, the throughput and energy efficiency both drop with increasing inter site distance. Again, not surprisingly, the throughput gain increases with the number of cooperating cells, denoted by $N_c$, as shown in Fig.~\ref{fig:FMF10b} (the bottom curve corresponds to $N_c=1$, the one in the middle to $N_c=2$, and the clustered group begins with $N_c=3$, and increases linearly with $N_c$; however, the gain saturates at $N_c=3$).
Surprisingly, CoMP may in fact lead to decreased energy, as shown in Fig.~\ref{fig:FMF10b} (the topmost curve corresponds to $N_c=3$ while the $N_c=1$ curve has an intermediate value for inter site distances less than 700 m, and is at the bottom for more than 700 m). Based on these results, it is concluded that the best results are obtained with $N_c=3$.
\ifCLASSOPTIONonecolumn
\begin{figure}[!t]
\begin{minipage}[b]{0.42\linewidth}
\centering
\includegraphics[height=60mm]{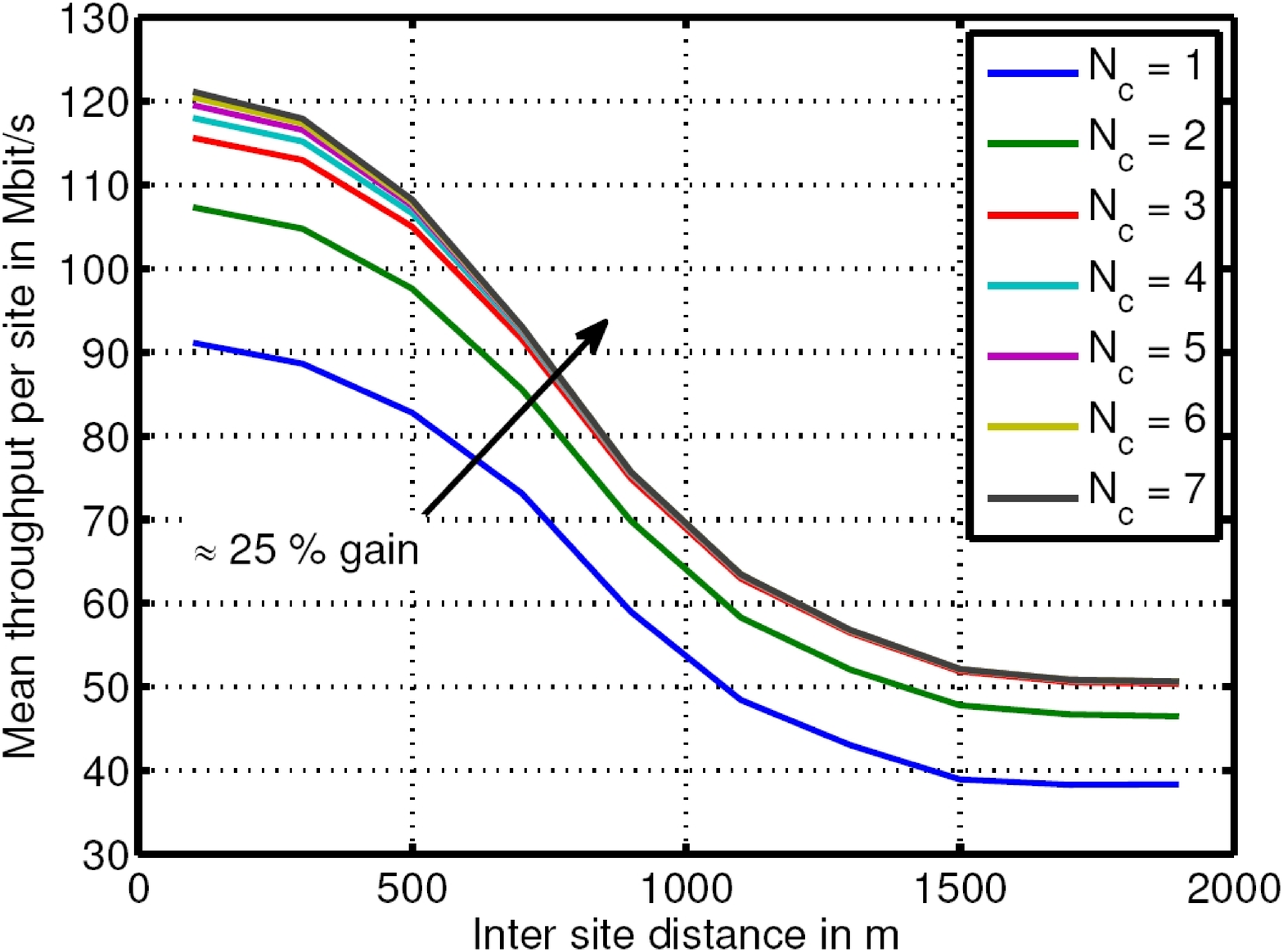}
\caption{Throughput per site for different inter site distances \cite{FMF10}.}
\label{fig:FMF10a}
\end{minipage}
\hspace{10mm}
\begin{minipage}[b]{0.42\linewidth}
\centering
\includegraphics[height=60mm]{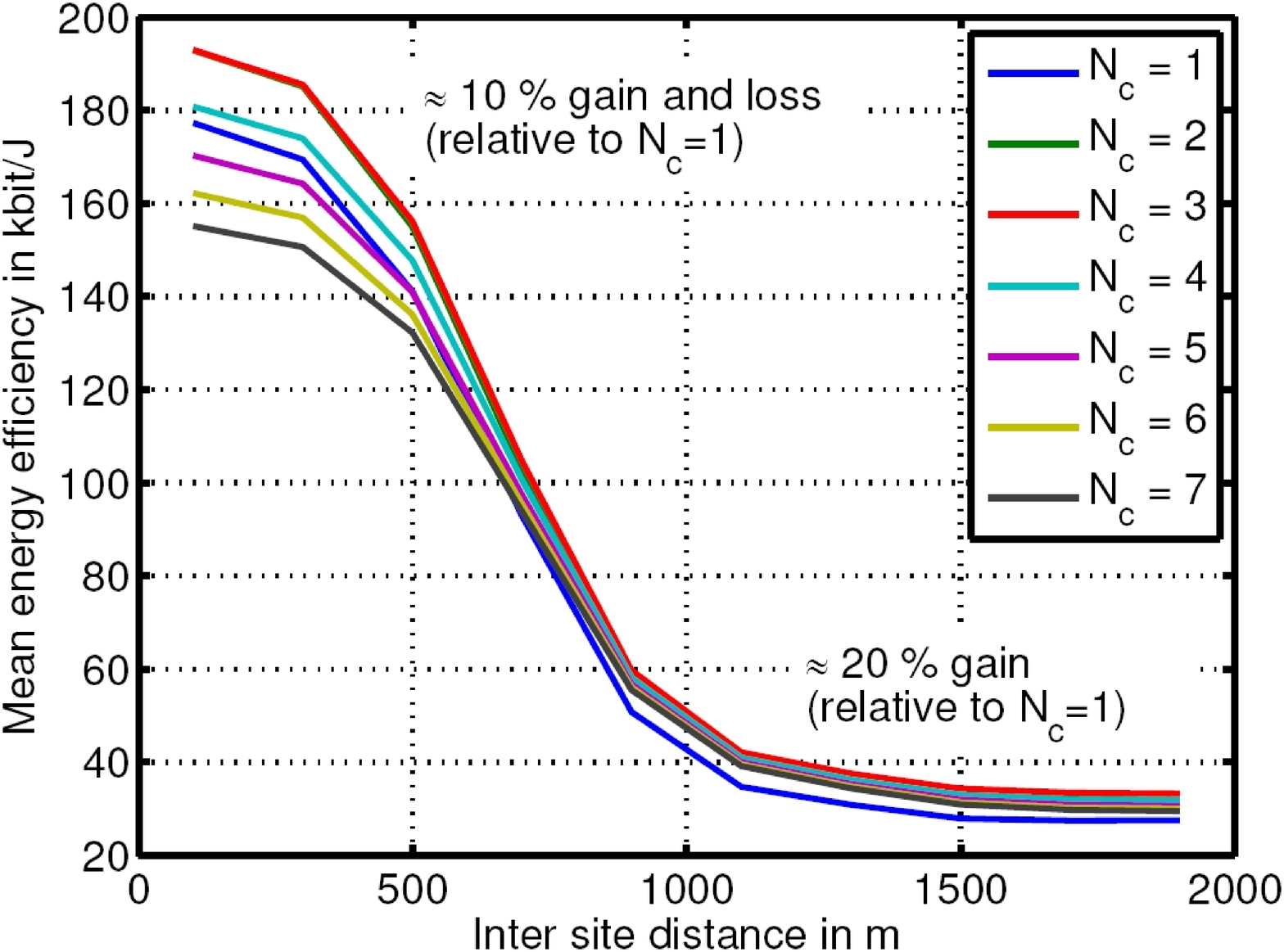}
\caption{Bit per Joule efficiency of CoMP schemes for different inter site distances \cite{FMF10}.}
\label{fig:FMF10b}
\end{minipage}
% These curves are from \cite{FMF10}
\end{figure}
\fi
\subsubsection{Massive MIMO}\label{sec:massivemimo}
We stated earlier that the improvement in spectral efficiency using regular MIMO, has been relatively modest. Typical MIMO installations use APs or RBSs with relatively few (e.g., less than 10) antennas. To achieve more dramatic gains than MIMO, a grander view of the MIMO concept envisions the use of orders of magnitude more antennas (e.g., 100 or more) at each AP or RBS, using a concept known as massive MIMO or Large-Scale Antenna Systems (LSAS). While the massive MIMO idea, first introduced in \cite{MM05,MM06}, seems very simple on the surface, there is a surprisingly rich mathematical theory that motivates its use. Asymptotic arguments can be used to establish that, as the number of antennas in a MIMO cell grows to infinity, the effects of uncorrelated noise and fast fading vanish, throughput and the number of terminals are independent of the size of the cells, spectral efficiency is independent of bandwidth, and the required transmitted energy per bit vanishes \cite{MM06}. Moreover, these advantages can often be achieved using relatively simple linear signal processing approaches such as Maximal Ratio Combining (MRC) and Maximal Ratio Transmission (MRT). For example, it was shown in \cite{MM06} that under realistic propagation assumptions, such approaches could in principle achieve an average per-user throughput of 17 Mb/s for 40 users in a 20 MHz channel in both the uplink and downlink directions, with a mean throughput per cell of 730 Mb/s and an overall spectral efficiency of 26.5 bps/Hz. Reference \cite{MM07} showed that it is possible to capture a significant amount of the gain with a finite number of antennas, depending on the degrees of freedom per user terminal that the channel offers. Simulations using measured channels taken from arrays of over 100 antennas at 2.6 GHz have confirmed that very high spectral efficiencies are possible in realistic propagation environments \cite{MM08,MM09}. Since the number of antennas is envisioned to be significantly larger than the number of active co-channel users, a large number of degrees of freedom are available. Most of the algorithms proposed for massive MIMO systems are kept simple in order to make the complexity of the system low.

Another advantage of the massive MIMO concept lies in its efficient handling of power generation/amplification
compared with a corresponding single-antenna system with the same aperture. In principle, each RT in the uplink of
a massive MIMO system requires only $1/N_t$ of the overall power to transmit, thereby significantly relaxing the
design requirement of the RT front-end PA, where $N_t$ is the number of transmit antennas.
%If the overall aperture size of the $N_t$ antennas is assumed to be the same as the aperture in the single-antenna system, and if the same overall transmit power is available,
A similar reasult can be expected to hold in the downlink. If this is the case, then in the massive MIMO case each individual PA needs to generate only $1/N_t$ of the output power. This leads to better energy efficiency.
Considering the very large number of antennas, the resulting gains in energy efficiency can be very substantial.
\subsection{Sleeping Mode for the Radio Base Station}
\label{sec:SM}
Based on the statistics of power consumption in RBSs, especially considering how much variation exists over time and how much power is used for transmitter idling, the idea of adding a sleeping mode to the RBSs is often suggested, e.g., \cite{Grant10,OKLN11,MCCM09,NWGY10,TRKP10,SE10}. Reference \cite{Grant10} mentions cutting off channels above 1.8 GHz at night. The study in \cite{HowMuchEnergy11} proposes an RBS power consumption model capturing the effects of each RBS component. This model shows that the power consumption at an RBS grows linearly with the number of transceiver chains. The authors also propose a long-term traffic model capturing the spatial and temporal traffic variations. Their simulations based on this traffic model show that by adding micro-sleep modes during idle transmit intervals, the system can achieve 15-20$\%$ energy savings without any performance degradation. Reference \cite{OKLN11} discusses turning off a number of RBSs in a cellular network so that the coverage of passive RBSs is assumed by other RBSs that remain active, or with the aid of range extension, multihop relay, or multipoint coordination. An interesting point made by \cite{OKLN11} is that if different service providers were to share traffic during off-peak times, they could all share a total of about an additional 30\% savings in total energy consumption. Reference \cite{MCCM09} reports results of simulations with a number of different network topologies. This study shows that when a sleeping mode is adopted, it can result in about 25-30\% savings of energy consumption. The study assumes the traffic is uniform and also, when a cell is switched off, its coverage can be filled by its neighbors. %We find both of these assumptions unrealistic.
In a similar fashion, \cite{NWGY10} also considers turning off cells in a cellular network and assumes that the neighboring cells can fill the coverage.

Reference \cite{TRKP10} states that for reasons such as meeting regulatory requirements, security, maintaining coverage, etc., completely turning off cells or BSs may not be possible. Instead, it proposes to turn off a number off carrier frequencies or high data rate services so that some savings in energy can be accomplished while the coverage is maintained. This reference states turning off a carrier frequency corresponds to turning off a card in an RBS and results in energy savings by reducing the transmit power and the energy needed for cooling. We would like to add to these the important contribution of DC power needed to keep the transmitter biased even though there may be no transmissions on this carrier frequency. The problem is formulated as an integer linear programming problem. The variables are as follows.
\begin{comment}
A day is divided into non-overlapping periods $\{1,2,\ldots,H\}$, $W_h$ is the percentage of the day of interval $h\in \{1,2,\ldots,H\}$, $d_i^h$ is the traffic demand in cell $i$ during the interval $h$, $C_i^f$ is the capacity of frequency $f$ at cell $i$, and $P_i^f$ is the power consumption in cell $i$ at frequency $f$, allowing for different values of $P_i^f$ due to potentially different cooling and transmission power requirements. The variable $x_i^{fh}$ is defined as a binary control variable where $x_i^{fh}=1$ if frequency $f$ is active in cell $i$ during $h$ and it is equal to 0 otherwise. Then the problem can be specified as an integer programming one as follows
\begin{equation*}
\min_{x_i^{fh}} \sum_{h=1}^HW_h\sum_{i=1}^N\sum_{f=1}^{F_i}P_f^ix_i^{fh}
\hspace{5mm} {\rm such\ that\ }\hspace{5mm}
\sum_{f=1}^{F_i}C_i^fx_i^{fh}  \ge d_i^h,\ x_i^{fh}  \in  \{0, 1\}\hspace{5mm}{\rm for\ all\ }h,i,f.
\end{equation*}
\end{comment}
%
A day is divided into non-overlapping periods $\{1,2,\ldots,H\}$, $W_h$ is the percentage of the day of interval $h\in \{1,2,\ldots,H\}$, $d_i^h$ is the traffic demand in cell $i$ during the interval $h$, $C_i^f$ is the capacity of frequency $f$ at cell $i$, and $P_i^f$ is the power consumption in cell $i$ at frequency $f$, allowing for different values of $P_i^f$ due to potentially different cooling and transmission power requirements. The variable $x_i^{fh}$ is defined as a binary control variable where $x_i^{fh}=1$ if frequency $f$ is active in cell $i$ during $h$ and it is equal to 0 otherwise. Furthermore, let $N$ denote the total number of cells and $F_i$ denote the frequencies assigned to cell $i$. Then the problem can be specified as an integer programming one as follows
\ifCLASSOPTIONonecolumn
\begin{equation*}
\min_{x_i^{fh}} \sum_{h=1}^HW_h\sum_{i=1}^N\sum_{f=1}^{F_i}P_f^ix_i^{fh}
\hspace{5mm} {\rm such\ that\ }\hspace{5mm}
\sum_{f=1}^{F_i}C_i^fx_i^{fh}  \ge d_i^h,\ x_i^{fh}  \in  \{0, 1\}\hspace{5mm}{\rm for\ all\ }h,i,f.
\end{equation*}
\else
\begin{align}
\min_{x_i^{fh}} & \sum_{h=1}^HW_h\sum_{i=1}^N\sum_{f=1}^{F_i}P_f^ix_i^{fh}\\
{\rm such\ that\ }
& \sum_{f=1}^{F_i}C_i^fx_i^{fh}  \ge d_i^h,\ x_i^{fh}  \in  \{0, 1\}\hspace{5mm}{\rm for\ all\ }h,i,f.\nonumber
\end{align}
\fi
\begin{table}[!t]
\ifCLASSOPTIONtwocolumn
\vspace{3mm}
\fi
\begin{center}
\caption{Energy Savings \cite{TRKP10}}
\label{tbl:sleep}
\begin{tabular}{|c|c|}
\hline
{Cell Type}&{Savings}\\
\hline
1&48\%\\ 2&45\%\\ 3&52\%\\ 4&65\%\\
\hline
\end{tabular}
\end{center}
\end{table}
\ifCLASSOPTIONtwocolumn
\vspace{-2mm}
\fi

By dividing the service area into business (1), residential (2), entertainment and shopping (3), and highway (4), and assigning somewhat realistic power consumption and traffic distributions, \cite{TRKP10} calculates the average energy savings per RBS type as in Table~\ref{tbl:sleep}. We find these savings very significant. Reference \cite{TRKP10} also studies turning off not only a number of carrier frequencies but also a number of high data rate services. Reported additional savings are only incremental. This fact, as well as the consideration that turning off services would not be desirable by customers, leads us to conclude that turning off services may not be a viable option.

Reference \cite{SE10} comes from the research organization of a service provider, France T\'el\'ecom, or Orange. It studies the problem under more realistic scenarios. Specifically, it uses data from a central part of Paris, France. Similarly to \cite{TRKP10}, it considers turning of a number of carrier frequencies at RBSs. It points out to the need for turn-on and turn-off times in shutting down frequencies and states the potential consequences of blocking call generation or handoff of some calls. Based on simulations of traffic in actual cells, it points to savings more significant than that reported in \cite{TRKP10}.

We conclude with the provision that completely shutting down an entire macrocell may indeed not be feasible in terms of satisfying user QoS, coverage, and regulatory requirements, whereas reduction of a number of carrier frequencies can lead to substantial energy savings for network operators. In order to accommodate nonzero turn-on and turn-off times, RTs can be equipped with the needed capability to switch carrier frequencies, in addition to during handoffs, also within a cell. The shutdown of frequencies should be coordinated among cells so that different frequencies are turned off in neighboring cells to aid in interference management. The potential to remove sectoring and employing omnidirectional antenna patterns to reduce power consumption at PAs as well as reduced cell sizes and hierarchical cells should also be investigated.

The use of sleeping modes for small cells operating under macrocells has come under investigation recently, see e.g., \cite{ABH11,SECC12,WQSR13}. Such hierarchical cellular structures will be described in more detail in the next subsection, but for now, let us just state that these structures are being considered mainly for increasing coverage or capacity. Since their constant operation will increase the energy use and therefore energy inefficiency, a sleeping mode for the small cells is being suggested. Hierarchical cellular structures with a sleeping mode is sometimes termed as cognitive small cell networks, see, e.g., \cite{WQSR13}. With a coordinated algorithm adjusting the macrocells, and with an increasing number of small cells, this sleeping mode approach has been reported to provide 10-60\% energy savings in the network, together with capacity increase \cite{ABH11}.
\subsection{Reduced Cell Size and Hierarchical Cells}
\label{sec:HCS}
Today's cellular network %, mainly consisting of cells deployed at approximately every 3-8 miles,
has its origins in a desire to provide voice service to mobile users. An important aspect of the technology is the seamless handover of a user from one RBS to another as it moves through different cells while in motion. As the technology and the applications are changing, there is no need to keep the same architecture. In our discussion of sources of inefficiency in cellular networks in Section~\ref{sec:losssources}, we observed that an RBS providing service to cell sizes currently deployed is highly inefficient. We made the observation that new technologies being introduced will actually make the inefficiencies in RBSs even worse. We also made the observation that users of wireless data services do not necessarily have high mobility patterns. According to a well-publicized study, while in 2008, 54\% of the total cellular network traffic originated indoors, in 2015 this number will increase to 74\% \cite{HB08}. Everyday experience would lead to the estimate that only a minority of the traffic generated outdoors originates from high mobility vehicles, most being originated by pedestrians or by users with no mobility. Since only high mobility users actually need the conventional, energy-inefficient cellular architecture, it makes sense to reevaluate this architecture for future needs.

Based on the discussion in Section~\ref{sec:losssources}, one can reach the conclusion that it makes sense to reduce cell sizes, except for users with high mobility who are moving at vehicular speeds. Users with vehicular speeds can be served by cells of conventional large sizes, whereas users who do not have high mobility can be served by an underlay cellular network, covering the same geographical area but with smaller size cells, as shown in Fig.~\ref{fig:umbrella}. Smaller size cells were considered previously, although not because of the concerns of energy efficiency, but because of the concerns of higher throughput needs in a given area. For the same reason, they are under consideration again, although with a different architecture. We will first explain a number of relevant terms. A macrocell is the conventional cell, served by an RBS and the antenna is installed on a mast or a building above average rooftop level. Its radius is specified over a wide range \cite{ITU-R94,ITU-R09}. Currently, this radius can be considered to be a few kms, although in LTE-Advanced, it is specified to be as low as 500 m for urban macrocells and 1.7 km for rural macrocells \cite{ITU-R09}. In the late 1980s, because of the expectation of increased traffic, the concept of smaller size cells were introduced \cite{SP85,Steele85,Greenstein92,Sarnecki93,SWW95}. These cells have antennas about the average roof level. They are considered to have radii smaller than macrocells, and are usually termed microcells. Then, for indoor applications, the concept of smaller cells with a coverage radius of about 50-100 feet was introduced. These cells are typically called picocells. Finally, due to the increase of data traffic originating from indoors, femtocells, which cover a residence or a small size business were introduced. Femtocells operate in the licensed spectrum of the service provider and are connected to the service provider's network via a broadband Internet connection through a cable modem or a Digital Subscriber Line (DSL) router. Currently femtocells provide service to the subscriber who purchases the service and the femtocell RBS product. They do this by setting up phone numbers allowed to use the cell during service setup. Some femtocells can be configured to provide open access. This can aid the service provider but unless there is any incentive, the femtocell owner would typically prefer closed access. In addition, there is no coordination between the macrocell and the femtocells and therefore handoffs become a challenge in the case of open access \cite{CAG08}. Femtocells can employ the same channels used by the macrocell which can cause interference. By using a number of techniques such as carrier selection, power calibration, or adaptive attenuation, it is possible to mitigate some of the interference, but interference issues remain one of the challenges in femtocell deployment \cite{CAG08}. Femtocells are a practical and inexpensive way for the system providers to offload the large traffic they are facing at a local level, without introducing any significant complexity into their networks. This way, they can still charge a subscriber and satisfy its high data rate needs, while using the same licensed band and keeping their customers. The absence of coordination with the macrocell is key to their operational advantage since they can be introduced without any hardware or software change in the macrocell.

On the other hand, the concept of a microcell underlay to serve users with low mobility and the use of macrocells for users with high mobility were introduced and studied earlier, see e.g., \cite{IGG93,YN96,MK00}. The presence of multiple tiers of networks such as a microcell underlay of macrocells is known as Hierarchical Cell Structure (HCS). Unlike femtocells, which are dumb devices as far as coordination with the macrocell is concerned, a cellular network with HCS requires coordination among cells in different tiers. Although descriptions of HCS are available in a number of cellular standards, and despite the large body of research on the subject, there have not been many commercial networks with large scale HCS systems \cite{Qualcomm07}.

Currently, a number of service providers support localization, or determination of the user's location. This is achieved by using one of three alternatives \cite{Localization101}. The first category employs network parameters and additional techniques based on signal strength or time offset information from the RBS to localize the user. The second set of techniques uses techniques such as triangulation from satellites as in the Global Positioning System (GPS) or from RBSs, employing angle of arrival, time of arrival, or similar techniques. The third category employs proximity information to a known reference, such as using known wireless LAN APs. By using these technologies or their combinations, service providers with both GSM and CDMA-based networks can provide localization today \cite{Localization101}. In fact, today, a large number of cell phones already have embedded GPS receivers. Furthermore, it is reported that while 56\% of the cell phone shipments in 2009 had GPS functionality and it was estimated that in the fourth quarter of 2011, this number would be about 80\% \cite{Shein10}. Various techniques can be used for assisting the GPS receiver in an RT to improve its performance in locations where the reception of the satellite signals is weak, e.g., \cite{Localization103,Localization1032}. Methods of localization are part of the WiMAX specifications \cite{VEBC09,WiMAXLBS}. For LTE, methods of localization are in the process of being incorporated into LTE-Advanced. As in WiMAX, the goal is to provide the same kind of location services available in GSM and CDMA networks and to enable next generation location based services which are considered to be revenue generators. The capabilities include localization by a number of methods, and in addition to the location, determination of the user velocity \cite{MSF09}. Of course, an accurate estimation of location with sufficiently frequent updates enables determination of user velocity. In addition, a number of methods to calculate user velocity from received signal parameters have been devised, see e.g., \cite{HM97,HM99,CLCS02,SMAT06,LSOH06} and their references. We can conclude that determination of user velocity is an achievable goal. This is important since we advocate segregating users on the basis of their mobility, or moving speeds.

It is quite clear that small cell deployments are a necessity for achieving power savings in cellular networks. Cell sizes that cover a city block or a residential neighborhood can efficiently provide high transmission rates to static and pedestrian users. Then, macrocells can be used for subscribers moving at vehicular speeds. A network using APs placed on lamp posts, providing Internet service is not an unrealistic goal. Such a network, known as Ricochet, was designed
and put in place by Metricom Inc. in 1994 and was active until 2006 \cite{Ricochet}. It provided wireless Internet access in 16 major metropolitan areas, serving about 51K subscribers in 2001. It had relatively slow transmission rates. Its APs %, shown in Fig.~\ref{fig:ricochet},
also functioned as repeaters, providing backhaul capability to transmit user data to an Internet point of presence. Backhaul is an important consideration in picocellular networks. We are now at a point new technological capabilities could provide broadband backhaul opportunities in such networks.
\begin{comment}
\begin{figure}
\begin{center}
\includegraphics[width=60mm]{ricochet.eps}
\caption{An AP and repeater employed by Metricom's Ricochet wireless Internet access network on a lamp post \cite{Ricochet}.}
\label{fig:ricochet}
\end{center}
\end{figure}
\end{comment}

Given smaller cells, and with a distributed algorithm to dynamically change used power based on traffic, substantial savings in energy use can be achieved. Depending on traffic, different channel allocations can be employed, and some APs can be turned off. A number of studies tried to quantify these savings. For example, \cite{CWP08} states about 60\% energy reduction is possible in urban areas for high data rate demand based on today's technology. Reference \cite{JHHJ10} calculates, only by using a power off feature, depending on the type of deployment, 50-80\% savings. Reference \cite{SOK10} estimates that, by using some intelligence and coordination among RBSs, 80\% savings during a weekday and 95\% during a weekend are possible. With the predicted high user transmission rates, these savings are likely to be even more substantial.

To briefly show the optimization problem one may encounter and solve, assume that one has a deployment of macrocells ${\cal M}$ and one wishes to build an overlay network of picocells ${\cal P}$. We assume the goal is to satisfy an Area Spectral Efficiency value of ${\rm ASE}_{{\cal M}\cup {\cal P}}({\bf r})$ (bits/s/Hz/m$^2$) as in \cite{AG99}, after the introduction of picocells. For a given location ${\bf r}$ and a given deployment of cells, the Signal-to-Interference-plus-Noise Ratio ${\rm SINR}_{{\cal M}\cup {\cal P}}({\bf r})$ can be analytically calculated. This leads to the bandwidth efficiency figure
\begin{equation}
\eta_{{\cal M}\cup {\cal P}} ({\bf r})=\log_2(1+{\rm SINR}_{{\cal M}\cup {\cal P}}({\bf r}))\ {\rm (bits/s/Hz)}
\label{eq:4}
\end{equation}
and by integration and proper scaling, to ${\rm ASE}(A)$ for an area $A$
\begin{equation}
\textrm{ASE}_{{\cal M}\cup {\cal P}}(A)=\frac{\int_A\eta_{{\cal M}\cup {\cal P}}({\bf r})p({\bf r})d{\bf r}}{\int_Ad{\bf r}}\ {\rm (bits/s/Hz/m^2)}
\label{eq:5}
\end{equation}
where $p({\bf r})$ is the probability density function for the location {\bf r}. Then, the optimization problem to solve is
\begin{equation}
\min_{\cal P} \mathbb{E} [{\cal M}\cup {\cal P}]\hspace{5mm} \textrm{subject to}\ \textrm{ASE}_{{\cal M}\cup {\cal P}}(A)\ge S_0\label{eq:6}
\end{equation}
where $\mathbb{E} [{\cal M}\cup {\cal P}]$ represents the energy consumption of the configuration ${\cal M}\cup {\cal P}$ and $S_0$ is the minimum acceptable ASE figure for the area. Although analytical expressions for  $\mathbb{E} [{\cal M}\cup {\cal P}]$ and $\textrm{ASE}_{{\cal M}\cup {\cal P}}(A)$ may be available, this is in general a difficult problem. But the limited number of locations for ${\cal P}$ enables the use of a greedy algorithm or a combinatorial optimization technique. We note that this general approach can be employed for deployment as well as the operation of the HCS. We also note that this general approach can be used for optimization based on measures other than ASE, as well as for different combinations of macrocell and picocell optimizations.

Smaller cells, such as femtocells, will reduce the power used by RBSs as well as RTs. We will now quantify this gain.
%Assume a user $u$ is connected to an RBS $s$.
Assume a user $u$ is connected to an RBS $s$. Let $c'$ and $u'$ denote the interfering RBSs in the downlink and interfering users in the uplink on subchannel $n$, respectively.
%There are other RBSs $c'$ and other users $u'$ that interfere with this communication.
Then, the uplink SINR can be written as
\begin{equation}
\gamma_{us}=\frac{h_{us}p_u}{I_s},
\hspace{5mm}
I_s = \sum_{c'}h_{c's}p_{c'}+\sum_{u'}h_{u's}p_{u'}+\sigma_s^2
\label{eqn-sinrus}
\end{equation}
where $I_s$ is the interference-plus-noise power at the RBS $s$, $p_u$, $p_{c'}$, and $p_{u'}$ are the power levels of the user $u$, RBSs $c'$, and other users $u'$, respectively, and $h_{us}$, and $h_{c's}$, and $h_{u's}$ are the corresponding channel parameters. Assume we wish to achieve a target SINR value of $\gamma_0$. Also, assume that we consider two cells $c$ and $s$ for user $u$ to be connected to. From (\ref{eqn-sinrus}), we can calculate the power levels the user terminal will need to operate at as $p_{us}=\gamma_0I_s/h_{us}$ and $p_{uc}=\gamma_0 I_c/h_{uc}$. Normally, the handover is made on the Reference Signal Received Power (RSRP) measurements. This is achieved by the RBS transmitting a reference signal at a given power level, the power level for this signal received at the user terminal is RSRP. Then,
\begin{equation}
{\rm RSRP}_{cu} = h_{cu}p_c\hspace{5mm}{\rm and}\hspace{5mm}{\rm RSRP}_{su}=h_{su}p_s
\end{equation}
are the RSRP values due to RBSs $c$ and $s$, at power levels $p_c$ and $p_s$, respectively. Conventionally, these values are used for a handover from one cell to another. So, if ${\rm RSRP}_{cu}>{\rm RSRP}_{su}$, then the user initiates a handover from RBS $s$ to $c$. We will call this criterion as the Strongest Cell Based (SCB) criterion. SCB is commonly used, but it does not take interference considerations or energy efficiency into account.

\begin{figure}[!t]
\centering
\includegraphics[width=72mm]{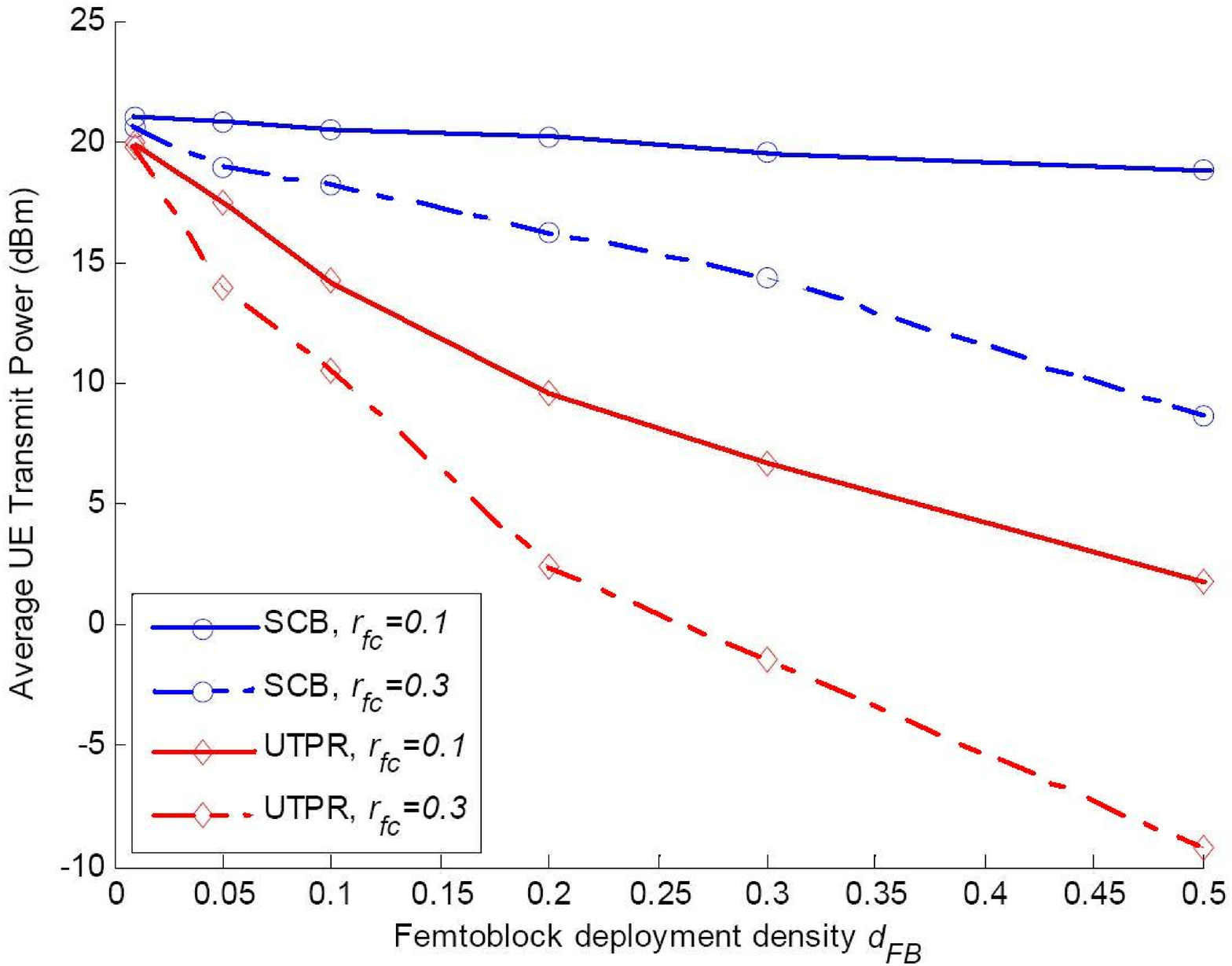}
\caption{Handover policies SCB vs. UTPR \cite{Xenakis}.}
\label{fig:utpr}
\ifCLASSOPTIONtwocolumn
\vspace{-5mm}
\fi
\end{figure}

We will now discuss a revised version of SCB that is based on minimizing the power dissipated by the user terminal, $p_{us}$ or $p_{uc}$. It can be assumed that $h_{uc}\cong h_{cu}={\rm RSRP}_{cu}/p_c$ and $h_{us}\cong h_{su}={\rm RSRP}_{su}/p_s$. Then, we can write
\begin{equation}
p_{us}=\frac{\gamma_0 p_s I_s}{{\rm RSRP}_{su}}\hspace{5mm}{\rm and}\hspace{5mm}p_{uc}=\frac{\gamma_0 p_c I_c}{{\rm RSRP}_{cu}}.
\end{equation}
Assume that the user $u$ is connected to RBS $s$ but there is an RBS $c$ with $p_{uc}<p_{us}$ while achieving $\gamma_0$. We will then require that the user $u$ hand over to RBS $c$. It can be shown that $p_{uc}< p_{us}$ implies
\begin{equation}
%{\rm RSRP}_{su} > {\rm RSRP}_{cu} \frac{p_c}{p_s} \cdot \frac{I_c-h_{uc}p_{uc}}{I_s+h_{us}p_{us}},\hspace{5mm}
\left({\rm RSRP}_{su}\right)_{\rm dB} > \left({\rm RSRP}_{cu}\right)_{\rm dB} +\left( \frac{p_c}{p_s} \cdot \frac{I_c-h_{uc}p_{uc}} {I_s}\right)_{\rm dB}.
\label{eqn:utpr}
\end{equation}
We will call this criterion User Transmit Power Reduction (UTPR) policy. Note that without the second term on the right hand side of (\ref{eqn:utpr}), it becomes equivalent to the SCB policy. A comparison of UTPR and SCB in a macrocell/femtocell environment, with $r_{fc}$ as the ratio of femtocells to macrocells and $d_{FB}$ as the density of femtoblock deployment, is given in Fig.~\ref{fig:utpr}. This figure is based on a simulation study using the criterion (\ref{eqn:utpr}) \cite{Xenakis}.
It assumes that the macrocells operate at a 5 MHz band, while the femtocells can uniformly pick one of the three 5 MHz bands. Two of these bands are adjacent to the band that the macrocell operates at, while the third coincides with it. This corresponds to a mixed strategy in which there is a shared spectrum that the macrocells and femtocells operate at and there are two protected bands for the femtocells.
An increased $r_{FC}$ or $d_{FB}$ corresponds to an increased number of femtocells and users. As expected, an increased $r_{fc}$ or $d_{FB}$ results in lower user transmit power. Note that the reduction in power level is substantial. This policy reduces the power dissipated by the RBS as well. This means that small cell densification can offer energy savings if required handover policy changes are made, such as the one in (\ref{eqn:utpr}). With these changes, service providers can leverage and justify the deployment costs for these small cells.

Smaller cells will reduce energy consumption substantially. Small cells also serve the purpose of providing higher capacity. We advocate the use of picocells on lamp posts, and to accommodate users with high mobility, we advocate an HCS. When similar considerations were made in the past, two obstacles were seen. The first was the backhaul limitations, and the second was the determination of user velocity. With operating frequencies of radios made with CMOS technology approaching 60 GHz where a tremendous amount of bandwidth exists, the backhaul problem can be solved. As elaborated earlier, determining user location and velocity is becoming increasingly more common with the goal of supporting location-based services.

We would like to add that currently small cells and HCS, under the name of heterogeneous networks, are under consideration by the research community and the standards bodies. However, the main interest is still in increasing throughput. We suggest that, for these architectures, energy efficiency and mobility management should be part of the investigation and standardization efforts.
\subsection{Energy Efficiency of Mobile Units}\label{sec:mobiles}
We stated earlier that 80\% of the total power in cellular networks is consumed at RBSs. For that reason, we concentrated our discussion on RBSs only, and so far ignored any efficiency issues with RTs. However, there are a number of considerations for RTs. According to Moore's law, processor capabilities double approximately every two years, which in turn causes more traffic transmitted and received by RTs. With increasing processor capabilities and increasing traffic, RTs have to consume more power. However, this situation immediately presents a dilemma. While, due to Moore's law, processor speeds increased by about two orders of magnitude in the recent ten years, during the same period, battery capacity has only increased by 80\% \cite{EnergyTrap}. Furthermore, during the same period, the output power level of a typical RT has doubled, with another doubling is expected within the next ten years \cite{EnergyTrap}. These two factors result in a power bottleneck for the mobile RTs. Consequently, users will have to constantly search for power outlets, a phenomenon not uncommon even today, and was termed as the ``energy trap'' in today's cellular networks \cite{FK06}. A number of energy-saving mechanisms such as turning off parts of the processor not being used can be introduced for some energy efficiency in RTs, but the energy trap is difficult to overcome by means of conventional approaches.
In what follows, we identify two types of solutions. One of them
mitigates the energy trap through protocol changes and the second one
through cognitive and cooperative solutions.
In Section~\ref{sec:losssources}, we briefly mentioned that DRX is used to introduce
micro-sleep cycles in the LTE standard. During DRX modes, an RT shuts down
most of its circuitry and does not transmit or receive any packets. In real
applications, DRX parameters are selected based on the application type
and QoS requirements. DRX parameters can be optimized to
introduce energy savings. In \cite{DRX09}, the authors present an
analytical framework to study the optimal DRX mechanisms for various
applications. The results in \cite{DRX09} indicate that 40-45\% energy
savings are possible for video and 60\% savings for voice-over-IP
applications are achievable through optimizing the DRX parameters.
This leads to a longer RT battery life which in turn helps to
mitigate the energy trap problem. The study in \cite{DRX09} also identifies the energy versus delay tradeoff such that the packet delay increases exponentially with the RT savings. This clearly shows that this tradeoff needs to be considered during the operational procedure.

In addition,
approaches based on cognitive radio \cite{SICR1,SICR2} and cooperation strategies \cite{SEA03} can be implemented in order to effectively combat the energy trap. To this end, three alternatives and possible architectures are discussed in \cite{EnergyTrap}. The first is a short-range cooperative network with collaborating cluster of RTs, serving as relays, that transmits to an RBS. While the RBS network is cellular, the RTs communicate among themselves using a short-range networking technology, such as Bluetooth. The second is the use of an HCS for power savings. This structure will be employed not only for energy efficiency, but also for optimizing QoS. In this structure, RT will handover between large and small cells where the handover algorithms will be designed for energy efficiency. The third architecture is a combination of the first and second ones, and is intended as an advanced or a long-term option. In all three architectures, multiple wireless devices are envisaged to take advantage of any underlying physical layer architecture, depending on energy efficiency as well as QoS needs.

We would like to note that although we touched upon the QoS issue together with the energy trap problem, it actually impacts all of the prior discussion. In other words, various algorithms discussed earlier for energy efficiency will clearly have to take the QoS needs of the applications into account. Due to space limitations, we did not elaborate on the QoS aspects of the energy efficiency solutions we discussed.

\section{Challenges Encountered, Lessons Learned, and Future Work}\label{sec:future}
In this section, we will discuss the state of green cellular wireless communications, particularly from the viewpoint of the potential solutions we have discussed in earlier sections. It is fair to start this discussion with the statement that from an academic and research viewpoint, the field is very active. It can also be stated that the level of consciousness on the subject is high in research circles. As discussed earlier, a number of equipment vendors as well as a number of service providers do pay attention to the subject. They do have publicized plans for making future systems more energy-efficient. Yet, as we attempted to discuss in detail in this paper, the gains to achieve are really tremendous, reaching the level of orders of magnitude improvement. Another way of making this statement is that the existing cellular wireless network is tremendously inefficient. As a result, a lot more emphasis to the detail of implementation of next generation of cellular wireless communication systems from hardware to software, from the physical layer of the communications protocol hierarchy to the higher layers, from the design of the network to, and in particular, its operation can be put in place. As we discussed in earlier sections, it is possible to approach this problem from the viewpoint of the adaptive control of an existing physical plant, in the face of time-varying parameters. Such an understanding can provide guidance in terms of the development of algorithms for operations such as traffic-adaptive cells, sleeping modes for the RBS, smaller cells, hierarchical cells, heterogenous wireless networks, etc. In earlier sections, we provided examples of what kinds of improvements are possible. In this section, this time, we will attempt to look at the other side of the coin, and try to explain the challenges encountered, what lessons can be deduced, and what can be done in order to achieve success in this area.

We will first discuss the issues that have to do with the physical layer. The advantages of OFDM for broadband communications, especially over time-varying wireless channels are well-understood. This is the reason OFDM has been chosen in all broadband wireless communications standards. Its disadvantages in terms of its high PAPR and power amplification are also well-understood, as discussed in Sections~\ref{sec:CEOFDM} and \ref{sec:classj}. Based on past investigation, CE-OFDM is a good solution to mitigate PAPR \cite{CL91,CC99,TS02,KMCSB05,Thompson05,TAPZG08}. However, the last reported work on this subject was in \cite{TAPZG08}. Based on its importance, we believe currently there is a good opportunity to understand its properties, potential further advantages, as well as limitations better. Such a study will result in the determination of its suitability as an  energy-efficient modulation technique, perhaps with some tradeoff in terms of bandwidth occupation. Other methods of PAPR reduction have been around for quite some time. Further research on this subject can be justified if substantial new gains are achieved.

A very significant issue with energy efficiency has to do with the PAs. There are newer architectures, such as Class J amplifiers, that provide a solution to the problem of inefficiency with RF PAs. However, these newer amplifiers require different semiconductor processes which have less yield, larger area, and therefore higher cost. It may be possible for equipment vendors to convince service providers that although these new technologies may result in higher CapEx, they will in the long run result in lower OpEx, and therefore can be desirable. Furthermore, they represent a better investment in terms of the environment, and therefore they serve the enlightened self interests of the service providers. We further note that, unlike the case for RTs, the RF PAs at RBSs need not be integrated with the rest of the system, and therefore the argument above can be made more easily.

As discussed in Section~\ref{sec:EETM}, link rate adaptation for energy efficiency can be extremely important. What is more important is the fact that energy efficiency is different than power efficiency. For green communications, the former is ultimately what counts although the latter seems to attract more attention in the literature. Reference \cite{KD10} showed that there is at least an order of magnitude improvement with a link adaptation algorithm designed for energy efficiency, where the model of the channel is an accurate one. We believe more research in this particular field will be extremely useful. For example, how much are the gains when there is a mismatch between the model and the actual channel? We believe this presents a rich set of problems with significant outcomes. We also believe that this research with potentially high impact can be carried out with relatively simple analysis, simulations, and most importantly, measurements based on actual implementations. In addition, when one takes QoS guarantees into consideration, the problem becomes even richer and more relevant. With gains of the order of a magnitude, we hold the belief that there is certainly room for useful work to be carried out in this area.

We will now summarize a number of algorithmic approaches that adapt the cellular wireless network infrastructure to the instantaneous traffic conditions. The techniques we study to this end can be considered to be adaptive algorithms that alter the parameters and characteristics of a physical plant, in this case the cellular wireless network, based on the statistics of its input. As such, the problem we discuss can be considered to be an adaptive control problem, albeit the plant to be controlled and the algorithms to be employed are very complex. Therefore, conventional adaptive control approaches will likely not work in this grand problem. Another scientific term used to describe the general system we are considering is a Cyber-Physical System (CPS), or a system of collaborating computational elements controlling physical entities. Currently, CPS is a highly active research field.

The first set of algorithms operate traffic-adaptive cells, or cells whose coverage area and frequency plans may change according to traffic, as we discussed in Section~\ref{sec:AFP}. In that section, we described a problem which can be cast as an LP formulation. We stated this LP in (\ref{eq:3}), as in \cite{BBDFGHHMPRSWW05}. The solution of this problem, according to \cite{BBDFGHHMPRSWW05}, is by means of solving the LP directly or by a distributed approximation to it. However, this is only one way of characterizing the problem and providing a solution for it. In general, there are a number of different ways this problem can be cast as, and there are a number of different solutions to it. A fundamental direction for the research community, therefore, is to investigate these solutions and the various tradeoffs among them.

The use of relays is an interesting, and we believe a challenging, subject regarding increasing energy efficiency in cellular networks. This topic is popular for increasing the capacity in cellular networks, but it is not clear how to use them for energy efficiency, although a number of publications, e.g., \cite{RDynamicResourceAssignment,RMadanEnergyEfficient,RHanzo,RHanGreen} have established that a gain can exist. We believe this subject requires a wider study to determine exactly when and how relays can aid in energy efficiency in cellular networks.

We discussed the importance of multiple antenna techniques in Section~\ref{sec:DA}. Multiple antennas can provide an energy efficiency gain equal to the combined gain of the RBS and RT antennas in both uplink and downlink. Due to space limitations, the number of antenna elements that can be placed on an RT is limited. However, the number of antenna elements that can be placed at the RBS is potentially very large. Therefore, it is possible to achieve extremely high energy efficiency gains by using a directional array at the RBS. In fact, as we discussed in Section~\ref{sec:massivemimo}, employing a very large number of antenna elements at the RBS is currently being investigated from a theoretical and implementation viewpoint under the general term Massive MIMO. An important property of this new architecture is the claim that with this technique there will be an enormous reduction in transmitted power. Whether this holds under realistic channels should be investigated and verified experimentally. Another form of a MIMO system under consideration in the standards is the CoMP technology. As discussed in Section~\ref{sec:comp}, the results presented in \cite{KD10} are quite optimistic for energy efficiency. However, whether they hold under realistic channel conditions should be studied theoretically and via simulations, but most importantly, by employing an experimental setup. Recent field trials on CoMP demonstrated the gap between the theoretical and practical gains \cite{IrmerCoMP11,FettweisJD12,FettweisCoMP11,FettweisWCNC11}. These studies outlined the current challenges that create this gap. In particular, finding the flexible set of cooperating base station clusters, synchronization, channel estimation and outdated CSI, backhaul latency and capacity issues, efficient feedback compression algorithms, and manageable complexity are the major challenges ahead. These issues need to be addressed in the near future in order to fully exploit the achievable gains.

The technique of sleeping modes for RBSs, where, due to light traffic, transmission by the RBS in a cell is either completely or partially turned off, will obviously generate significant energy savings. A number of references, such as \cite{TRKP10} and \cite{SE10}, suggest that by partially turning off RBSs, depending on traffic characteristics energy savings of 45-65\% or even more is possible. An LP formulation for the solution exists. There is no doubt that this technique will result in significant savings. The further study required on this subject has to do with how to implement the technique in existing platforms to take advantage of the traffic, and more importantly, how to design the new generation of RBS equipment so that maximum advantages can be realized.

Currently, the deployment of small cells in HCSs, or heterogenous networks, has gained importance due to the potential capacity increases it offers. As we discussed in Section~\ref{sec:HCS}, these deployments can also be used to improve the energy efficiency of cellular wireless networks. We have described potential research directions in this area in Section~\ref{sec:HCS}. For example, mobility-related issues such as determining the location and the velocity of users needs to be included in network design. In general, this area is very rich in terms of its research potential and it is highly recommendable to researchers interested in the subject of green cellular communications.

We note that although the bulk of the energy inefficiency in a cellular wireless communication system resides with the RBSs, it is important to minimize energy inefficiencies within RTs. The main reason for this is not environmental factors such as the release of greenhouse gases, but the mismatch of the speed of increase in complexity of the processors versus that of the lifetime of batteries. This difference is known as the energy trap and if necessary measures are not taken, it will make the operational time of RTs exponentially shorter with time, forcing users recharge their batteries exponentially more often. As we identified in Section~\ref{sec:mobiles}, protocol changes such as introducing discontinuous transmit or receive cycles, and cooperative or cognitive transmissions need to be studied in order to mitigate the energy trap problem in RTs.

Video traffic is  expected to be dominant in future cellular wireless networks \cite{Cisco13}. When this is the case, there is a natural question as to whether it is possible to handle it in an energy-efficient way. This question was investigated since the early 2000s. Early works employed techniques from source coding, channel resource allocation, error concealment, and energy allocation with the main purpose of reducing the energy consumption in RTs, see e.g., \cite{LEWG03,KZEB05,KEZBP05}. There is some more recent work in this general area, see \cite{MHPS13} and its references. As a recent example employing this general approach, we refer the reader to \cite{RP13} which states their method reduces energy consumption by 50\% compared to conventional H.264/AVC (Advanced Video Coding). This general approach and similar techniques to improve energy efficiency of RTs are important to mitigate the energy trap problem as described above. However, as discussed earlier, the big majority of the energy inefficiency in cellular networks is in RBSs. Therefore, the question of if it would be possible to attain major energy savings in RBSs for video traffic is important. There are three recent approaches we will discuss to that end. The first such approach is based on the development of energy-efficient video transmission schemes, taking into account a power model for circuit elements, studied in \cite{LRC09}. Three techniques are described in \cite{LRC09}, we will concentrate on what is described as Client-Buffer-Related Energy-Efficient Video Transmission (CBEVT) since it provides the largest RF energy savings among the three. The work models power consumption in various analog blocks both at the transmitter and the receiver in the transmit and idle/receive modes and calculates the dissipated energy in terms of transmission parameters, such as modulation characteristics. By carefully adjusting the occupancy in the buffer and by employing transmission parameters that minimize the RF energy per bit, CBEVT is reported to achieve energy savings of up to 85\%. Clearly, this is a very significant gain. The simulation results are for an 802.11b network. Investigation of a similar technique for next generation cellular wireless systems is an interesting research direction. Another aspect of future research in this area involves a quantification of its computational complexity and its effect on the net energy consumption gain. The second approach employs user mobility to increase energy-efficiency for video and is presented in \cite{KFP13}. The main idea is that the transmissions are most efficient when the RT is close to the RBS. Based on this observation, \cite{KFP13} proposes two techniques which monitor the buffer occupancy and the actual playback time of the video stream in order to make most of the transmission when the RT is close to the RBS. Some simulation results show that as one approaches the cell center, there are significant power efficiency gains. We believe a number of questions remain as to, for example, mobility patterns or the measurement of RT location with respect to that of the RBS. However, the basic idea is an interesting one. As in \cite{LRC09}, this technique is a cross-layer approach, and therefore the complexity increase can be a concern. Finally, we would like to discuss the results in \cite{KVT13}. This reference provides a study of multi-cell versus single-cell transmission, specifically for video over LTE and LTE-Advanced. It introduces a system model based on a Finite-State Markov-Chain (FSMC) to evaluate average service data rates, at the same time introducing new bandwidth and energy performance measures. It employs video broadcasting services  standardized in LTE, known as evolved Multimedia Multicast/Broadacst Service (eMBMS). The main concern of the paper is to evaluate the cell-edge transmission rate and energy gain performance of users in the single-cell case versus multi-cell case where the multi-cell case employs COMP, described earlier. Based on simulation results, the paper concludes that significant gains are possible, although they are not uniform. There are cell-edge locations with two times transmission rate improvement and there are cell locations where with 80\% energy gain. As a conclusion of the results in these three references, we can state that it is possible to introduce new techniques to transmit video in an energy-efficient way in future cellular wireless networks.

It can be expected that QoS considerations, which we did not discuss in detail, will bring a very rich set of alternatives to these algorithms. In addition to the QoS issues, studies of the performance, accuracy, and stability of the algorithms will be intellectually interesting. We would like to see at least some of these algorithms to be implemented in real networks and deliver their theoretical gains.

In this paper, we have discussed a number of techniques to increase the energy efficiency of cellular wireless networks. Clearly, these techniques have the potential to improve the energy efficiency tremendously. Since the field is at its infancy, it can be expected that better algorithms with more energy efficiency gain will be developed. All of this is very impressive. However, an important question remains: Will these techniques be implemented? If no, why not? If yes, how? By their very nature, it is extremely difficult to give accurate answers to these questions. However, we can make a number of general, perhaps somewhat realistic observations. First of all, let us discuss potential roadblocks in general terms. One leading such roadblock has to do with economics. A number of the proposals we discussed may have significant costs. Examples are new power amplifiers employing expensive semiconductor processes or massive MIMO, deployed by using hundreds of antennas as well as their RF chains. We believe that the expected enormous increases in traffic can justify service providers to invest in new equipment. In fact, the industry is beginning to discuss rolling out a new generation of equipment around the year 2020. Some equipment vendors and service providers already call this new generation of equipment Fifth Generation, or 5G. What exactly 5G will be is under investigation today. With the expectation of orders of magnitude more traffic, and with the expectation that about 2020 is the year when 4G equipment will become obsolete, service providers will need to invest in new, much higher capacity equipment. We believe that the discontinuity that will take place at that time will likely justify the costs associated with a number of new capabilities to be introduced. For example, based on the unpublished presentations in \cite{GC13-1,GC13-2,GC13-3}, the industry seems to have already accepted the concept of massive MIMO and its associated costs. This discontinuity also provides an opportunity to overcome problems of deployment issues and operational stability. Another roadblock may be the fact that the new algorithms will result in additional complexity. The complexity increase may be absorbed by more capable integrated circuits of the future, but it may be stated that there will be an increased energy dissipation due to these algorithms and it may be possible that these will offset the gains due to the result of the algorithms. Our response is based on a comparison of the energy inefficiencies in today's networks we discussed versus the increase in complexity and its consequent energy dissipation. We believe that the former can be expressed in terms of orders of magnitude, while the latter appears to be incremental. Yet, the final determination will have to be based on a quantitative analysis. The time to carry out this analysis is when all of the algorithms have been fine tuned, however. Next few years provide an opportunity to engage in this algorithm development and its analysis so that it can be determined whether the algorithms can be implemented in 5G. In reality, there are several questions that need to be answered before an algorithm is adopted for implementation and deployment. In this paper, we have only been able to discuss the first results of a number of algorithms. We will list some of the subsequent questions as potential research directions next. While the algorithms become more mature, questions such as them will have to be answered next. What is the dynamic behavior, e.g., delay, stability temporal or mobility-based algorithms such as sleep modes? What are the QoS impacts of the algorithms? Under what scenarios the algorithms become beneficial and which scenarios make them non-beneficial? At this point we do not have the answers to these questions, but would like to point out that these and similar questions are the relevant ones for adaption and deployment of the algorithms by the industry. 
\section{Conclusion}\label{sec:conclusion}
ICT industry generates 2-4\% of all of the Carbon footprint generated by human activity, equal to about 25\% of all car emissions and approximately equal to all airplane emissions. With the increased use of the Internet, this trend is going to increase. For wireless cellular networks, the expected increase in traffic is even more than the Internet as a whole, doubling approximately every year. Due to the proliferation of smart phones, tablets, social networks, and mobile video, this trend is expected to hold for a long time.

Designs of cellular wireless networks were based on large user throughput and high service provider capacity, without any considerations for power or energy efficiency. In order to address these concerns, we identified the sources of energy inefficiency in these networks. To that end, it is known that most of the energy inefficiency is in the RBSs. In RBSs, the PA is a large source of inefficiency. A major source of inefficiency in current cellular wireless networks is the large link distances. This structure, although originally designed for mobile users and telephony, is increasingly being used for users in fixed locations but with large traffic demands. A two-tier network where mobile users with vehicular speeds are supported by means of a large cell and users in fixed locations or at pedestrian speeds by means of smaller underlay cells can be solution for this problem. Algorithms for the operation and handoffs should be designed for such networks. Another major source of energy inefficiency that can be exploited is due to daily and weekly traffic patterns. These variations are major, showing major differences between say, day and night. It can be speculated that a number of algorithms, such as turning off RBSs with low usage and handing off their users to RBSs with large radii, can be devised to remove this inefficiency. Finally, we note that while the expectations of traffic increase in cellular networks is by about 66\% per year, the expected data revenue increase is only about 6-11\%. For this reason, reduction of OpEx by improving the energy efficiency in such networks will be extremely important for service providers.

It is relatively straightforward to calculate that by considering all of the energy gains considered in the paper, it is possible to achieve several orders of gain in energy efficiency. We conservatively estimate this to be at least two orders of magnitude. Obviously, this is a substantial figure. We strongly believe that, the engineering community, and especially the standards bodies, should strive to achieve the maximum possible gain in the face of exponentially increasing traffic demands from cellular networks and the subsequent potential danger of more damage to our environment due to the release of substantial amounts of greenhouse gases.

\section{Acknowledgement}
The authors would like to thank the authors who have given permission to use figures from their publications, as indicated by citations in figure captions of this paper. They would also like to thank
the Editor-in-Chief Dr. Ekram Hossain and six anonymous reviewers. Their comments and suggestions have augmented and improved our original manuscript. 

\bibliographystyle{mybibstyle}
\small
\bibliography{IEEEabrv,bib/GreenComm}
\end{document}